\documentclass[aps,prx,twocolumn,superscriptaddress,floatfix,longbibliography,10pt]{revtex4-2}

\usepackage[utf8]{inputenc}
\usepackage[T1]{fontenc}
\usepackage{microtype}
\usepackage{amsmath,amsfonts,amssymb,mathtools}
\usepackage{graphicx}
\usepackage{bm}
\usepackage{siunitx}
\usepackage{setspace}
\usepackage{lineno}
\usepackage{enumitem}

\newcommand{\ddr}[2]{\frac{\delta #1}{\delta #2}}
\newcommand{\tavg}[1]{\left\langle #1 \right\rangle}

\newcommand{\ldravg}[1]{\left\langle #1 \right\rangle_{\bm{L},\bm{D},\bm{R}}}
\newcommand{\pin}[1]{\int \mathcal{D} #1}
\newcommand{\pinh}[1]{\int \mathcal{D} \hat{#1}}

\newcommand{\spt}{\star}
\newcommand{\ldrmavg}[1]{
  \left\langle #1 \right\rangle_{
    \begin{subarray}{l}
      \bm{L}, \bm{R}, \\
      \bm{D} \setminus \{D_{\hat{0}}, D_{\hat{0}'}\}
    \end{subarray}
  }
}
\newcommand{\conj}{\hspace{-1in}\text{conjugate: }}

\usepackage{hyperref}
\newcommand{\figref}[2]{Fig.~\hyperref[#1]{\ref*{#1}(#2)}}
\newcommand{\Figref}[2]{Figure~\hyperref[#1]{\ref*{#1}(#2)}}
\newcommand{\secref}[1]{Sec.~\ref{#1}}
\newcommand{\eref}[1]{Eq.~\eqref{#1}}
\newcommand{\Eref}[1]{Equation~\eqref{#1}}
\newcommand{\aref}[1]{Appendix~\ref{#1}}

\begin{document}

\title{Connectivity Structure and Dynamics of Nonlinear Recurrent Neural Networks}

\author{David G. Clark}
\thanks{Contact author: \href{mailto:dgclark@fas.harvard.edu}{dgclark@fas.harvard.edu}}
\affiliation{Zuckerman Institute, Columbia University, New York, New York 10027, USA}
\affiliation{Kavli Institute for Brain Science, Columbia University, New York, New York 10027, USA}
\thanks{Present address: Kempner Institute for the Study of Natural and Artificial Intelligence, Harvard University, Cambridge, Massachusetts 02138, USA}

\author{Owen Marschall}
\affiliation{Zuckerman Institute, Columbia University, New York, New York 10027, USA}

\author{Alexander van Meegen}
\affiliation{Center for Brain Science, Harvard University, Cambridge, Massachusetts 02138, USA}
\thanks{Present address: School of Life Sciences and School of Computer and Communication Sciences, École Polytechnique Fédérale de Lausanne, 1015 Lausanne, Switzerland}

\author{Ashok Litwin-Kumar}
\thanks{Contact author: \href{mailto:a.litwin-kumar@columbia.edu}{a.litwin-kumar@columbia.edu}}
\affiliation{Zuckerman Institute, Columbia University, New York, New York 10027, USA}
\affiliation{Kavli Institute for Brain Science, Columbia University, New York, New York 10027, USA}

\date{\today}

\begin{abstract}
Studies of the dynamics of nonlinear recurrent neural networks often assume independent and identically distributed couplings, but large-scale connectomics data indicate that biological neural circuits exhibit markedly different connectivity properties. These include rapidly decaying singular-value spectra and structured singular-vector overlaps. Here, we develop a theory to analyze how these forms of structure shape high-dimensional collective activity in nonlinear recurrent neural networks. We first introduce the random-mode model, a random-matrix ensemble related to the singular-value decomposition that enables control over the spectrum and right-left mode overlaps. Then, using a novel path-integral calculation, we derive analytical expressions that reveal how connectivity structure affects features of collective dynamics: the dimension of activity, which quantifies the number of high-variance collective-activity fluctuations, and the temporal correlations that characterize the timescales of these fluctuations. We show that connectivity structure can be invisible in single-neuron activities while dramatically shaping collective activity. Furthermore, despite the nonlinear, high-dimensional nature of these networks, the dimension of activity depends on just two connectivity parameters---the variance of the couplings and the effective rank of the coupling matrix, which quantifies the number of dominant rank-one connectivity components. We contrast the effects of single-neuron heterogeneity and low dimensional connectivity, making predictions about how z-scoring data affects the dimension of activity. Finally, we demonstrate the presence of structured overlaps between left and right modes in the \textit{Drosophila} connectome, incorporate them into the theory, and show how they further shape collective dynamics.
\end{abstract}

\maketitle

\section{Introduction}
\label{sec:introduction}

The collective activity of a high-dimensional nonlinear system is determined by the structure of the interactions between its elements, but the mapping from structure to activity is generally analytically tractable only for limited forms of structure. In systems neuroscience, structure in synaptic connectivity controls the function of a neural system by determining what patterns of activity are produced by populations of neurons. This paper studies this mapping in the context of nonlinear recurrent neural networks, used widely in neuroscience as models of neural-circuit dynamics \cite{barak2013fixed, mante2013context, sussillo2015neural} and in machine learning as systems for sequence processing \cite{hochreiter1997long, sutskever2014sequence, cho2014learning, krishnamurthy2022theory}. While these models omit many aspects of actual neural circuits, they capture several of their fundamental features, including their large scale, nonlinear units, and recurrent interactions. Functionally, a key property of such systems is that, like real neural circuits, they can generate rich time-varying activity in the absence of external input \cite{rajan2010stimulus, engelken2023lyapunov}.

Theoretical studies of nonlinear recurrent neural networks, including the pioneering work of Sompolinsky \textit{et al.} (Ref.~\cite{sompolinsky1988chaos}), often focus on the case of independent and identically distributed (i.i.d.) couplings, or simple variants thereof \cite{marti2018correlations}. Such disordered networks produce high-dimensional chaotic activity, reminiscent of asynchronous cortical activity observed \textit{in vivo} \cite{sompolinsky1988chaos}. Initializing recurrent neural networks so that they exhibit chaotic dynamics has also been shown to facilitate subsequent learning \cite{sussillo2009generating, laje2013robust, aljadeff2015transition, depasquale2018full, Asabuki2025}. 

Experimental access to large-scale connectivity structure creates an opportunity and imperative to develop theories that explain models beyond those with i.i.d.\ couplings. Recently, large-scale synaptic reconstructions of neural circuits, or ``connectomes,'' have become available in multiple species, including the full brain of \textit{Drosophila melanogaster} \cite{scheffer2020connectome} and increasingly large portions of mammalian circuits \cite{loomba2022connectomic,shapson2024petavoxel,microns2025functional,tavakoli2025light}. These large-scale connectomes reveal that neural circuits deviate substantially from the i.i.d.\ assumption, as previously demonstrated via the presence of small-scale motifs \cite{song2005highly}. In particular, in accordance with real-world networks more generally \cite{thibeault2024low}, neural circuits appear to exhibit low-rank structure in which the connectivity is well described by fewer rank-one components than the network size (we demonstrate this explicitly for the \textit{Drosophila} connectome in \secref{sec:fly_analysis}). In neural-network models, recurrent or otherwise, such structure emerges naturally in several contexts, including when connections depend on distance in physical or feature spaces \cite{akjouj2024complex, wang2023geometry, pezon2024linking}, or when networks are trained on tasks with low dimensional structure \cite{schuessler2020interplay, martin2021implicit, clark2025symmetries}. 

In parallel with the development of large-scale connectivity maps, recent advances in recording technologies, including silicon probes \cite{trautmann2025large} and large-scale calcium imaging \cite{stringer2019high, stringer2019spontaneous}, now enable the activities of many neurons to be monitored simultaneously, with the latest datasets including hundreds of thousands of neurons or more \cite{manley2024simultaneous}. This technological progress has enabled researchers to characterize collective properties of neural activity rather than focusing solely on single-neuron responses. These collective properties capture how neurons interact and reveal distributed computational processes that are visible only with high-yield recording technologies \cite{chung2021neural}. Such population-level features are typically analyzed using dimensionality reduction techniques like principal components analysis \cite{cunningham2014dimensionality, gao2015simplicity, trautmann2019accurate, meshulam2024statistical}.

\subsection{Relating connectivity and collective-activity structure analytically.}
We currently lack analytical tools to predict relationships between connectivity structure, described by connectomics or other datasets, and collective activity, described by large-scale neural recordings. Dynamical mean-field theory (DMFT) is a theoretical tool from statistical physics widely used to analytically characterize activity in large nonlinear recurrent neural networks. Although recent advances in DMFT have extended these techniques to describe collective features in networks producing high-dimensional chaotic activity \cite{clark2023dimension}, these calculations have been limited to i.i.d.\ connectivity, or connectivity with simple correlations between reciprocal couplings. Extending such calculations to more complicated connectivity structures rapidly becomes unwieldy. Consequently, such analyses have not been performed for networks whose connections are constrained by the statistics of connectomic or other datasets, limiting our ability to leverage these datasets for theoretical insight. 

One tractable approach has been to study networks with \textit{very} low dimensional connectivity and thus activity (specifically, in the limit where the rank of the coupling matrix remains finite while the network becomes large; see \secref{sec:comp_to_lowrank}). However, connectomes, like that of \textit{Drosophila} considered in \secref{sec:fly_analysis}, do not support this finite-rank assumption. Furthermore, while neural activity recorded in experiments is often low dimensional, this may be inherited from the low dimensionality of experimental tasks; in more complicated tasks or spontaneous states, activity is typically higher dimensional \cite{stringer2019high, stringer2019spontaneous, manley2024simultaneous}.

To address these limitations and challenges, this paper makes two contributions.
\begin{enumerate}[leftmargin=1em, itemsep=-0.2em]
    \item {A tractable model of neural-network coupling matrices that we call the random-mode model.} Motivated by the \textit{Drosophila} connectome considered in \secref{sec:fly_analysis}, we introduce the random-mode model in \secref{sec:def_of_rmm}, in which the couplings are generated as a sum of rank-one outer products of random left and right modes with component-specific strengths. Unlike well-studied low-rank recurrent neural-network models, we scale the rank with the network size (\secref{sec:comp_to_lowrank}). We show in \secref{sec:svd} that parametrizing these strengths provides control over the singular-value spectrum of the coupling matrix.
    \item {A new path-integral calculation of a specific four-point function of network activity,} originally defined by Clark \textit{et al.} (Ref.~\cite{clark2023dimension}), that captures features of collective activity including its dimensionality. We review these and other summary statistics in \secref{sec:model_and_summary_stats}. In 
    Secs.~\ref{sec:relating}--\ref{sec:soln_of_random_mode_model},
    we describe the calculation, based on fluctuations around the saddle point in a path integral, and apply it to the random-mode model. In \secref{sec:interp}, we use the solution to study how features of the coupling matrix control collective activity.
\end{enumerate}

Additionally, in \secref{sec:low_eff_rank}, we analyze notable limiting cases of the random-mode model. In \secref{sec:heterogeneity}, we study a generalization of the model featuring heterogeneity among single-neuron properties and contrast these effects with those of low-rank structure in connectivity. Finally, in \secref{sec:LR_correlations}, we incorporate structured overlaps between the left and right modes into the model, demonstrating their presence in the \textit{Drosophila} connectome and showing how they further shape collective dynamics.

\section{Random-mode model}
\label{sec:rmm}

\subsection{Spectral structure of neural-circuit connectivity}
\label{sec:fly_analysis}

\begin{figure*}
    \centering
    \includegraphics[width=7in]{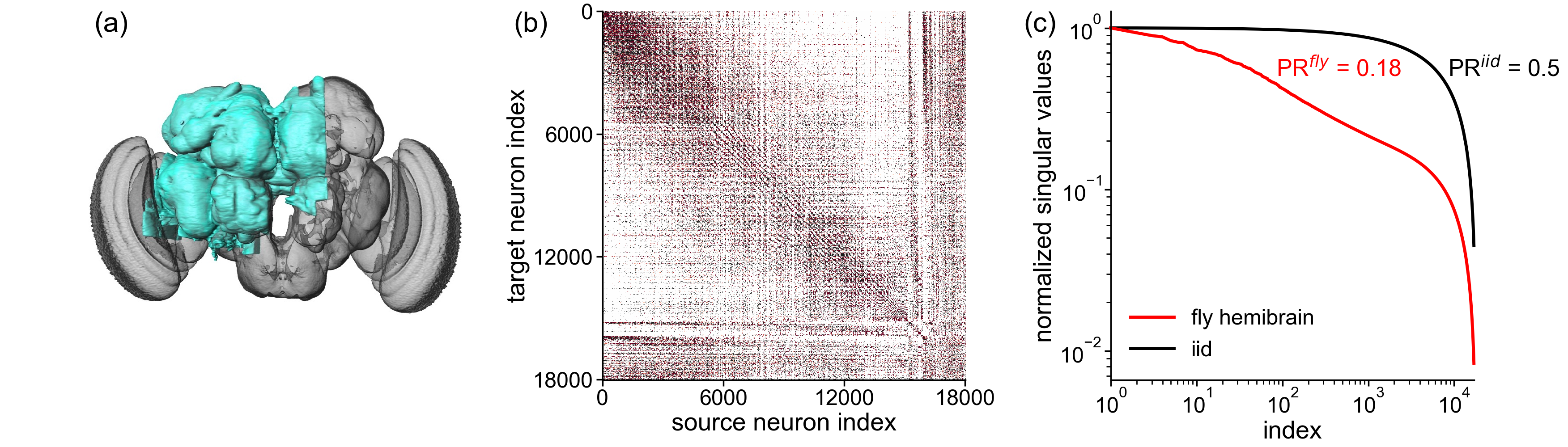}
    \caption{Analysis of \textit{Drosophila} central-brain connectome. (a) Volume of fly brain for which reconstruction was performed (blue; reproduced from \cite{scheffer2020connectome}). (b) Normalized coupling matrix (elements summed within $10 \times 10$ blocks to aid visualization). (c) Singular-value spectra of the normalized coupling matrix (red) and an i.i.d.\ random matrix (gray). The fly connectome exhibits a smooth spectrum that decays quickly, corresponding to a reduced participation ratio. $N = 18028$ neurons.}
    \label{fig:fly_fig_1}
\end{figure*}

Let $\bm{J}$ denote an $N \times N$ weighted, directed coupling matrix among $N$ neurons. Many theoretical studies have examined single-neuron \cite{sompolinsky1988chaos} and collective \cite{clark2023dimension} properties of activity in nonlinear recurrent neural networks with i.i.d.\ $J_{ij}$. However, the coupling matrices of real-world networks, including neural circuits, often exhibit approximate low-rank structure that deviates dramatically from the i.i.d.\ assumption. To illustrate this, we analyzed a central-brain connectome of the fruit fly \textit{Drosophila melanogaster} using the singular-value decomposition (SVD), which decomposes any matrix as a sum of rank-one components, with each component's strength given by the corresponding singular value [Figs.~\ref{fig:fly_fig_1}(a) and \ref{fig:fly_fig_1}(b); details in \aref{sec:fly_appdx}]. Specifically, the SVD decomposes $\bm{J}$ in the form
\begin{equation}
    \bm{J} = \sum_{a=1}^M S_a \bm{u}_a \bm{v}_a^T,
    \label{eq:svd}
\end{equation}
where, for $a = 1, \ldots, M$, $S_a > 0$ are the singular values; $\bm{u}_a$ and $\bm{v}_a$ are $N$-dimensional left and right singular vectors, respectively; and $M$ is the rank of $\bm{J}$.

The ranked singular values of the \textit{Drosophila} connectome decay much more rapidly than those of an i.i.d.\ matrix of the same size, which follow a universal distribution independent of single-element statistics [\figref{fig:fly_fig_1}{c}; similar analysis appears in Thibeault \textit{et al.} (Ref.~\cite{thibeault2024low})]. We quantify this rapid decay using the participation ratio (PR) of the squared singular-value spectrum, $\text{PR}^S = (\sum_{a=1}^M S_a^2)^2 / \sum_{a=1}^M S_a^4$. The connectome's participation ratio is 0.18, substantially lower than the value 0.5 characteristic of i.i.d.\ matrices, confirming that this biological network exhibits a much more concentrated low-rank structure than would arise from i.i.d.\ couplings.

\subsection{Definition of the random-mode model}
\label{sec:def_of_rmm}

\begin{figure*}
    \centering\includegraphics[width=7in]{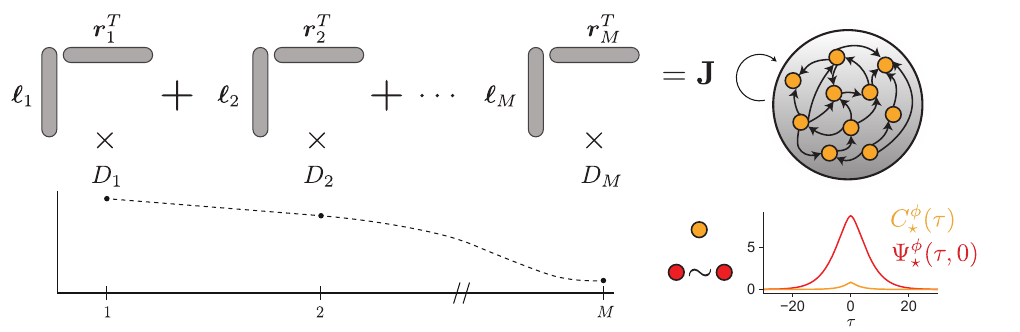}
    \caption{Schematic of the random-mode model. Upper: couplings $\bm{J}$ are generated as a sum of outer products, ${\bm \ell}_a {\bm r}^T_a$, with component strengths $D_a$. Lower: the two-point function $C^\phi_\spt(\tau)$ and four-point function $\Psi^\phi_\spt(\tau)$ are calculated in terms of the statistics of $D_a$. The two-point function depends only on the effective gain $g_\text{eff}$, while the four-point function depends on both $g_\text{eff}$ and $\text{PR}^D$, the effective dimension of the connectivity determined by the $D_a$ distribution.}
    \label{fig:ldr_cartoon}
\end{figure*}

Both the singular-value spectrum and, as we will show later, the overlaps of the left and right singular vectors of the \textit{Drosophila} connectome deviate from the predictions of an i.i.d.\ coupling matrix. We therefore introduce the random-mode model, a generative model for coupling  matrices that allows us to capture these properties within a random-matrix ensemble (Fig.~\ref{fig:ldr_cartoon}).

The random-mode model has the same mathematical form as an SVD, but is a statistical generative process for $\bm{J}$ rather than a matrix factorization. The model generates a coupling matrix of the form
\begin{equation}
    \bm{J} = \sum_{a=1}^M D_a \bm{\ell}_a \bm{r}_a^T,
    \label{eq:rmm_sum_def}
\end{equation}
where, for each $a = 1, \ldots, M$, $D_a > 0$ are the component strengths; $\bm{r}_a$ and $\bm{\ell}_a$ are $N$-dimensional right and left modes, respectively; and $M$ is the number of components, where $\text{rank} \leq \text{min}(N, M)$. We now describe the properties of these quantities. See \aref{sec:glossary_of_terms} for a glossary of terms. 

\subsubsection{Component strengths.} The component strengths $D_a$ are analogous to the singular values $S_a$ of $\bm{J}$. We model these quantities deterministically with the requirement that they are defined for arbitrarily large $M$ and that empirical averages $M^{-1} \sum_{a=1}^M f(D_a)$ converge to limiting values, denoted $\tavg{f(D)}_D$, as $M \to \infty$.

\subsubsection{Left and right modes.} The left and right modes $\bm{\ell}_a$ and $\bm{r}_a$ are analogous to the left and right singular vectors $\bm{u}_a$ and $\bm{v}_a$ of $\bm{J}$. We assume they are sampled i.i.d.\ across the neuron index $i$:
\begin{equation}
    P(\{\bm{\ell}_a, \bm{r}_a\}_{a=1}^M) = \prod_{i=1}^N P(\{\ell_{ai}, r_{ai}\}_{a=1}^M),\label{eq:iid_over_i}
\end{equation}
where $P(\{\ell_{a}, r_{a}\}_{a=1}^M)$ is the joint distribution over the $2M$ mode components, which must be specified. We assume that this distribution has the following moments fixed:
\begin{align}
    &\tavg{\ell_a} = \tavg{r_a} = 0, \label{eq:zero_mean_assum} \\
    &\tavg{\ell_a \ell_b} = \tavg{r_a r_b} = \frac{1}{N}\delta_{ab}. \label{eq:sets_scaling}
\end{align}
This yields orthonormality of modes in expectation,
\begin{equation}
\begin{split}
    \tavg{\bm{\ell}_a^T \bm{\ell}_b} &= \delta_{ab}, \\
    \tavg{\bm{r}_a^T \bm{r}_b} &= \delta_{ab},
\end{split}\label{eq:mode_second_moments}
\end{equation}
in analogy to singular vectors, for which orthonormality holds exactly \footnote{This random structure, rather than exact, deterministic orthonormality, offers analytical tractability by eliminating the need to enforce orthonormality---a complicated global constraint; but see also Ref.~\cite{ingrosso2020optimal}, which handles random orthonormal matrices in a regression setting.}.\nocite{ingrosso2020optimal} Individual mode components are $\mathcal{O}(1/\sqrt{N})$. Note that we have not constrained second moments that mix left and right mode components.

\subsection{Rank scaling and comparison to low-rank recurrent neural networks}
\label{sec:comp_to_lowrank}

A crucial modeling choice is how the number of modes $M$ scales relative to the network size $N$. The two main possibilities are $M = \mathcal{O}(N)$ or $M = \mathcal{O}(1)$, corresponding to extensive and intensive rank scaling, respectively. The random-mode model is characterized by extensive rank scaling, where both $M$ and $N$ approach infinity while maintaining a fixed ratio:
\begin{equation}
    \alpha = \frac{M}{N} = \mathcal{O}(1).
    \label{eq:alpha_def}
\end{equation}
By choosing $\alpha$ to be small, or by choosing the spectrum of ranked component strengths to decay rapidly, we can construct connectivity that is low rank or well described by a small number of rank-one components, compared to the number of neurons.

The alternative approach of intensive rank scaling, where $M$ remains finite while $N \to \infty$, leads to what has been termed ``low-rank recurrent neural networks'' in the literature \cite{mastrogiuseppe2018linking, schuessler2020dynamics, beiran2021shaping, dubreuil2022role}. In some instances, the intensive-rank component is added to an i.i.d.\ matrix modeling unstructured ``background'' connectivity \cite{mastrogiuseppe2018linking, schuessler2020dynamics}. Note that, in this paper, when we use ``low rank'' or ``low dimensional,'' we refer to quantities that are small as a fraction of $N$, even if they are extensive; this differs from some prior works (such as those on low-rank recurrent neural networks) where these terms refer to intensive quantities.

The distinction between extensive- and intensive-rank scaling has important implications for specifying the distribution over the $2M$ mode components $P(\{\ell_a, r_a\}_{a=1}^M)$. For finite $M$, one can directly specify this distribution. At large $N$, such a distribution encodes the relative geometry of the left and right modes, since the inner products between modes converge to ($N$ times) the second moments of $P(\{\ell_a, r_a\}_{a=1}^M)$, with negligible fluctuations. A common choice is a $2M$-dimensional multivariate Gaussian distribution. An important insight from prior work on the intensive-rank case is that these overlaps are crucial for shaping the dynamics and computations implemented by the network, as overlaps between connectivity modes dictate how  activity modes interact \cite{mastrogiuseppe2018linking, schuessler2020dynamics, beiran2021shaping, dubreuil2022role}.

Our goal is to go beyond the intensive-rank setup, which produces correspondingly low dimensional activity, to model more sophisticated, higher dimensional dynamics. However, specifying the full distribution $P(\{\ell_a, r_a\}_{a=1}^M)$ in the random-mode model, where $M \to \infty$, is more complicated. First, it must be well defined for all $M$ to allow taking the limit. More fundamentally, we would like to use the random-mode model to construct generative models that capture certain essential spectral features of coupling matrices, over which we have systematic control. Using a joint Gaussian distribution would require specifying $M^2$ parameters which, since $M = \mathcal{O}(N)$, is comparable to defining the full $N \times N$ coupling matrix.

For this reason, this paper specializes to a case where $P(\{\ell_{a}, r_{a}\}_{a=1}^M)$ factorizes across the mode index $a$:
\begin{equation}
    P(\{\ell_{a}, r_{a}\}_{a=1}^M) = \prod_{a=1}^M P(\ell_{a}, r_{a}),
    \label{eq:modes_distribution}
\end{equation}
where $P(\ell_{a}, r_{a})$ must be specified. Moreover, until \secref{sec:LR_correlations}, we further specialize to the case where the left and right modes are fully independent: $P(\ell_{a}, r_{a}) = P(\ell_a) P(r_a)$, with each marginal factor $P(\ell_{a})$ or $P(r_{a})$ being a univariate Gaussian with mean zero and variance $1/N$ \footnote{The Gaussian assumption for individual components is not crucial---our results depend on only first and second moments for large $N$ due to the central limit theorem; higher-order cumulants (scaling appropriately with $1/N$) are subdominant in the path-integral saddle-point approximation.}. In this case, all $2MN$ components specifying the modes are i.i.d.\ with mean zero and variance $1/N$; we refer to this case as ``i.i.d.\ modes.'' In \secref{sec:LR_correlations} we use a nonfactorized form of $P(\ell_{a}, r_{a})$ to specify correlations between $\ell_a$ and $r_a$, introducing $M$ deterministic parameters (in addition to the $M$ component strengths) to specify the left-right correlation for each $a$.

\subsection{Relationship to singular-value decomposition}
\label{sec:svd}

In the SVD, the left and right singular vectors are orthonormal: $\bm{u}_a^T \bm{u}_b = \bm{v}_a^T \bm{v}_b = \delta_{ab}$. In contrast, the left and right modes are orthonormal only in expectation [\eref{eq:mode_second_moments}];
once sampled, there are random $\mathcal{O}(1/\sqrt{N})$ overlaps,
\begin{equation}
\begin{split}
\bm{\ell}_a^T \bm{\ell}_b &= \delta_{ab} + \mathcal{O}(1/\sqrt{N}), \\
\bm{r}_a^T \bm{r}_b  &= \delta_{ab} + \mathcal{O}(1/\sqrt{N}). 
\end{split}\label{eq:lr_approx_orthogonality}
\end{equation}
When the effective rank (defined below) is small and $N$ is large, the component strengths and modes approximate singular values and vectors, respectively, since the effects of the $\mathcal{O}(1/\sqrt{N})$ deviations from orthonormality in the sampled modes are negligible. Indeed, a well-known result in high-dimensional geometry is that finite numbers of random vectors with i.i.d.\ components approximate orthonormal bases at large $N$ \cite{vershynin2018high}. When the effective rank is not small, overlaps between different modes produce a discrepancy between the component strengths $D_a$ and singular values $S_a$. We now characterize this discrepancy analytically.

We denote the $n$th moments and participation ratio of the component-strength distribution by
\begin{align}
    r_n &=\tavg{D^n}_D, \label{eq:D_moment} \\
     \text{PR}^D &= \frac{r_2^2}{r_4}, \label{eq:PRD_def}
\end{align}
from which we define the ``effective rank'' of the couplings:
\begin{equation}
    \text{effective rank} = \alpha \text{PR}^D.
    \label{eq:eff_rank_def}
\end{equation}

The effective rank is related to the participation ratio of the squared singular-value spectrum at large $N$ \footnote{This follows from using Wick's theorem to evaluate the expectations of $\text{tr}\left(\bm{L}\bm{D}\bm{R}^T \bm{R}\bm{D}\bm{L}^T\right)$ and $\text{tr}\left(\bm{L}\bm{D}\bm{R}^T \bm{R}\bm{D}\bm{L}^T \bm{L}\bm{D}\bm{R}^T \bm{R}\bm{D}\bm{L}^T\right)$ for the numerator and denominator, respectively, of $\text{PR}^S$.} through
\begin{equation}
    \text{PR}^S = \frac{\alpha \text{PR}^D}{1 + 2\alpha \text{PR}^D}.
    \label{eq:svd_connection}
\end{equation}
Expanding in small $\alpha\text{PR}^D$, $\text{PR}^S = \alpha \text{PR}^D + \mathcal{O}\bm{(}(\alpha \text{PR}^D)^2\bm{)}$, demonstrating that the component strengths are closely related to the singular values in the low dimensional regime. While we express the analytic results in this paper in terms of $\alpha\text{PR}^D$, the above equation can always be used to translate between $\alpha\text{PR}^D$ and $\text{PR}^S$.

We further quantify this discrepancy in the case where $D_a = 1$ for all $a$. Using methods from free probability theory \cite{pennington2017resurrecting, pennington2018emergence} (\aref{sec:free_prob_calc}; Fig.~\ref{fig:free_prob}), the $M$ singular values are distributed over a range with boundaries
\begin{equation}
    S_{\pm} = \sqrt{ 1 + \frac{5\alpha}{2} - \frac{\alpha^{2}}{8} 
    \pm  \left( 1 + \frac{\alpha}{8} \right)^{3/2} \sqrt{8\alpha} }.
    \label{eq:free_prob}
\end{equation}
For small $\alpha$, $S_{\pm} = 1 \pm \sqrt{2\alpha} + \mathcal{O}(\alpha)$. Thus, while the random-mode model spreads the nonzero singular values over a range, this spread becomes negligible for small $\alpha$. More generally, the distributions of $D_a$ and $S_a$ coincide for small $\alpha \text{PR}^D$.

Having presented a generative model of connectivity, we now present the network model that transforms this connectivity into neural activity. We then define summary statistics that characterize the structure of this activity, allowing us to analyze its dependence on the connectivity.

\section{Network model and summary statistics}
\label{sec:model_and_summary_stats}

\subsection{Recurrent neural network model}
\label{sec:dynamic_model}

We study a recurrent neural network of $N$ neurons. Each neuron $i \in \{1,\ldots,N\}$ is characterized by its preactivation $x_i(t)$ and activation $\phi_i(t) = \phi\bm{(}x_i(t)\bm{)}$, where $\phi(\cdot)$ is a scalar nonlinearity; we use $\phi(x) = \text{erf}\left({\sqrt{\pi} x}/{2}\right)$, which is sigmoid-shaped and has $\phi'(0) = 1$ \footnote{This behaves similarly to the more conventional $\phi(x) = \tanh(x)$ but allows for analytical evaluation of Gaussian integrals.}. The network dynamics are governed by
\begin{subequations}
\begin{equation}
    T[x_i](t) = \sum_{j=1}^N J_{ij} \phi_j(t), \label{eq:eoms}
\end{equation} \vspace{-2em}
\begin{equation}
    \text{where} \quad T[x](t) = (1 + \partial_t)x(t). \label{eq:canonical_choice}
\end{equation}
\end{subequations}
Here, $J_{ij}$ denotes the synaptic coupling from neuron $j$ to neuron $i$, and $T[\cdot]$ is a causal functional that specifies the single-neuron dynamics for which \eref{eq:canonical_choice} is a canonical choice. We expect our results to be agnostic to the specific choice of $T[\cdot]$. We next introduce correlation functions that capture single-neuron and collective activity properties. 

\subsection{Two-point functions and duality of neuronal and temporal covariances}
\label{sec:order_parameters}

\begin{figure}
    \centering
\includegraphics[width=3in]{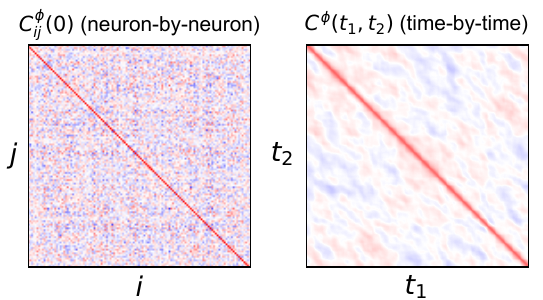}
    \caption{Duality of neuron-by-neuron and time-by-time covariances and its relation to the dimension of activity. Both plots are based on the same simulation of a network of $N=2500$ neurons with $g=2.25$ and i.i.d.\ couplings. The dimension can be computed either by computing the statistics of the off diagonals of the neuron-by-neuron covariance $C^\phi_{ij}(0)$ (left), or by computing the fluctuations away from the translation-invariant mean-field form of the time-by-time covariance $C^\phi(t_1,t_2)$ (right).}
    \label{fig:nrn_time}
\end{figure}

Let $a \in \{x, \phi\}$ denote either preactivations or activations. We define two complementary covariance functions, averaging over either time or neurons,
\begin{align}
    C^a_{ij}(\tau) &= \tavg{ a_i(t) a_j(t + \tau) }_t, \label{eq:tac} \\
    C^a(t_1, t_2) &= \frac{1}{N}\sum_{i=1}^N  a_i(t_1) a_i(t_2).
    \label{eq:nac}
\end{align}
These functions exhibit a form of duality: for any lag $\tau$, the matrix $\bm{C}^a(\tau)$ with elements $C^a_{ij}(\tau)$ indexed by $(i,j)$ and the matrix with elements $C^a(t_1, t_2+\tau)$ indexed by $(t_1, t_2)$ have the same eigenvalue spectrum, up to scaling (to see this, note that both matrices can be computed from a time-by-neuron activity matrix $\bm{A}$ as $\bm{A}^T \bm{A}$ and $\bm{A} \bm{A}^T$, respectively, which have identical spectra) \footnote{This duality is reminiscent of ergodicity theorems, which relate temporal and spatial averages, but differs in a variety of ways, including that it holds even for finite-time intervals, whereas ergodicity concerns infinite-time limits.}.

We also define the neuron-averaged response function, describing the propagation of infinitesimal perturbations:
\begin{equation}
    S^a(t_1, t_2) = \frac{1}{N}\sum_{i=1}^N \left. \frac{\delta a_i(t_1)}{\delta I_i(t_2)} \right|_{\bm{I} = 0},
\end{equation}
where $I_i(t)$ is a source term added to the right-hand side of the network dynamics: $T[x_i](t) = \sum_{j=1}^N J_{ij}\phi_j(t) + I_i(t)$.
Finally, we define ``stationarized'' counterparts,
\begin{equation}
    C^a(\tau) = \tavg{C^a(t,t\pm \tau)}_t, \:\:\:\: S^a(\tau) = \tavg{S^a(t, t-\tau)}_t. \label{eq:stationarized}
\end{equation}

\subsection{Four-point functions, dimension of activity, and principal-component timescales}
\label{sec:pr_sec}

In contrast to the above two-point functions, which characterize single-neuron activity, four-point functions describe features of collective activity. The same four-point functions were used in Ref.~\cite{clark2023dimension}. Here, we motivate these functions in more detail and express them in two ways via the duality between neuronal and temporal covariances. 

To motivate their definition, consider the dimension of activity, which can be quantified as the participation ratio of the spectrum of the equal-time covariance matrix $\bm{C}^a(0)$ \cite{rajan2010inferring, abbott2011interactions, gao2015simplicity, gao2017theory, litwin2017optimal, recanatesi2019dimensionality, engelken2023lyapunov, recanatesi2020scale, hu2022spectrum}. Given the eigenvectors $\bm{v}^a_k$ and eigenvalues $\lambda^a_k$ for $k = 1, \ldots, N$, as one would compute as part of principal components analysis, the participation ratio is
\begin{equation}
    \text{PR}^a = \frac{1}{N} \frac{(\sum_{k=1}^N \lambda^a_k )^2}{\sum_{k=1}^N (\lambda^a_k)^2}.
\end{equation}

To see why $\text{PR}^a$ provides a meaningful measure of dimension, consider the case where $D$ eigenvalues equal a positive constant and the remaining eigenvalues are zero. Then, $\text{PR}^a = D/N$. More generally, when the spectrum exhibits a smooth decay rather than a hard cutoff, the participation ratio identifies the characteristic decay scale (divided by $N$) \cite{gao2017theory}. We compare the participation ratio to alternative measures of effective dimensionality in \aref{sec:appdx_alt_dim}. Without the normalization factor $1/N$, this quantity would vary between $1$ and $N$; with normalization, it varies between $1/N$ and $1$. 

In the limit $N \to \infty$, this quantity could have three qualitatively different behaviors: 
\begin{itemize}[leftmargin=1em, itemsep=-0.2em]
    \item {Subextensive dimensionality:} $\text{PR}^a = 0$. The number of dimensions filled by activity grows sublinearly with $N$. 
    \item {Nontrivial extensive dimensionality:} $0 < \text{PR}^a < 1$. The number of dimensions filled by activity grows linearly with $N$ and not all dimensions are filled equally. 
    \item {Trivial extensive dimensionality:} $\text{PR}^a = 1$. All dimensions are filled equally by activity. 
\end{itemize}

An advantageous property of the participation ratio is that it can be expressed as the ratio of the squared trace and Frobenius norm of $\bm{C}^a(0)$, both of which can be further expressed in terms of matrix elements $C_{ij}^a(0)$,
\begin{align}
    \text{PR}^a &= \frac{\left(\text{tr}\bm{C}^a(0)\right)^2}{\lVert \bm{C}^a(0) \rVert_F^2} \nonumber \\
    &= \frac{C^a(0)^2}{\frac{1}{N}\sum_{i=1}^N C^a_{ii}(0)^2 + \frac{1}{N}\sum_{i \neq j}C^a_{ij}(0)^2 },
\end{align}
where $C^a(0)$ is given by \eref{eq:stationarized}. Assuming that diagonal elements are uniform with negligible fluctuations, i.e., $C^a_{ii}(0) = C^a(0)$ for all $i$ (as occurs, for example, under i.i.d.\ connectivity \cite{sompolinsky1988chaos}), this simplifies to
\begin{equation}
    \text{PR}^a = \frac{C^a(0)^2}{C^a(0)^2 + \psi^a(0,0)}, \label{eq:pr_expr_one}
\end{equation}
where, following Ref.~\cite{clark2023dimension}, we define the four-point function
\begin{equation}
    \psi^a(\bm{\tau}) = \frac{1}{N} \sum_{i\neq j} C^a_{ij}(\tau_1) C^a_{ij}(\tau_2),
    \label{eq:lower_case_def}
\end{equation}
where $\bm{\tau} = (\tau_1, \tau_2)$. The three scaling behaviors of dimensionality outlined above correspond in the limit $N \to \infty$ to ${\psi^a(0,0) \to \infty}$, ${\psi^a(0,0) \to \text{const.}}$, or ${\psi^a(0,0) \to 0}$, respectively. Since $\psi^a(0,0)$ is given by a sum over $\mathcal{O}(N^2)$ squared cross-covariances times $1/N$, its magnitude is $N$ times that of an individual squared cross-covariance. Both i.i.d.\ matrices and the extensive-rank matrices generated by the random-mode model lead to $\mathcal{O}(1/\sqrt{N})$ cross-covariances. Thus, $\psi^a(0,0) \to \text{const.}$ as $N \to \infty$, leading to nontrivial extensive dimensionality, $0 < \text{PR}^a < 1$.

In this paper, we focus on a related function $\Psi^a(\bm{\tau})$ that includes the diagonal terms,
\begin{equation}
    \Psi^a(\bm{\tau}) = \frac{1}{N} \sum_{i,j=1}^N  C^a_{ij}(\tau_1) C^a_{ij}(\tau_2).
    \label{eq:psi-def-nrn}
\end{equation}
As per the duality between neuronal and temporal covariances, $\Psi^a(\bm{\tau})$ can also be expressed in terms of the time-by-time covariance $C^a(t_1,t_2)$ as
\begin{equation}
    \Psi^a(\bm{\tau}) = N \tavg{ C^a(t_1, t_2) C^a(t_1 + \tau_1, t_2 + \tau_2) }_{t_1, t_2}.
    \label{eq:psi-def-time}
\end{equation}
The neuron-by-neuron [\eref{eq:psi-def-nrn}] and time-by-time [\eref{eq:psi-def-time}] definitions of $\Psi^a(\bm{\tau})$ form the basis of the cavity and path-integral calculations of this function, respectively.
Finally, the dimension of activity is given by
\begin{equation}
    \text{PR}^a = \frac{C^a(0)^2}{\Psi^a(0,0)},
    \label{eq:PR_C_Psi}
\end{equation}
which, unlike \eref{eq:pr_expr_one}, holds even when single-neuron variances $C^a_{ii}(0)$ are nonuniform across neurons. 

To study timescales of collective activity, we consider the principal components of activity,
\begin{equation}
    p^a_k(t) = \frac{1}{\sqrt{\lambda^a_k}} (\bm{v}^a_k)^T \bm{a}(t).
\end{equation}
These principal components are the basis of much of modern analysis of high-dimensional neural data \cite{cunningham2014dimensionality, gao2015simplicity, trautmann2019accurate, meshulam2024statistical}. They all have unit variance due to the $1/\sqrt{\lambda^a_k}$ normalization, but potentially very different characteristic timescales [\figref{fig:Psi_timescale_match}{a}]. To extract the timescales of just the leading components, we weight them by the squares of their corresponding eigenvalues. This gives
\begin{align}  
    \frac{1}{N}\sum_{k} (\lambda^a_k)^2 \tavg{  p^a_k(t) p^a_k(t+\tau) }_t &= \Psi^a(\tau,0),
\end{align}
showing that $\Psi^a(\bm{\tau})$ captures the temporal structure of the leading principal components of activity [Figs.~\ref{fig:Psi_timescale_match}(c) and \ref{fig:Psi_timescale_match}(d)]. Note that weighting by the eigenvalues themselves, rather than their squares, gives $N^{-1} \sum_k \lambda^a_k \tavg{ p^a_k(t) p^a_k(t+\tau) }_t = C^a(\tau)$, recovering single-neuron information [Figs.~\ref{fig:Psi_timescale_match}(b) and \ref{fig:Psi_timescale_match}(d)].

\section{Path-integral analysis of the random-mode model}
\label{sec:path_integral_analysis}

We aim to compute the summary statistics $C^a(t_1, t_2)$, $S^a(t_1, t_2)$, and $\Psi^a(\bm{\tau})$ in the limit $N\to \infty$ under the disorder average. For a general function $\mathcal{F}$, we use the subscript $\spt$ to denote such limiting values:
\begin{equation}
    \mathcal{F}_\spt =  \lim_{N \to \infty} \tavg{\mathcal{F}}_{\bm{J}}.
\end{equation}
These values also correspond to the saddle point of a path integral. The summary statistics we have defined are self-averaging, meaning that the same values are obtained, up to $\mathcal{O}(1/\sqrt{N})$ fluctuations, when they are computed based on a single realization of a large network. In this paper, we are interested in statistically stationary states such that, for $N \to \infty$, the two-point functions depend on time differences only and thus can be replaced by their stationarized counterparts [\eref{eq:stationarized}], i.e., for all $t$,
\begin{equation}
    C^a_\spt(t, t \pm \tau) = C^a_\spt(\tau), \:\:\:\: S_\spt^a(t, t-\tau) = S^a_\spt(\tau).
\end{equation}

Calculating two-point functions is well-established ``standard DMFT'' \cite{crisanti2018path, helias2020statistical, zou2024introduction}. As we review in \aref{sec:computing}, four-point functions were previously calculated for networks with i.i.d.\ couplings using a two-site version of the cavity method \citep{clark2023dimension}. In this section, we introduce a new path-integral approach to calculating four-point functions that allows for the rapid solution of the i.i.d.\ model and generalizes to more complicated models, including the random-mode model.

\subsection{Relating \texorpdfstring{$\Psi^a$}{the four-point function} to temporal covariance fluctuations}
\label{sec:relating}

To cast the problem in the language of path integrals, we turn to the time-by-time definition of $\Psi^a(\bm{\tau})$ [\eref{eq:psi-def-time}]. If the time-by-time covariance $C^a(t_1, t_2)$ were equal to its limiting value, $C^a_{\spt}(t_2 - t_1)$, we would have
\begin{equation}
    {\Psi_{\spt}^a(\bm{\tau}) \overset{?}{=} N \tavg{C^a_{\spt}(t_2-t_1) C^a_{\spt}(t_2-t_1+\tau_2 - \tau_1)}_{t_1,t_2} 
    = 0} \nonumber 
\end{equation}
since, for large time differences, $C^a_{\spt}(t_2 - t_1)$ decays exponentially in $|t_2 - t_1|$ \citep{sompolinsky1988chaos}. However, based on the neuron-by-neuron definition [\eref{eq:psi-def-nrn}], $\Psi_{\spt}^a(\bm{\tau})$ is clearly a nonzero, order-one quantity. This apparent contradiction is resolved by considering the $\mathcal{O}(1/\sqrt{N})$ fluctuations around the limiting value. The necessity of these fluctuations becomes clear when considering the duality between neuronal and temporal covariances: the presence of nonzero cross-covariances $C^a_{ij}(0)$ indicates low dimensional structure in the system; equivalently, this structure manifests as the system revisiting similar states over time, resulting in nonzero temporal covariance $C^a(t_1,t_2)$ even for large time differences $|t_2-t_1| \gg 1$ (Fig.~\ref{fig:nrn_time}). To capture this behavior, we define fluctuations around the saddle point as
\begin{equation}
    \delta C^a(t_1,t_2) = C^a(t_1, t_2) - C^a_{\spt}(t_2-t_1).
\end{equation}
We express $\Psi^a(\bm{\tau})$ purely in terms of these fluctuations as
\begin{equation}
    \Psi^a(\bm{\tau}) = N \tavg{\delta C^a(t_1, t_2)\delta C^a(t_1+\tau_1, t_2+\tau_2)}_{t_1, t_2},
\end{equation}
noting that terms involving $C^a_{\spt}(t_2-t_1)$ vanish at large time differences. Our task, therefore, is to compute the covariance of the fluctuations, $\delta C^a(t_1, t_2)$, captured asymptotically by
\begin{equation}
    \tavg{N \delta C^a(t_1,t_2) \delta C^a(t_3,t_4)}_{\spt}. \label{eq:asymto_cov}
\end{equation}
Then, to compute $\Psi_\spt^a(\bm{\tau})$, we take in \eref{eq:asymto_cov} the limits
\begin{align}
    t_3 &= t_1 + \tau_1, &  |t_2 - t_1| &\to \infty, \label{eq:temporal-limits} \\
    t_4 &= t_2 + \tau_2, &  \tau_1, \tau_2 &\sim \mathcal{O}(1). \nonumber
\end{align}

\subsection{Calculation of two- and four-point functions}
\label{sec:calculating_pathint}

Details of the calculation are given in \aref{sec:path-integral}. Briefly, we compute the covariance of fluctuations [\eref{eq:asymto_cov}] using the Martin-Siggia-Rose-Janssen–de Dominicis (MSRJD) path-integral formalism \cite{martin1973statistical, de1978dynamics, hertz2016path, crisanti2018path, helias2020statistical, zou2024introduction}. For any fixed coupling matrix $\bm{J}$, the path integral is
\begin{multline}
    Z[\bm{J}] = \pin{\bm{x}} \pinh{\bm{x}} \\
    \exp \Bigg\{i\sum_{i=1}^N  \int dt \hat{x}_i(t) \Bigg( T[{x}_i](t)
    - \sum_{j=1}^N  J_{ij} \phi_j(t)\Bigg) \Bigg\}.
\end{multline}
The auxiliary field $\hat{x}_i(t)$ is conjugate to $x_i(t)$ and enforces the network's equations of motion [\eref{eq:eoms}] through the $\delta$-function representation $(2\pi)^{-1}\int d\hat{x} \, e^{i \hat{x} x} = \delta(x)$. Upon introduction of source terms (that are not needed for our purposes), this path integral serves as a generating functional for correlation and response functions. 

In principle, to compute $\Psi^a_\spt(\bm{\tau})$ in this framework, after integrating out $\bm{J}$ from $Z[\bm{J}]$ to obtain a statistical field theory governing two-point functions, we (1) find the saddle-point solution for two-point functions by extremizing the action (i.e., standard DMFT); (2) compute the time$^2$-by-time$^2$ Hessian matrix of the action describing fluctuations around this saddle point; (3) invert this Hessian and extract the sub-block corresponding to the covariance of $\delta C^\phi(t_1, t_2)$ [\eref{eq:asymto_cov}]; and (4) apply the temporal limits [\eref{eq:temporal-limits}] to this sub-block to obtain $\Psi^\phi_\spt(\bm{\tau})$.

Direct inversion of the time$^2$-by-time$^2$ Hessian is analytically intractable. However, we show that the temporal limits commute with Hessian inversion, allowing us to invert a low dimensional frequency-dependent Hessian instead (\aref{sec:sep_limits}). This readily yields the Fourier-space function $\Psi^\phi_\spt(\bm{\omega})$, where $\bm{\omega} = (\omega_1, \omega_2)$. A few additional steps yield $\Psi^x_\spt(\bm{\omega})$. This approach allows for the efficient calculation of two- and four-point functions for a given model, requiring only a few lines of computation directly from the action of the theory.

As a demonstration of this approach, we apply it to a generalization of the i.i.d.\ model with correlated reciprocal couplings, recovering results of Ref.~\cite{clark2023dimension} (\aref{sec:partially_symmetric_appdx}).

\subsection{Solution of the random-mode model}
\label{sec:soln_of_random_mode_model}

Applied to the random-mode model, the self-consistent equations that determine $C^\phi_{\spt}(t_1,t_2)$ are (\aref{sec:two_pt_fn_rmm})
\begin{subequations}
\begin{align}
    T[x](t) &= \eta^x(t),  \\
     \eta^x &\sim \mathcal{GP}(0, \alpha r_2 C^\phi_\spt), \label{eq:gp_expr} \\
    C_{\spt}^\phi(t_1, t_2) &= \tavg{\phi(t_1) \phi(t_2)}_{\spt} \label{eq:twopt-rmm},
\end{align}
\end{subequations}
where \eref{eq:gp_expr} indicates that $\eta^x(t)$ is a Gaussian field with zero mean and correlation function $\alpha r_2 C^\phi_\spt(t,t')$. These equations are equivalent to those for an i.i.d.\ model [Eqs.~\eqref{eq:basic_ssp}--\eqref{eq:basic_scc}] with coupling strength
\begin{equation}
    g_\text{eff} = \sqrt{\alpha r_2}.
    \label{eq:g_eff_rmm}
\end{equation}
The four-point function, our main object of interest, is (\aref{sec:rmm_hes_subsec})
\begin{equation}
    \Psi_{\spt}^\phi = \frac{1 + \frac{1}{\alpha \text{PR}^D} | g_\text{eff}^2 S_{12}^\phi|^2}{|1 - g_\text{eff}^2 S_{12}^\phi|^2} C_{12}^\phi, \label{eq:path-int-result-rmm}
\end{equation}
where we have simplified notation by suppressing frequency variables via the shorthand $C^\phi_{12} = C^\phi_{\spt}(\omega_1) C^\phi_{\spt}(\omega_2)$ and $S^\phi_{12} = S^\phi_{\spt}(\omega_1) S^\phi_{\spt}(\omega_2)$. The same solution is obtained through a more involved, but intuitive, two-site cavity calculation (\aref{sec:cav_appdx}; Fig.~\ref{fig:cavity_schematic}).

This solution for the random-mode model is one of our main results, which we now spend some time interpreting. 

\subsection{Interpreting the solution}
\label{sec:interp}

\begin{figure*}
    \centering
    \includegraphics[width=3.35in]{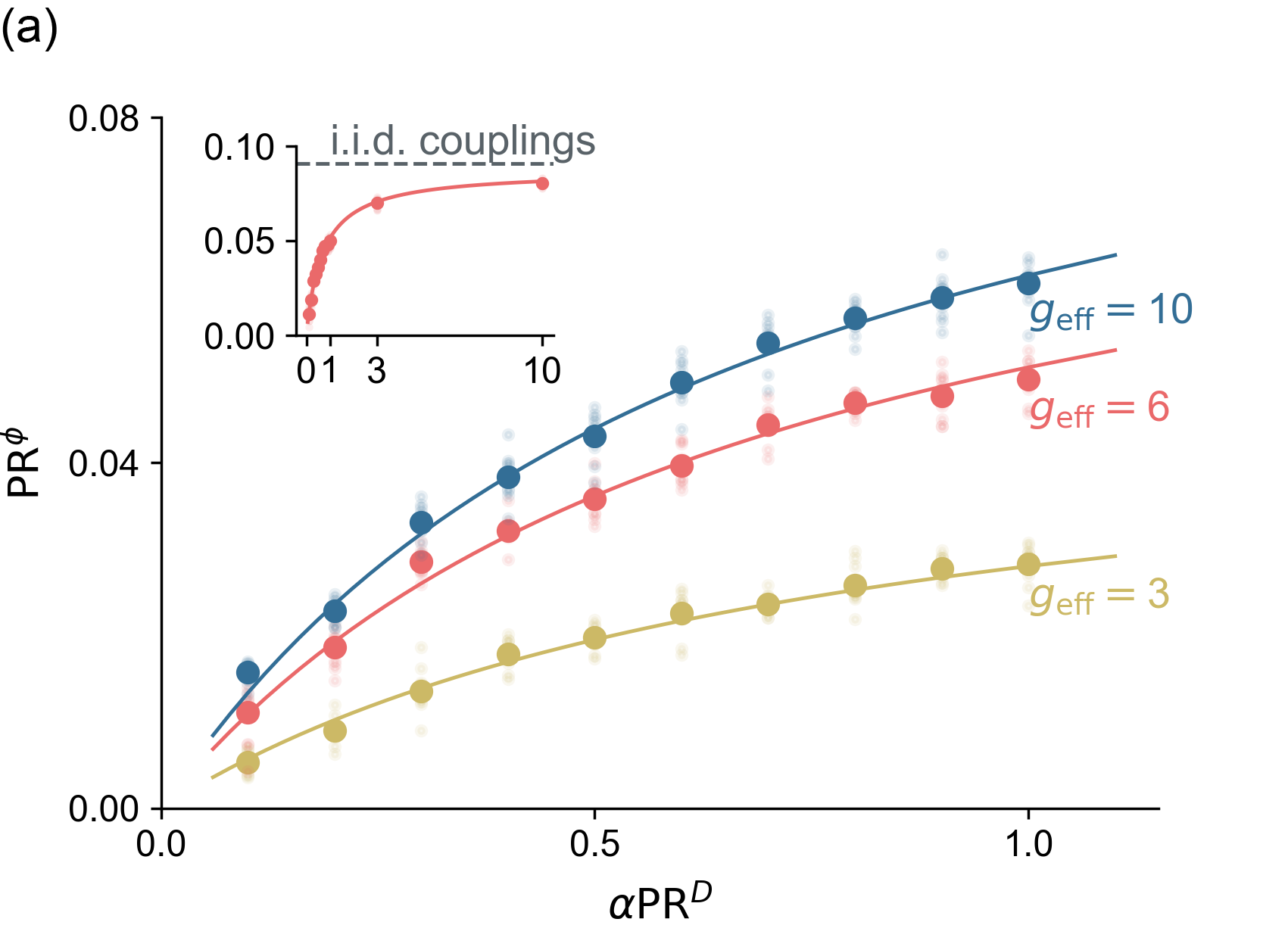}\includegraphics[width=3.35in]{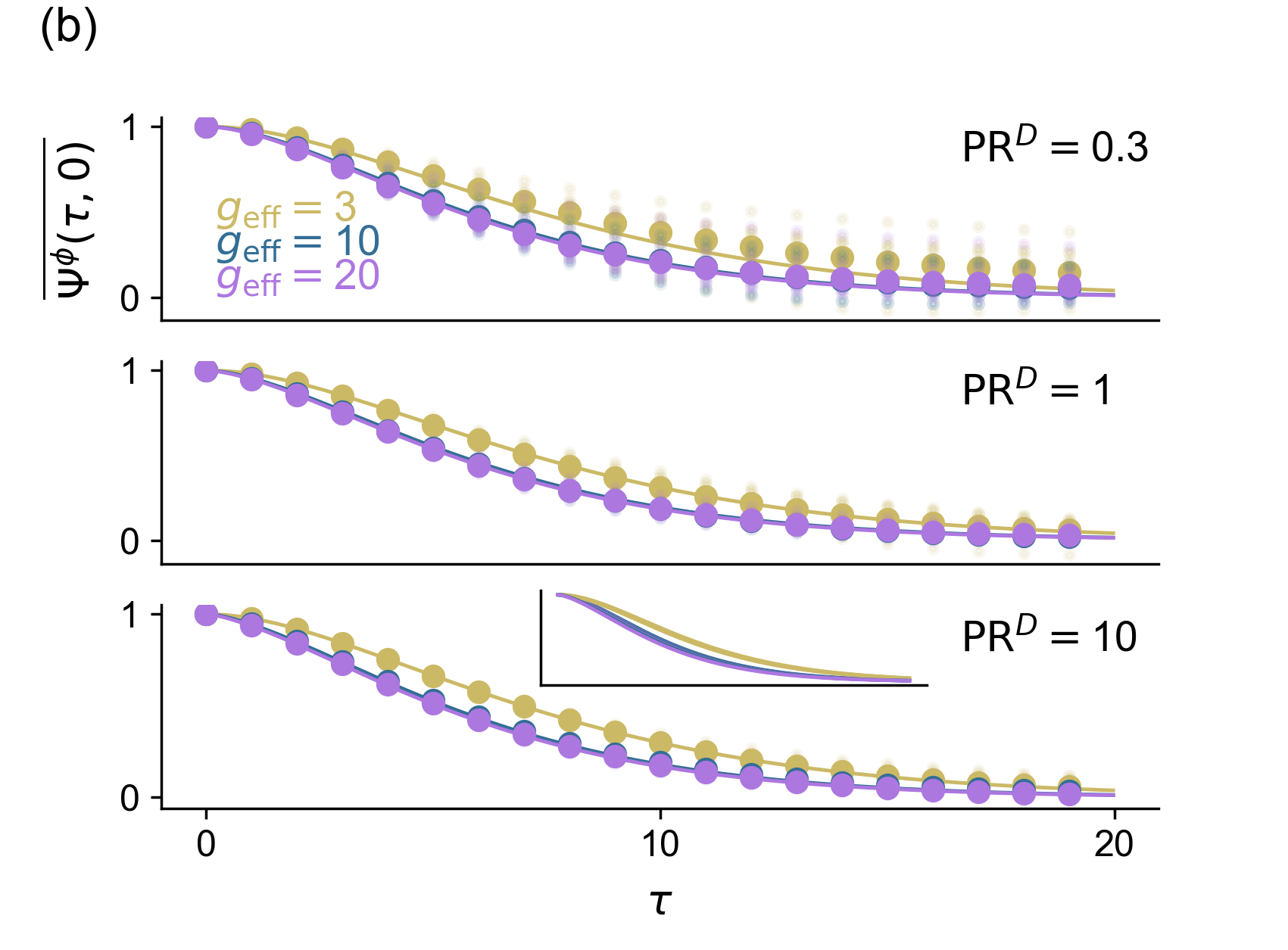}
    \caption{Dimension of activity and collective timescales in the random-mode model. (a) Activity dimension $\text{PR}^\phi$ versus effective rank $\alpha \text{PR}^D$ for various coupling strengths $g_\text{eff}$. Thin dots, individual simulations; thick dots, means over ten simulations; lines, theoretical predictions. Inset: extended $g_\text{eff} = 6$ case, showing convergence to i.i.d.\ coupling behavior with growing $\alpha$. (b) Normalized four-point function $\overline{\Psi^{\phi}(\tau, 0)} \equiv \Psi^{\phi}(\tau, 0)/\Psi^{\phi}(0, 0)$ for various $g_\text{eff}$ and $\alpha \text{PR}^D$. Inset: theory curves for all $\alpha \text{PR}^D$, demonstrating relative invariance of collective timescales to the effective rank. Simulations use $N=5000$ neurons.}
    \label{fig:LDR_PRD_match}
\end{figure*}

Our analysis reveals two features of connectivity that characterize how it shapes activity, namely, the effective coupling strength $g_\text{eff}$ and the effective rank $\alpha \text{PR}^D$. The former characterizes local structure (the magnitudes of individual couplings), while the latter characterizes global structure (the number of large connectivity components). The above equations determining $C_\spt^\phi(\tau)$ and $\Psi^\phi_\spt(\bm{\tau})$ immediately reveal that global structure in the connectivity is invisible at the single-neuron level but shapes collective activity. Specifically, the correlation function $C^\phi_\spt(\tau)$ is fully determined by $g_{\text{eff}}$ and is insensitive to the effective rank of connectivity. This has important implications for neural data analysis. In particular, analyses of single-neuron activity properties may miss signatures of structured connectivity that become apparent only when analyzing collective activity properties among simultaneously recorded neurons. 

What is the nature of the dependence of collective-activity properties, such as the dimension of activity, on the effective rank? To plot this relationship and validate the theory, we simulated networks and computed the participation ratio from the empirical equal-time covariance matrix of the activations (see \aref{sec:numerical_details} for numerical details). We used a component-strength spectrum $D_a \propto \exp(-\beta_D a/M)$, which yields $\text{PR}^D = \beta^{-1}_D\tanh(\beta_D)$.
To attain a desired effective rank $\alpha\text{PR}^D < 1$, we set $\alpha = M/N = 1$ and solved for $\beta$; for $\alpha\text{PR}^D > 1$, we fixed $\text{PR}^D = 1$ and increased $\alpha$. We adjusted $g_\text{eff}$ by rescaling $D_a$ uniformly. 
We find excellent theory-simulation agreement (Fig.~\ref{fig:LDR_PRD_match}). \Figref{fig:LDR_PRD_match}{a} illustrates how the dimension of activity, $\text{PR}^\phi$, varies with $\alpha \text{PR}^D$ for different values of $g_\text{eff}$. Consistent with i.i.d.-coupling networks, $\text{PR}^\phi$ increases monotonically with $g_\text{eff}$ \cite{clark2023dimension}. It also increases monotonically with $\alpha \text{PR}^D$.

Taking $\alpha\text{PR}^D \to \infty$ for fixed $g_{\text{eff}}$ in \eref{eq:path-int-result-rmm} recovers the formula for i.i.d.\ couplings [\eref{eq:psi_phi_sol}] of Ref.~\cite{clark2023dimension}. Indeed, in this limit, $\bm{J}$ approaches an i.i.d.\ matrix. Correspondingly, $\text{PR}^\phi$ approaches the activity dimension of a network with i.i.d.\ couplings [\figref{fig:LDR_PRD_match}{a} inset]. Comparing \eref{eq:psi_phi_sol} to the corresponding i.i.d.\ formula ($1/\alpha \text{PR}^D = 0$) and noting that single-neuron properties are unaffected by $\alpha \text{PR}^D$, one sees that any structure beyond i.i.d.\ connectivity ($1/\alpha \text{PR}^D$ finite) strictly {decreases} the dimension of activity.

The dimension of activity depends on $\Psi^{\phi}_{\spt}(\bm{\tau})$ at $\bm{\tau} = (0,0)$; we now examine the temporal profile of this function. \Figref{fig:LDR_PRD_match}{b} shows a normalized version of $\Psi^{\phi}(\tau, 0)$, which is related to the correlation functions of leading principal components, for various values of $g_\text{eff}$ and $\alpha \text{PR}^D$. The decay timescale of $\Psi^{\phi}_{\spt}(\tau, 0)$ decreases with increasing $g_\text{eff}$, approaching a limiting behavior at $g_\text{eff} \approx 10$. While $\alpha \text{PR}^D$ affects the overall scale of this function and, consequently, the dimension of activity, it has little effect on the decay timescale and, hence, little effect on the timescales of leading principal components (\figref{fig:LDR_PRD_match}{b}, inset). Thus, the effective rank primarily influences the dimension of activity rather than the collective temporal structure. 

A remarkable feature of these results is that only two connectivity statistics, $g_{\text{eff}}$ and $\alpha \text{PR}^D$, are sufficient to determine the two- and four-point functions. Because of the nonlinear, high-dimensional nature of the network, this reduction is unexpected. This reduction breaks down when structured mode overlaps are introduced (\secref{sec:LR_correlations}), in which case the formula for the four-point function requires knowledge of the full joint distribution of component strengths and overlaps.

\section{Limiting behavior of the random-mode model}
\label{sec:low_eff_rank}

Here we examine the behavior of the random-mode model in two limiting cases: as the system approaches the transition to chaos from above ($g_\text{eff} \to 1^+$) and in the limit of low effective rank ($\alpha \text{PR}^D \to 0^+$).

We first consider the limit $g_\text{eff} \to 1^+$ for arbitrary $\alpha \text{PR}^D$. The relevant expansion parameter is $0 <g_\text{eff}-1 \ll 1$. For networks with i.i.d.\ couplings with coupling strength $g$, it is known that, as $g \to 1^+$, $\Psi^a_{\spt,\text{i.i.d.}}(\bm{\tau}) \sim 1/(g-1)$ and $\text{PR}^a_\text{i.i.d.} \sim (g-1)^3$ \cite{clark2023dimension}. For the random-mode model, to leading order in $1/(g_\text{eff}-1)$,
\begin{equation}
    \Psi^a_\spt(\bm{\tau}) = \left( 1 + \frac{1}{\alpha \text{PR}^D} \right) \underbrace{\Psi^a_{\spt,\text{i.i.d.}}(\bm{\tau})}_{\sim 1/(g_\text{eff}-1)}.
    \label{eq:psi_near_transition}
\end{equation}
Consequently, to leading order in $g_\text{eff}-1$, the dimension of activity behaves as
\begin{equation}
     \text{PR}^a = \left( 1 + \frac{1}{\alpha \text{PR}^D} \right)^{-1} \underbrace{\text{PR}^a_\text{i.i.d.}}_{\sim (g_\text{eff}-1)^3}.
     \label{eq:pr_lim_g}
\end{equation}
Thus, as the network approaches the transition to chaos from above, structured connectivity reduces the dimension by a factor of $( 1 + 1/\alpha \text{PR}^D )^{-1}$ compared to a network with i.i.d.\ couplings. Taking the further limit of $\alpha \text{PR}^D \to 0^+$ yields the simple relation
\begin{equation}
    \text{PR}^a = \alpha \text{PR}^D \text{PR}^a_\text{i.i.d.}.
    \label{eq:simple_relation}
\end{equation}
The difference between $\alpha \text{PR}^D$ and  $\text{PR}^S$ is a higher-order correction, so we also have
$\text{PR}^a = \text{PR}^S \text{PR}^a_\text{i.i.d.}$;
that is, in the limits of both low effective rank and small effective coupling strength, the dimension of activity is equal to the dimension of the singular-value spectrum multiplied by the dimension of activity for an i.i.d.\ network.

We next consider the limit $\alpha \text{PR}^D \to 0^+$, for arbitrary $g_\text{eff}$. In this limit, the component strengths closely approximate the singular values. To leading order in $\alpha \text{PR}^D$, 
\begin{align}
    \text{PR}^a &= K^a(g_\text{eff}) \alpha \text{PR}^D, \\
    K^a(g_\text{eff}) &= { C^a_\spt(0)^2}\Bigg{[}\int d\bm{\omega} \left|\frac{g_\text{eff}^2S^a_{12}}{1 - g_\text{eff}^2S^\phi_{12}}\right|^2 C^\phi_{12}\Bigg{]}^{-1}. \label{eq:K_a}
\end{align}
We verify this limiting behavior in simulations using $\alpha = 1$ and a step function for the component strengths: $D_a = \text{const.}$ for $a/N \leq \text{PR}^D$ and $D_a = 0$ otherwise. We varied $\text{PR}^D$, observing linear relationships with $\text{PR}^\phi$ and $\text{PR}^x$ in the $\text{PR}^D \to 0^+$ limit [\figref{fig:spectrum_shapes}{a}].

Focusing on $K^\phi(g_\text{eff})$, note that its expression [\eref{eq:K_a}] resembles the formula for the dimension of activity in an i.i.d.\ network with coupling strength $g_\text{eff}$ [\eref{eq:psi_phi_sol}], but includes an additional factor $|g_\text{eff}^2 S^\phi_{12}|^2$ in the integral. Numerical evaluation confirms that, as expected from \eref{eq:simple_relation}, $K^\phi(g_\text{eff}) \to \text{PR}^\phi_\text{i.i.d.}$ as $g_\text{eff} \to 1^+$. As $g_\text{eff}$ increases, $K^\phi(g_\text{eff})$ increases monotonically to approximately $1.53 \, \text{PR}^\phi_{\text{i.i.d.}}$ [\figref{fig:spectrum_shapes}{a}, inset]. Thus, for networks with low effective rank, $\text{PR}^\phi \sim \alpha \text{PR}^D \text{PR}^\phi_{\text{i.i.d.}}$ up to an order-one fudge factor that becomes exactly one as $g_\text{eff} \to 1^+$.

Finally, we consider the dependence of the full eigenvalue spectrum $\lambda^a_i$ of $\bm{C}^a(0)$ on the spectrum of component strengths $D_a$. The latter leads to the former via a highly nonlinear relationship. Indeed, using the same step-function component-strength spectrum, the ranked eigenvalues $\lambda^\phi$ and $\lambda^x$ of the covariance matrix decay smoothly rather than exhibiting a cutoff [\figref{fig:spectrum_shapes}{b}]. Nevertheless, certain properties of these connectivity and activity spectra are linked. In particular, in the limit $\alpha \text{PR}^D \to 0^+$, $\alpha\text{PR}^D$ and $\text{PR}^a$ exhibit a linear relationship [with proportionality factor $K^a(g_\text{eff})$], implying that the ratios of the squared second moments and fourth moments of these spectra are linked (despite, even in this limit, the full spectra not coinciding). 

\section{Single-neuron heterogeneity}
\label{sec:heterogeneity}

\begin{figure*}
    \centering
    \includegraphics[width=7in]{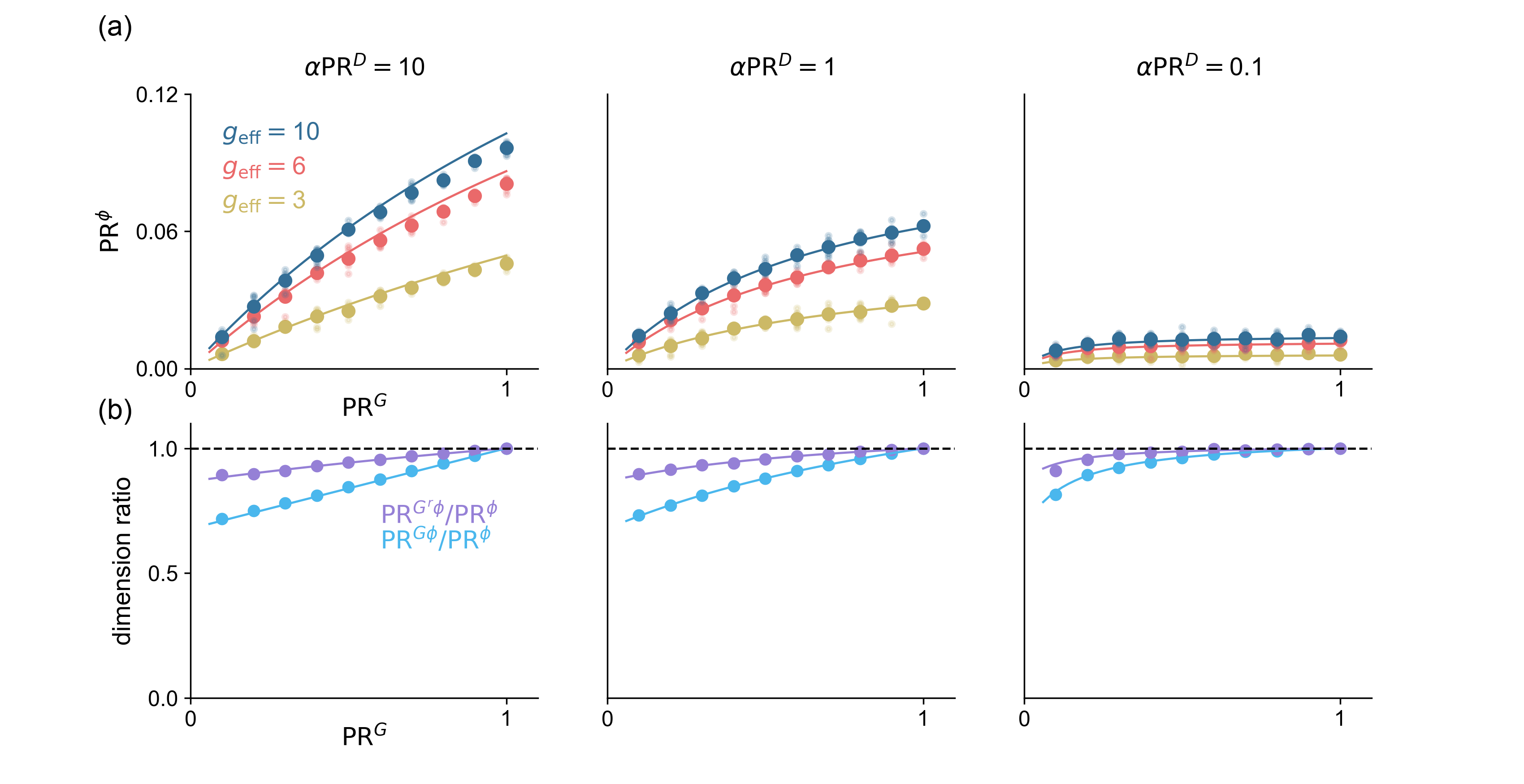}
    \caption{Effect of single-neuron heterogeneity on dimension of activity in the random-mode model. (a) Dimension of normalized activity $\text{PR}^\phi$ versus participation ratio of gain distribution $\text{PR}^G$ for various coupling strengths $g_\text{eff}$ and effective ranks $\alpha \text{PR}^D$. Thin dots, individual simulations; thick dots, means over ten simulations; lines, theoretical predictions. (b) Fractional reduction in dimension for weighted activations relative to $\text{PR}^\phi$. Blue, $G_i \phi_i$; purple, $G^\text{readout}_i \phi_i$. $g_\text{eff} = 10$. Thick dots, means over ten simulations, averaging before taking the ratio; lines, theoretical predictions. All simulations use $N=5000$ neurons.}
    \label{fig:LDR_PRG_match}
\end{figure*}

Cortical neurons display broad distributions of firing rates \cite{hromadka2008sparse}. This heterogeneity among neurons is analogous to the heterogeneity of component strengths in the random-mode model, but operates in the neuron basis rather than the component basis. We now compare how low dimensional connectivity structure and heterogeneous neuronal properties affect collective activity by extending our framework to include single-neuron heterogeneity.

In a general formulation described in \aref{sec:het_appdx}, we extend the nonlinearity of neuron $i$ to $\Phi_{\bm{\theta}_i}(x)$, which depends on a vector of neuron-specific parameters $\bm{\theta}_i$, and solve for the resulting two- and four-point functions. Here, to model firing-rate heterogeneity specifically, we take $\bm{\theta}_i$ to consist of a single gain parameter $G_i$, with $\Phi_{\bm{\theta}_i}(x) = G_i \phi(x)$. In analogy with the component-strength distribution, we define
\begin{align}
    q_n &= \tavg{G^n}_G, \\ 
    \text{PR}^G &= \frac{q_2^2}{q_4}.
\end{align}
The single-site picture that determines the two-point functions is largely similar to \eref{eq:twopt-rmm}.

We analyze collective activity (four-point functions) for (1) unnormalized activations $\Phi_{\bm{\theta}_i}(t) = G_i \phi\bm{(}x_i(t)\bm{)}$ and (2) normalized activations $\phi_i(t) = \phi\bm{(}x_i(t)\bm{)}$, the latter being analogous to $z$-scored firing rates in neural recordings, where $z$ scoring removes single-neuron heterogeneity while preserving structure from cross-neuron and temporal correlations. The four-point functions for the unnormalized and normalized variables are found to be
\begin{subequations}
\begin{align}
    \Psi^\Phi_\spt &= \frac{\frac{1}{\text{PR}^G} + \frac{1}{\alpha \text{PR}^D}|g_\text{eff}^2 S^\phi_{12}|^2}{|1-g_\text{eff}^2S^\phi_{12}|^2} C^\Phi_{12}, \label{eq:psi_G_unnorm} \\
    \Psi^\phi_\spt &= \frac{1 + \left(\frac{1}{\text{PR}^G} + \frac{1}{\alpha \text{PR}^D} - 1\right)|g_{\text{eff}}^2 S^\phi_{12}|^2 }{|1 - g_{\text{eff}}^2 S^\phi_{12}|^2} C^\phi_{12},
    \label{eq:psi_G_normalized}
\end{align}
\end{subequations}
respectively, where $C^\Phi_{12}  = q_2^2 C^\phi_{12}$ and we have updated the definition of the effective coupling strength to $g_\text{eff} = \sqrt{\alpha r_2 q_2}$.
Equation~\eqref{eq:psi_G_normalized} demonstrates that reductions in effective rank, described by $\alpha \text{PR}^D$, and reductions in the heterogeneity of gains, described by $\text{PR}^G$, have symmetric effects on the dimension of normalized activity, with their combined effect captured by the harmonic mean $\left(\frac{1}{\alpha \text{PR}^D} + \frac{1}{\text{PR}^G} \right)^{-1}$ \footnote{This symmetry arises because the dynamics of the network are determined by an effective coupling matrix $\bm{L}\bm{D}\bm{R}^T \bm{G}$, where $\bm{D} = \text{diag}(D_1,\ldots,D_M)$ and $\bm{G} = \text{diag}(G_1,\ldots,G_N)$. Swapping the distributions of $D_a$ and $G_i$ does not change the spectrum when $\bm{L}$ and $\bm{R}$ are i.i.d., since $\bm{L}\bm{D}\bm{R}^T \bm{G}$ and $\bm{R}^T \bm{G} \bm{L}\bm{D}$ have the same nonzero eigenvalues.}. One might expect that normalizing the activations before computing the dimension of activity would remove the effect of heterogeneous gains on the dimension. This is not the case, however, since neuronal gains affect the recurrent dynamics, not merely the observed firing rates.

To further explore these effects, we compare the activity dimensions for three sets of activations: unnormalized activations and normalized activations (1 and 2 above); and (3) normalized activations multiplied by an independent set of ``readout'' gains, $\Phi^\text{readout}_i(t) = G^\text{readout}_i \phi_i(t)$, where the readout gains have the same distribution as the actual gains. Case (3), whose four-point function is given by \eref{eq:psi_G_readout}, considers variables with the same distribution of firing rates as the unnormalized activations, but where this heterogeneity is unrelated to the recurrent dynamics. 

To validate our theoretical predictions [Eqs.~\eqref{eq:psi_G_unnorm}, \eqref{eq:psi_G_normalized}, and \eqref{eq:psi_G_readout}], we simulated networks with component strengths and gains given by $D_a \propto \exp(-\beta_D a/M)$ and $G_i \propto \exp(-\beta_G i/N)$ (Fig.~\ref{fig:LDR_PRG_match}). The dimension of activity is determined by the coupling strength $g_\text{eff}$, the effective rank $\alpha \text{PR}^D$, and the participation ratio of the gain distribution $\text{PR}^G$. Increases in each of these parameters lead to a higher activity dimension for all three sets of activations.

Normalized activations exhibit the highest dimension. While scaling normalized activations by heterogeneous factors reduces dimension as expected, the magnitude of this reduction depends on whether $G_i$ or $G^\text{readout}_i$ is used, even when their distributions are identical. This occurs because neurons with the largest gains preferentially participate in the leading modes of the normalized activations. Further scaling these already-dominant neurons by their gains results in overrepresentation of these neurons compared to using random gains. Thus, the dimension of unnormalized activations is lowest, with the dimension of random-readout activations falling between those of the normalized and unnormalized activations.

In experimental data analysis, it is common to normalize single-neuron activities to control for differences in firing rates across neurons. Our analysis implies that such normalization may not eliminate the effects of single-neuron heterogeneity on collective-activity properties. The persistence of these effects through normalization provides an experimental signature for distinguishing between heterogeneity that affects recurrent dynamics and heterogeneity that affects only the observed firing rates, for example, in a circuit that is not recurrent.

\section{Mode overlaps}
\label{sec:LR_correlations}

\begin{figure*}
    \centering
    \includegraphics[width=7in]{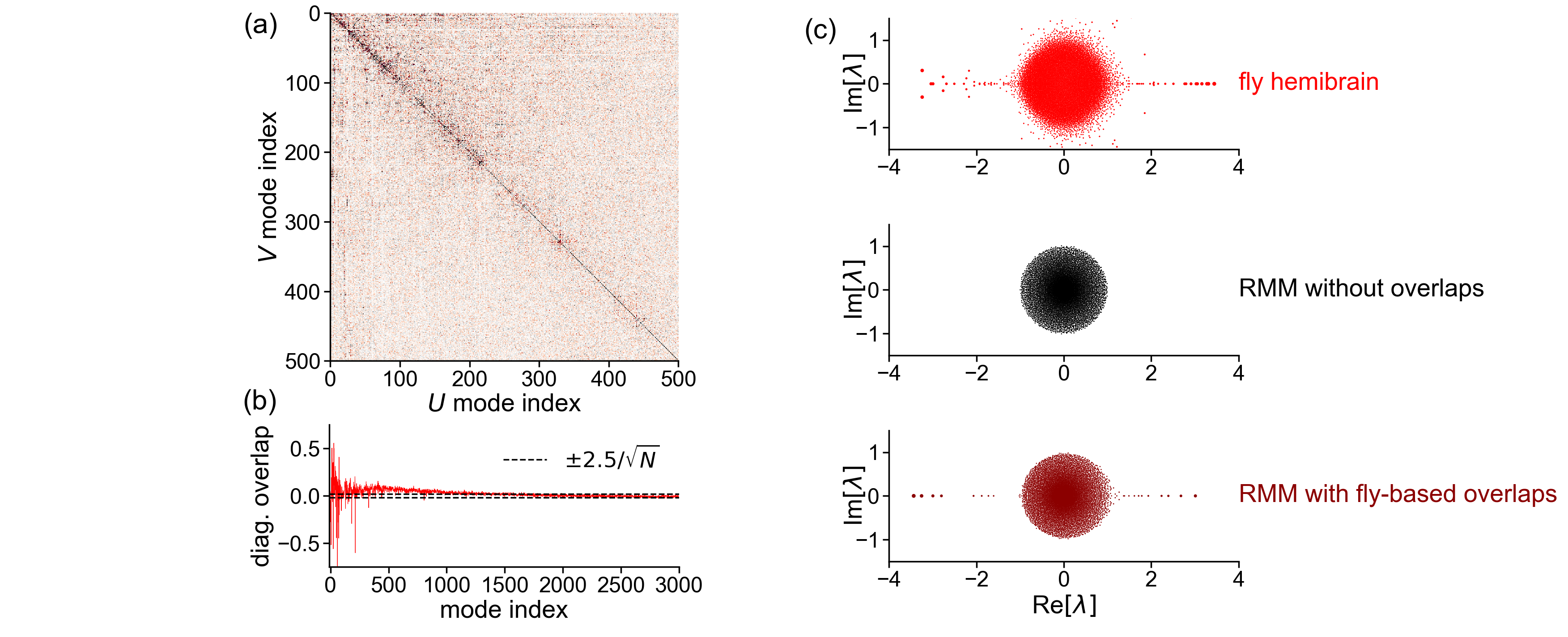}
    \caption{Right-left mode overlaps and eigenvalue spectra of the fly-brain connectome. (a) Overlap matrix between right and left singular vectors of the connectome. First 500 modes are shown. (b) First 3000 diagonal elements of the overlap matrix from (a). Horizontal lines indicate an approximate cutoff for $\mathcal{O}(1/\sqrt{N})$ random overlaps.
    (c)
    Eigenvalue spectra of various connectome-based matrices. Top: actual hemibrain spectrum, exhibiting large outliers along the real axis. Middle: eigenvalues from a realization of the random-mode model (RMM), using singular values from the connectome as component strengths $D_a$, resulting in a circularly symmetric spectrum. Bottom: random-mode model realization with right-left mode overlaps incorporated, where overlaps are given by the inner products of left and right singular vectors from the fly connectome SVD. Data from \citet{scheffer2020connectome}, $N = 18028$ neurons.}
    \label{fig:fly_fig_2}
\end{figure*}

Thus far, we have assumed i.i.d.\ modes, with the factorization $P(\ell_a, r_a) = P(\ell_a) P(r_a)$ implying that left-right overlaps $\bm{\ell}_a^T \bm{r}_a$ are random and $\mathcal{O}(1/\sqrt{N})$. The same holds for $\bm{\ell}_a^T \bm{r}_b$ with $a \neq b$ due to the more fundamental assumption that $P(\{\ell_a, r_a\}_{a=1}^M)$ factorizes across $a$. However, computations in biological circuits require structured interactions between modes, suggesting stronger, nonrandom overlaps between left and right modes. To test this, we return to the \textit{Drosophila} connectome and compute the overlap matrix between left and right singular vectors, $O_{ab} = \bm{u}_a^T \bm{v}_b$ [\figref{fig:fly_fig_2}{a}]. Contrary to the small, unstructured overlaps expected under the i.i.d.\ assumption, the connectome exhibits large, structured overlaps, particularly along the diagonal [\figref{fig:fly_fig_2}{b}].

Motivated by the presence of these large diagonal overlaps, we extend the random-mode model to include correlations between corresponding left and right modes. Specifically, we take $P(\ell_a, r_a)$ to be a zero-mean bivariate Gaussian with marginal variance $1/N$ and covariance $\rho_a/N$, where $\rho_a$ is the correlation between the $a$th mode pair. Both $D_a$ and $\rho_a$ are treated deterministically, and we assume that $M^{-1} \sum_a f(D_a, \rho_a)$ converges to a limiting value $\tavg{f(D, \rho)}_{D,\rho}$ as $N \to \infty$. In this setting, $\bm{\ell}_a^T \bm{r}_a = \rho_a$, with negligible $\mathcal{O}(1/\sqrt{N})$ fluctuations around this structured overlap, while $\bm{\ell}_a^T \bm{r}_b$ for $a \neq b$ remains unstructured and $\mathcal{O}(1/\sqrt{N})$. This additional structure produces eigenvalue spectra resembling that of the connectome, including large real outliers [\figref{fig:fly_fig_2}{c}] \cite{rider2014extremal}.

This purely diagonal overlap structure, however, can generate \textit{only} real eigenvalue outliers. The presence of complex outliers in the connectome, together with the full overlap matrix itself [\figref{fig:fly_fig_2}{a}], implies the presence of richer correlations. Capturing such features would require relaxing the full factorization of $P(\{\ell_a, r_a\})$ across $a$ while retaining $\mathcal{O}(M)$ parameter scaling---for example, via blockwise factorization or Markovian correlations across $a$. We leave such generalizations for future work. We also note that by setting $\rho_a = 1$ for all $a$, the model becomes a Hopfield network storing an extensive set of Gaussian patterns. In such networks, each mode can be ``condensed'' [$\mathcal{O}(1)$ overlap with activity] or ``uncondensed'' [$\mathcal{O}(1/\sqrt{N})$ overlap with activity]. Our analysis assumes no condensed modes, but sufficiently strong positive overlaps break this assumption. The Gaussian case, however, is unable to generate multiple multistable attractors as in a memory system \cite{beiran2021shaping}. Note that the form of the single-element density is relevant for the condensed patterns, but not for the uncondensed patterns. Allowing for condensed patterns within our analysis would require specifying this density. See Ref.~\cite{clark2025transient} for a DMFT analysis of generalized Hopfield dynamics in which these ideas are implemented. 

In \aref{sec:LR_appdx}, we compute the two- and four-point functions. As in previous cases, the dynamics reduce to single-site processes,
\begin{subequations}
\begin{align}
    T[x](t) &= \eta^x(t) + \int^t dt' R_\spt(t,t') \phi(t'), \label{eq:sym_ssp} \\
    \eta^x &\sim \mathcal{GP}(0, Q_\spt),  \\
    C_{\spt}^\phi(t_1, t_2) &= \tavg{\phi(t_1) \phi(t_2)}_{\spt}, \\
    S^\phi_\spt(t_1, t_2) &= \tavg{\frac{\delta \phi(t_1)}{\delta I(t_2)}}_{\spt}.
\end{align}
\end{subequations}
Assuming temporal stationarity, the kernels are given in Fourier space by
\begin{align}
    Q_\spt(\omega) &= \alpha\tavg{\frac{D^2}{|1 - D\rho S^\phi_\spt(\omega)|^2}}_{D,\rho} C^\phi_\spt(\omega), \label{eq:sym_Q_expr} \\
    R_\spt(\omega) &= \alpha\tavg{\frac{D\rho}{1 - D\rho S^\phi_\spt(\omega)}}_{D,\rho}. \label{eq:sym_R_expr}
\end{align}
Unlike the ``i.i.d.\ modes'' case, the single-site picture is not equivalent to that of an i.i.d.\ network for some $g_\text{eff}$.

To better understand the self-coupling kernel $R_\spt(\omega)$, we expand in powers of $\rho$. The coefficients encode connectivity motifs of increasing orders,
\begin{equation}
\begin{split}
    &R_\spt(\omega) = \tavg{J_{ii}}_\spt +  \tavg{NJ_{ij} J_{ji}}_\spt \: S^\phi_\spt(\omega) + \cdots, \text{  where} \\
    &\tavg{J_{ii}}_\spt = \alpha \tavg{D\rho}_{D,\rho}, \:\:
    \tavg{N J_{ij} J_{ji}}_\spt = \alpha \tavg{(D\rho)^2}_{D,\rho}.
\end{split}
\end{equation}
Here, $\tavg{J_{ii}}_\spt$ corresponds to deterministic self-couplings or autapses. To avoid such biologically implausible $\mathcal{O}(1)$ self-connections, we set $\tavg{D\rho}_{D,\rho} = 0$, yielding $J_{ii} = \mathcal{O}(1/\sqrt{N})$ \footnote{If model neurons represent clusters of biological neurons, strong self-connections may instead be interpreted as within-cluster coupling \cite{stern2014dynamics, litwin2012slow}.}.\nocite{stern2014dynamics, litwin2012slow} The second term in the expansion (note that $i\neq j$) reflects reciprocal correlations in an otherwise i.i.d.\ network (\aref{sec:partially_symmetric_appdx}). Nonzero higher-order terms indicate that this structure cannot be reduced to either of these simpler cases. Notably, if both $\tavg{NJ_{ii}}_\spt$ and $\tavg{N J_{ij} J_{ji}}_\spt$ vanish, all higher-order terms vanish as well, making reciprocal covariance necessary and sufficient for a nonzero self-coupling kernel in the single-site description. 

The four-point function, meanwhile, is given by
\begin{align}
    \Psi^\phi_\spt &= \frac{V - 1 + \left\{\frac{(1+X_{12})(1+X_{21}^*)}{1 - W^*} + \text{H.c.}\right\}}{|1-U|^2} C^\phi_{12}, \label{eq:lr_psi_result} \\
    U &= \alpha\tavg{D^2 \Sigma_1 \Sigma_2}_{D,\rho}, \:
    V = \alpha\tavg{D^4 |\Sigma_1\Sigma_2|^2}_{D,\rho}, \nonumber \\
    W &= \alpha\tavg{D^2 \rho^2 \Sigma_1^* \Sigma_2}_{D,\rho}, \:
    X_{12} = \alpha\tavg{D^3 \rho \Sigma_1 |\Sigma_2|^2}_{D,\rho}, \nonumber \\
    \Sigma_k &= \frac{S^\phi_k}{1 - D\rho S^\phi_k} \quad (k = 1,2). \nonumber
\end{align}

\begin{figure*}
    \centering
    \includegraphics[width=7in]{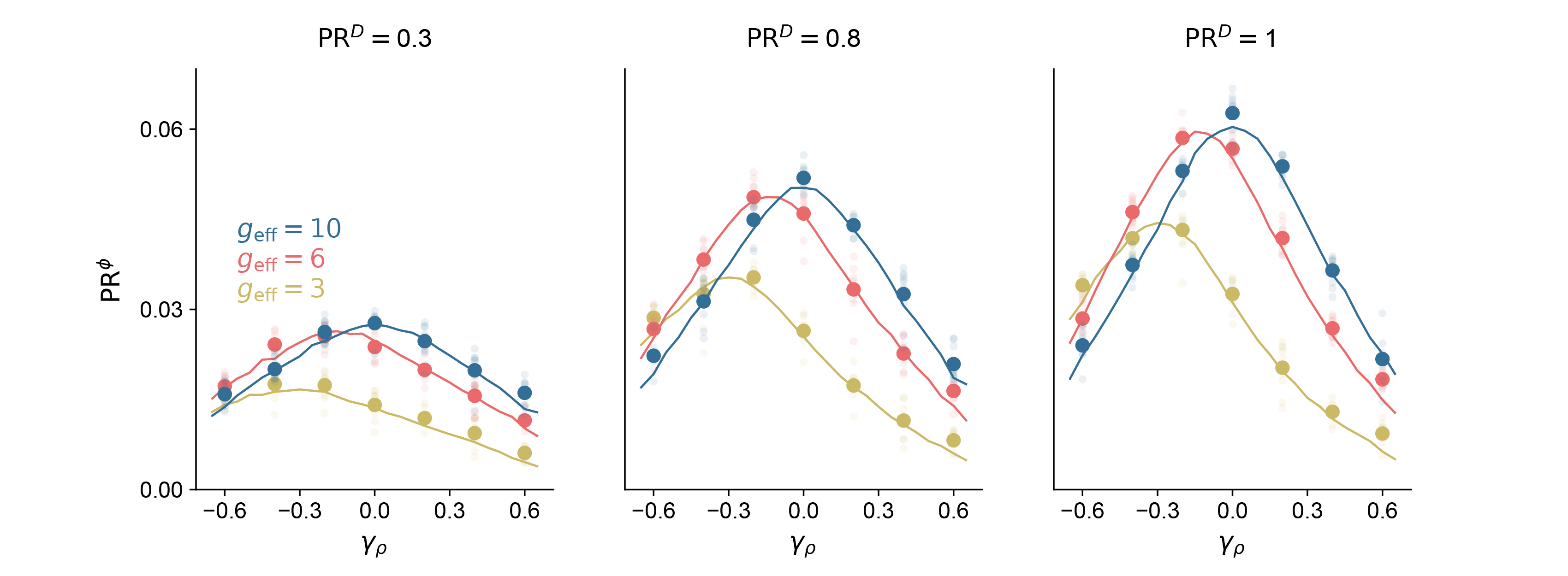}
    \caption{Effect of right-left mode correlations on dimension of activity in the random-mode model. Dimension of activity $\text{PR}^\phi$ versus correlation parameter $\gamma_\rho$ in the $(D,\rho)$ joint distribution, for various coupling strengths $g_\text{eff}$ and effective ranks $\alpha \text{PR}^D$. Thin dots, individual simulations; thick dots, means over ten simulations; lines, theoretical predictions. All simulations use $N=5000$ neurons.}
    \label{fig:LDR_sym_match}
\end{figure*}

We validate this theory using $\alpha = 1$ and $D_a \propto \exp(-\beta_D a/N)$, as before. For overlaps, we use a sigmoidal spectrum for $\rho_a$ qualitatively inspired by \figref{fig:fly_fig_2}{b}, with multiplicative prefactor $\gamma_\rho$ and other parameters chosen such that $\tavg{D\rho}_{D,\rho} = 0$ (\aref{sec:sigmoid_form}). 
Across values of $g_\text{eff}$ and $\text{PR}^D$, sufficiently strong overlaps, of either sign, reduce the dimension of activity (Fig.~\ref{fig:LDR_sym_match}). The maximum dimension occurs for $\gamma_\rho < 0$, where dynamics most strongly suppress activity aligned with dominant connectivity modes. This effect weakens as $g_\text{eff}$ increases.

In summary, our analysis reveals how connectivity structure shapes collective activity and its dimensionality. When mode overlaps are unstructured, connectivity structure reduces to two parameters: the effective rank $\alpha \text{PR}^D$ and effective coupling strength $g_\text{eff}$. Structured left-right mode overlaps break this simplification: Collective dynamics then depend on the full joint distribution of component strengths and overlaps, providing an additional mechanism for controlling activity dimensionality.

\section{Discussion}
\label{sec:discussion}

\subsection{Motifs versus global structure}
Neuronal connectivity is often described using statistics of motifs: local patterns like chains or cycles among small groups of neurons \cite{milo2002network, song2005highly, hu2013motif, aceituno2019universal}. These can be estimated from partial observations of a weight matrix which, until recently, set the limits of experimental measurement. Advances in whole-brain connectome reconstruction now allow access to global connectivity, including spectral features. Our parametrization of coupling matrices represents this global structure, with control over both the spectrum and mode overlaps. These spectral features determine how activity modes interact dynamically and thus may more directly dictate neural computations compared to motifs.

That said, in some studies the primary goal is to capture motifs or graph-theoretic features (e.g., degree distributions) over which the random-mode model does not provide control (although some of these features have spectral signatures that could be incorporated). When such features are the focus, other connectivity models are better suited---for instance, maximum-entropy random graph models constrained by fixed structural features (e.g., the configuration model for matching degree distributions).

\subsubsection{Linking connectivity and activity datasets.}  
The random-mode model, together with its dynamical theory, could aid in the interpretation of large-scale neural activity recordings in the context of anatomical connectivity, such as in \textit{Drosophila}, where both the connectome and whole-brain activity data are available \cite{schaffer2023spatial}. In the current work, we used this connectome to characterize the spectral structure of an actual large neural circuit, stopping short of considering the activity that it generates. Future work could progressively incorporate richer connectome constraints into the model to identify which features most strongly shape collective activity. One possible sequence would be to add the spectrum of component strengths, then diagonal mode overlaps, and finally off diagonal overlaps, assessing at each step how much the correspondence between modeled and observed activity improves. This is similar to approaches in statistical physics that aim to recover higher-order correlations by constraining lower-order statistics \cite{schneidman2006weak, maoz2020learning, meshulam2024statistical}.

\subsection{Learning}
Learning alters collective dynamics by modifying connectivity, the core principle of machine learning. Our analysis focuses on structured but untrained networks that produce chaotic activity. Nevertheless, the extensive low-rank connectivity structures we study provide a starting point for understanding trained networks.

Large networks trained on simple tasks often show intensive low-rank modifications to their initial couplings, resulting in a finite number of task-relevant eigenvalues \cite{schuessler2020interplay, schuessler2020dynamics}. Existing theoretical tools can already describe such systems by taking the trained weights, performing a low-rank decomposition, modeling mode overlaps statistically, and studying the resulting low-rank DMFT \cite{dubreuil2022role}. By contrast, empirical studies of large networks trained on real-world tasks suggest the emergence of extensive low-rank structure, with smooth spectral profiles with many significant components \cite{martin2021implicit}. This structure resembles that modeled by the random-mode model. Whether mode overlaps extracted from such networks can be incorporated into the random-mode model framework, and thereby yield a working theory for the extensive-rank case, is an open question.

A more ambitious direction is to develop a DMFT for networks whose weights are learned from data, rather than first training networks and then retrofitting a statistical model. Such a theory would describe not only how data shape the weights, as addressed by classical works \cite{gardner1988space}, but also how data interact with recurrent network dynamics. The path-integral formalism is well suited to this, as it allows derivation of DMFT equations for coupling matrices drawn from a Gibbs ensemble with an energy function given by a task loss. While this approach has been applied to low dimensional tasks \cite{bordelon2025mean}, applying it to high-dimensional tasks could yield a theory of how such tasks generate extensive-rank weights that, in turn, produce extensive-dimensional activity whose single-neuron and collective properties the theory would describe.

Another route to extensive-rank structure is via networks that perform many tasks, which could be modeled, for example, by representing mode overlaps through blockwise factorization of $P(\{\ell_a, r_a\}_{a=1}^M)$ across $a$. This multitask scenario is increasingly relevant in both neuroscience and machine learning \cite{yang2019task, driscoll2024flexible}.

Even in standard recurrent neural network training, initializing couplings with low-rank structure could improve procedures such as FORCE learning, which rely on suppressing chaos \cite{sussillo2009generating, laje2013robust}. Lower-dimensional attractors may be easier to control, potentially leading to faster convergence.

\subsection{Modeling functional consequences of single-neuron heterogeneity}
A central question in neuroscience is how neuronal diversity shapes collective dynamics and computation \cite{marder2006variability}. \aref{sec:het_appdx} presents a general framework that assigns each neuron a parameter vector $\bm{\theta}$ and enables the study of how different forms of heterogeneity affect properties such as dimensionality. Section~\ref{sec:heterogeneity} focuses on gain heterogeneity to model firing-rate diversity, but this framework could also incorporate more intricate cell-specific features, such as electrophysiological characteristics \citep{teeter2018generalized}, linking single-neuron heterogeneity to functional consequences for population activity.

\subsection{Transient dynamics and excitation-inhibition} 
Our analysis has focused on stationary states with translation-invariant temporal correlations. Many neural computations, however, rely on transient dynamics that break stationarity and depend on absolute time. Extending the DMFT accordingly is numerically tractable for two-point functions but would become unwieldy for four-point functions, $\Psi^a_\spt(t_1,t_2,t_3,t_4)$. Representing and manipulating such a large object may require approximations or low-rank decomposition.

One notable transient phenomenon is transient amplification, linked to non-normal coupling matrices that naturally arise from excitatory-inhibitory segregation (Dale’s law) \cite{murphy2009balanced, hennequin2014optimal}. Although the random-mode model does not enforce Dale’s law, the resulting matrices are nevertheless non-normal and may exhibit transient amplification; such effects may be amplified through structured off diagonal overlaps between left and right modes. Non-normality also decouples the highest-variance dimensions from the slowest modes. Relatedly, extending the random-mode model to respect Dale’s law and incorporating a nonsaturating $\phi(x)$---achieving stability via excitation-inhibition balance rather than saturation \cite{rajan2006eigenvalue, kadmon2015transition}---could enable more direct comparisons to both connectivity and activity data.

\subsection{Random basis property}
The random-mode model assumes that the $2M$ mode components associated with each neuron are sampled i.i.d.\ for each neuron [\eref{eq:iid_over_i}]. This leads to neuronal permutation symmetry of the distribution over $\bm{J}$. Furthermore, when the marginal distributions over left and right mode components are isotropic Gaussians, as we have assumed, the embedding into neuronal space is, by definition, random and Gaussian. This ``random basis property'' may underlie heterogeneous tuning observed in neural circuits \cite{clark2025symmetries}, but in our case it imposes modeling limitations: While the random-mode model has configurable spectral properties, it precludes configuration of spatial or anatomical properties.

This limitation could be addressed by conditioning the distribution over the $2M$ mode components on neuron-local properties such as brain region or spatial location, provided the number of effectively distinct possibilities is intensive. This would yield a regionally or spatially embedded random-mode model, apt for connectome studies.

Because we assume zero-mean mode components [\eref{eq:zero_mean_assum}], $\tavg{J_{ij}} = 0$ for $i \neq j$. Thus, the low-rank structure is \textit{not} inherited from the expected weight matrix. This differs from the mechanism described by \citet{thibeault2024low}, who analyze cases where low-rank structure arises because $\tavg{\bm{J}}$ is low rank, with the random part treated as noise. By Weyl’s inequality, the singular values of $\bm{J}$ differ from those of $\tavg{\bm{J}}$ by, at most, the spectral norm of the noise. This mechanism is fundamentally different from the random-mode model's zero-mean, randomly embedded structure. Which of these mechanisms is more prevalent in real-world networks is an open question.

\subsection{Alternative generative models}
\citet{tiberi2023hidden} proposed a generative model for coupling matrices that parametrizes the eigendecomposition of $\bm{J}$ rather than its SVD. Their formulation does not capture relationships between eigenvalues and eigenvectors. Our SVD-like approach allows simultaneous control over component strengths and their corresponding mode overlaps. Whereas \citet{tiberi2023hidden} studied noise-driven linear networks within a random-matrix-theory framework, our DMFT formulation applies to nonlinear dynamics.

Connectivity parametrizations resembling the random-mode model have been used to study the correspondence between spiking and rate-based recurrent neural networks \cite{schmutz2023emergent} and to expedite training spiking networks \cite{lin2024spiking}.

\section{Conclusion} The random-mode model and associated analytical techniques provide tools for exploring the relationship between connectivity and collective dynamics. Our parametrization of connectivity is a middle ground between intensive-rank models and i.i.d.\ random networks. Building on this framework stands to further bridge neuronal connectivity, activity, and function.

\section*{Acknowledgments}
We thank L. F. Abbott for feedback. A. v. M. was supported by the Swartz Foundation and the Kempner Institute for the Study of Natural and Artificial Intelligence. D. G. C, O. M., and A. L.-K. were supported by the Kavli Foundation and Gatsby Charitable Foundation (Grant No. GAT3708). D. G. C. was also supported by the National Institutes of Health (Grant No. T32NS064929). A. L.-K. was also supported by the National Science Foundation (Grant No. 2443158) and the National Institutes of Health (Grant No. RF1DA060772).

\begin{widetext}

\section*{Appendix}
\appendix

\section{Glossary of terms}
\label{sec:glossary_of_terms}

\hangindent=2em Number of neurons ($N$): Total number of neurons in the network.

\hangindent=2em Number of modes ($M$): Number of components in the random-mode model.

\hangindent=2em (Effective) coupling strength ($g$, $g_\text{eff}$): Standard deviation of the couplings times $\sqrt{N}$ in the i.i.d.\ connectivity model, random-mode model [\eref{eq:g_eff_rmm}], or variants thereof [e.g., see just below Eqs.~\eqref{eq:psi_G_unnorm} and \eqref{eq:psi_G_normalized}].

\hangindent=2em Random-mode model: Network connectivity model: $\bm{J} = \sum_{a=1}^M D_a \bm{\ell}_a \bm{r}_a^T$ [\eref{eq:rmm_sum_def}].

\hangindent=2em Connectivity component ($\bm{\ell}_a \bm{r}_a^T$): Outer product of output and right modes.

\hangindent=2em Right mode ($\bm{r}_a$): Neuronal pattern in the $a$th component onto which activity patterns are projected.

\hangindent=2em Left mode ($\bm{\ell}_a$): Neuronal pattern in the $a$th component along which the projected activity pattern is expanded.

\hangindent=2em Component strength ($D_a$): Scaling factor for the $a$th component.

\hangindent=2em Ratio of modes to neurons ($\alpha$): Given by $\alpha = M/N$ [\eref{eq:alpha_def}].

\hangindent=2em Participation ratio of component strengths ($\text{PR}^D$): Given by $\text{PR}^D = r_2^2 / r_4$ [\eref{eq:PRD_def}] where $r_n = \langle D^n \rangle_D$ [\eref{eq:D_moment}].

\hangindent=2em Effective rank ($\alpha \text{PR}^D$): Connectivity dimensionality measure [\eref{eq:eff_rank_def}].

\hangindent=2em Two-point function: Correlation function given by $C^\phi(t_1,t_2) = (1/N) \sum_{i=1}^N  \phi_i(t_1) \phi_i(t_2)$ [\eref{eq:nac}].

\hangindent=2em Four-point function: Higher-order correlation function given by $\Psi^\phi(\tau_1, \tau_2) = (1/N) \sum_{i,j=1}^N  C^\phi_{ij}(\tau_1) C^\phi_{ij}(\tau_2)$ [\eref{eq:psi-def-nrn}]. See also time-by-time definition [\eref{eq:psi-def-time}].

\section{Preprocessing of fly hemibrain connectome}
\label{sec:fly_appdx}

We use the dataset from \citet{scheffer2020connectome}, in which synaptic counts are reported for each connection. These counts determine the magnitudes of the coupling matrix elements. Neurotransmitter probabilities for each neuron are obtained from a machine-learning analysis of the electron microscopy data \cite{eckstein2024neurotransmitter}. Each neuron is assigned its most probable neurotransmitter, which is then mapped to a synaptic sign (excitatory or inhibitory) according to \hyperref[tab:table1]{Table~\ref{tab:table1}} \citep{liu2013glutamate,molina2019selectivity}.

\begin{table}
\centering
\begin{tabular}{lc}
\hline
{Neurotransmitter} & {Synaptic sign} \\
\hline
GABA & $-1$ (Inhibitory) \\
Acetylcholine & $+1$ (Excitatory) \\
Glutamate & $-1$ (Inhibitory) \\
Serotonin & Ignored \\
Octopamine & Ignored \\
Dopamine & Ignored \\
Neither & Ignored \\
\hline
\end{tabular}
\caption{Mapping of neurotransmitters to synaptic signs.}
\label{tab:table1}
\end{table}

Synaptic counts do not directly correspond to effective synaptic strengths, so it is reasonable to choose ``fudge factors'' to relate the two. Our primary goal in choosing these factors is to standardize the spectrum of the coupling matrix so that it is roughly confined to the unit disk. At the same time, we want to preserve key structural features of the original connectivity.

Specifically, the preprocessing is designed to preserve:
\begin{enumerate}[label=(\roman*), itemsep=-0.2em]
\item the sparsity pattern (for which the connections are zero versus nonzero);
\item the sign of each connection (excitatory versus inhibitory);
\item approximately, the relative magnitudes of same-sign inputs to each neuron; and
\item approximately, the relative magnitudes of same-sign outputs from each neuron.
\end{enumerate}
The last two properties are preserved only approximately because they are enforced simultaneously by the iterative normalization procedure described below.

We normalize the coupling matrix using an iterative scaling method analogous to the Sinkhorn-Knopp algorithm. First, for each row, we rescale the positive elements by a row-specific factor so that their $L2$ norm equals $1/\sqrt{2}$. We then rescale the negative elements, using a separate factor, so that their $L2$ norm also equals $1/\sqrt{2}$. This yields rows with a total $L2$ norm of one and balanced excitatory and inhibitory contributions. A small number of neurons with only excitatory or only inhibitory inputs retain an $L2$ norm of $1/\sqrt{2}$. 

Next, we apply the same rescaling procedure to each column, which disrupts the previous row normalization. We alternate the row and column scaling steps until convergence. The result can be interpreted as the projection of the original matrix onto the set of matrices that are properly normalized along both rows and columns, while keeping the same nonzero elements and signs.

This iterative normalization provides a principled way to choose the required scale factors while achieving our goal of a standardized spectrum.

\section{Free probability calculation}
\label{sec:free_prob_calc}

We wish to determine the limiting density of the singular values of
\begin{equation}
    \bm{J} = \bm{L} \bm{R}^T,
\end{equation}
where $\bm{L}$ and $\bm{R}$ are independent $N \times M$ random matrices with $\alpha = M/N$ and entries of variance $1/N$.
Equivalently, we can compute the eigenvalues of
\begin{equation}
    \bm{J}^T \bm{J} = \bm{R} \bm{L}^T \bm{L} \bm{R}^T
\end{equation}
and then take the square root to obtain the singular values. These eigenvalues are the same as those of
\begin{equation}
    \bm{L}^T \bm{L}  \bm{R}^T \bm{R},
\end{equation}
which is the product of two independent Wishart matrices,
\begin{equation}
    \bm{L}^T \bm{L}  \bm{R}^T \bm{R} = \bm{W}_L \bm{W}_R.
\end{equation}

The Stieltjes transform $G(z)$ of a spectral density $\rho(\lambda)$ is defined as
\begin{align}
    G(z) &= \int_{-\infty}^\infty  \frac{d\lambda \rho(\lambda)}{z - \lambda} , 
    \quad z \in \mathbb{C} \setminus \mathbb{R}, \\
    \rho(\lambda) &= -\frac{1}{\pi} \lim_{\epsilon \to 0^+} \mathrm{Im} G(\lambda + i\epsilon).
\end{align}
In a two-dimensional electrostatics analogy, the real and imaginary parts of $G(z)$ correspond to the Cartesian components of the electric field generated by a line charge density $\rho(\lambda)$ on the real axis. The inversion formula is then analogous to Gauss's law, relating the discontinuity of the normal field component to the local charge density.

The moment-generating function and $S$ transform are defined by
\begin{align}
    M(z) &= z G(z) - 1, \\
    S(z) &= \frac{1+z}{z M^{-1}(z)},
\end{align}
where $M^{-1}(z)$ is the functional inverse of $M(z)$. For a Marchenko-Pastur distributed Wishart matrix $\bm{W}$ with aspect ratio $\alpha$,
\begin{equation}
    S(z) = \frac{1}{1 + \alpha z}.
\end{equation}
A key fact in free probability is that {the $S$ transforms of free variables multiply.} Thus, for $\bm{W}_1 \bm{W}_2$, the $S$ transform is
\begin{equation}
    S(z) = \frac{1}{(1 + \alpha z)^2}.
\end{equation}
From the definition of the $S$ transform, the inverse moment-generating function is
\begin{equation}
    M^{-1}(z) = \frac{1+z}{z S(z)} 
              = \frac{(1+z)(1 + \alpha z)^2}{z}.
\end{equation}
Therefore, $M(z)$ satisfies
\begin{equation}
    z = \frac{[1+M(z)][1 + \alpha M(z)]^2}{M(z)}.
\end{equation}
Using $M(z) = z G(z) - 1$, we obtain a cubic equation for $G(z)$,
\begin{equation}
    -\alpha^{2} z^{3} G(z)^{3} 
    + 2 z^{2} (\alpha^{2} - \alpha) G(z)^{2} 
    + z \big( z + 2\alpha - \alpha^{2} - 1 \big) G(z) 
    - z = 0.
\end{equation}
In principle, this can be solved using the cubic formula, but the explicit form is cumbersome. For the support boundaries of $\rho(\lambda)$, we need only the points where $\mathrm{Im} G(z)$ changes from zero to nonzero. This occurs when the discriminant of the cubic vanishes.
The discriminant is
\begin{equation}
    \Delta = \alpha^{2} \lambda^{7} 
    \left[ 4 \lambda^{2} 
    + (\alpha^{2} - 20\alpha - 8) \lambda 
    + (-4\alpha^{3} + 12\alpha^{2} - 12\alpha + 4) \right],
\end{equation}
where we replaced $z$ with $\lambda$ to indicate restriction to the real axis. Setting $\Delta = 0$ gives $\lambda = 0$ or the nontrivial quadratic equation
\begin{equation}
    4 \lambda^{2} 
    + (\alpha^{2} - 20\alpha - 8) \lambda 
    + (-4\alpha^{3} + 12\alpha^{2} - 12\alpha + 4) = 0.
\end{equation}
The roots of this quadratic are
\begin{equation}
    \lambda_\pm 
    = 1 + \frac{5\alpha}{2} - \frac{\alpha^{2}}{8} 
    \pm  \left( 1 + \frac{\alpha}{8} \right)^{3/2} \sqrt{8\alpha}.
\end{equation}
The singular values correspond to the square roots of the eigenvalues, given by \eref{eq:free_prob}. We confirm this formula numerically in Fig.~\ref{fig:free_prob}.

\begin{figure*}
    \centering
    \includegraphics[width=\linewidth]{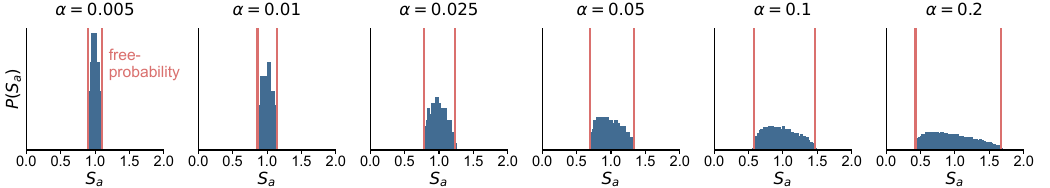}
    \caption{Distribution of singular values in the random-mode model. Histograms show singular values of $\bm{L}\bm{D}\bm{R}^T$ with $D_a = 1$ for various $\alpha$. For $\alpha \ll 1$, the singular values concentrate at 1. Vertical lines, prediction of \eref{eq:free_prob}. $N = 5000$.}
    \label{fig:free_prob}
\end{figure*}

\section{Alternative measures of effective dimensionality}
\label{sec:appdx_alt_dim}

Alternative measures of effective dimensionality, which could be applied to both connectivity and activity in place of the participation ratio, have been proposed in the literature. \citet{roy2007effective} define effective rank as the exponential of the entropy of a categorical probability distribution obtained by normalizing the eigenvalues to sum to 1, i.e., $p_i = \lambda_i / \sum_{j=1}^N \lambda_j$. Written in terms of this distribution, the participation ratio is $\text{PR} = \left(\sum_{i=1}^N  p_i^2\right)^{-1}$, while the exponentiated entropy is $\exp (H) = \exp\left(-\sum_{i=1}^N  p_i \log p_i\right)$. By Jensen's inequality, $\text{PR} \leq \exp(H)$. Exact equality holds in certain cases, including when exactly $K \leq N$ eigenvalues are equal and the rest are zero, yielding $\text{PR} = \exp(H) = K$. Furthermore, for smoothly decaying spectra as discussed above [for which $p_i = f(i/Nw)/\sum_{j=1}^N f(j/Nw)$, with $w \ll 1$], both measures have $\text{PR} = c_1 \: Nw$ and $\exp(H) = c_2 \: Nw$, where $c_1$ and $c_2$ are $N$- and $w$-independent constants depending on the specific form of $f$. Thus, while the measures differ in detail, they provide similar characterizations of effective dimensionality. This entropy-based definition has been used in neuroscience applications, for example, in Ref.~\cite{christodoulou2022regimes} to analyze transient amplification in neural networks. Both measures capture the number of dominant modes in the system. Thibeault \textit{et al.} (Ref.~\cite{thibeault2024low}) call (a metric closely related to) the the participation ratio-based metric the ``srank'' (the difference being whether the maximum eigenvalue or sum over all eigenvalues is used as a normalizing factor) and call the entropylike metric the ``erank.''

\section{Review: Calculating two- and four-point functions for i.i.d.\ couplings}
\label{sec:computing}

We now review how to compute these two- and four-point functions in a classic network model with i.i.d.\ couplings $\bm{J}$. The first- and second-order coupling statistics are
\begin{align}
    \tavg{J_{ij}} = 0, 
    \hspace{1.5em} \tavg{J_{ij}^2} = \frac{g^2}{N}. \label{eq:g}
\end{align}
The magnitude of a typical coupling is $g / \sqrt{N}$. We refer to $g$ as the coupling strength. For the canonical dynamics $T[x](t) = (1+\partial_t)x(t)$, the network is quiescent for $g < 1$ and chaotic for $g > 1$, with a sharp phase transition as $N \to \infty$. In this paper we assume that the network is nonquiescent.

The two-point function $C^a_{\spt}(t_1, t_2)$ can be computed through a single-site picture that describes the dynamics of a typical neuron embedded in the rest of the network, with preactivation $x(t)$ and activation $\phi(t)$. The single-site dynamics are given by
\begin{equation}
    T[x](t) = \eta^x(t), \label{eq:basic_ssp}
\end{equation}
where $\eta^x(t)$ is a Gaussian field with mean zero and covariance $g^2 C^\phi_\spt(t_1, t_2)$. We denote this by
\begin{align}
     \eta^x \sim \mathcal{GP}(0, g^2 C^\phi_\spt).
     \label{eq:eta-stats}
\end{align}
$C^\phi_\spt(t_1, t_2)$ is determined self-consistently by enforcing
\begin{equation}
    C^\phi_{\spt}(t_1, t_2) = \tavg{\phi(t_1) \phi(t_2)}_{\spt}, \label{eq:basic_scc}
\end{equation}
where $\tavg{\cdots}_{\spt}$ denotes an average within this single-site process, i.e., with respect to the Gaussian distribution of $\eta^x(t)$. Once $C^\phi_\spt(t_1, t_2)$ has been determined, $C^x_\spt(t_1, t_2)$ follows easily. This single-site problem can be derived through either a single-site cavity calculation (see Ref.~\cite{mezard1987spin}) or a saddle-point condition in a path integral (see Refs.~\cite{crisanti2018path,helias2020statistical}). For the i.i.d.\ couplings considered here, there is a simpler heuristic derivation: In the neuronal input $\sum_{j=1}^N J_{ij}\phi_j(t)$, the correlations between the couplings $J_{ij}$ and dynamic variables $\phi_j(t)$ can be safely neglected to leading order in $1/N$, yielding both Gaussianity of $\eta^x(t)$ by the central limit theorem and the second-order statistics of \eref{eq:eta-stats}.

\citet{clark2023dimension} first computed $\Psi_{\spt}^a(\bm{\tau})$ using a dynamic, two-site version of the cavity method, based on the neuron-by-neuron definition [\eref{eq:psi-def-nrn}]. This method finds $\psi^a_{\spt}(\bm{\tau})$, the off diagonal contribution to $\Psi^a_{\spt}(\bm{\tau})$, noting that the on-diagonal contribution is simply $C^a_{\spt}(\tau_1) C^a_{\spt}(\tau_2)$. A cavity is first created by removing two neurons from the network and allowing the rest of the network, the reservoir, to generate dynamic activity. The cavity neurons are then introduced, and their effect on the reservoir is treated perturbatively. This yields a pair of coupled mean-field equations for the cavity units, generalizing the single-site picture discussed above to a two-site picture. Finally, self-consistency conditions are constructed by recognizing that the cavity pair is statistically equivalent to any reservoir pair. This calculation results in expressions for the four-point function in Fourier space, given by
\begin{equation}
    \Psi_{\spt}^\phi(\bm{\omega}) = \frac{C_{\spt}^\phi(\omega_1) C_{\spt}^\phi(\omega_2)}{|1 - g^2 S_{\spt}^\phi(\omega_1) S_{\spt}^\phi(\omega_2)|^2}
    \label{eq:psi_phi_sol}
\end{equation}
for the activations, and 
\begin{subequations}
\begin{align}
    \Psi_{\spt}^x(\bm{\omega}) &= C_{\spt}^x(\omega_1)C_{\spt}^x(\omega_2) + |U(\bm{\omega})|^2 C_{\spt}^\phi(\omega_1)C_{\spt}^\phi(\omega_2) \nonumber \\
    &+ U(\bm{\omega}) C_{\spt}^{x\phi}(\omega_1)C_{\spt}^{x\phi}(\omega_2) + \text{H.c.}
    \label{eq:psi-x-formula}, \\
    U(\bm{\omega}) &= \frac{g^2 S_{\spt}^x(\omega_1)S_{\spt}^x(\omega_2)}{1 - g^2 S_{\spt}^\phi(\omega_1)S_{\spt}^\phi(\omega_2)},
    \label{eq:Psi_x_result_cavity_U}\end{align}\label{eq:Psi_x_combined}\end{subequations}
for the preactivations, where $\bm{\omega} = (\omega_1, \omega_2)$ and $C^{x\phi}_\spt(\omega)$ is a cross-covariance between the preactivation and activation. Note that, if the joint distribution of preactivations were Gaussian, $\Psi^x_\spt(\bm{\tau})$ and $\Psi^\phi_\spt(\bm{\tau})$ would differ only by a proportionality constant due to Price's theorem \cite{price1958useful}. The more complex relationship observed here reflects the non-Gaussian joint statistics across different neurons, which is relevant because the network is nonlinear.

\section{Path-integral calculation of two- and four-point functions for i.i.d.\ couplings}
\label{sec:path-integral}

\begin{figure*}
    \centering
    \includegraphics[width=\linewidth]{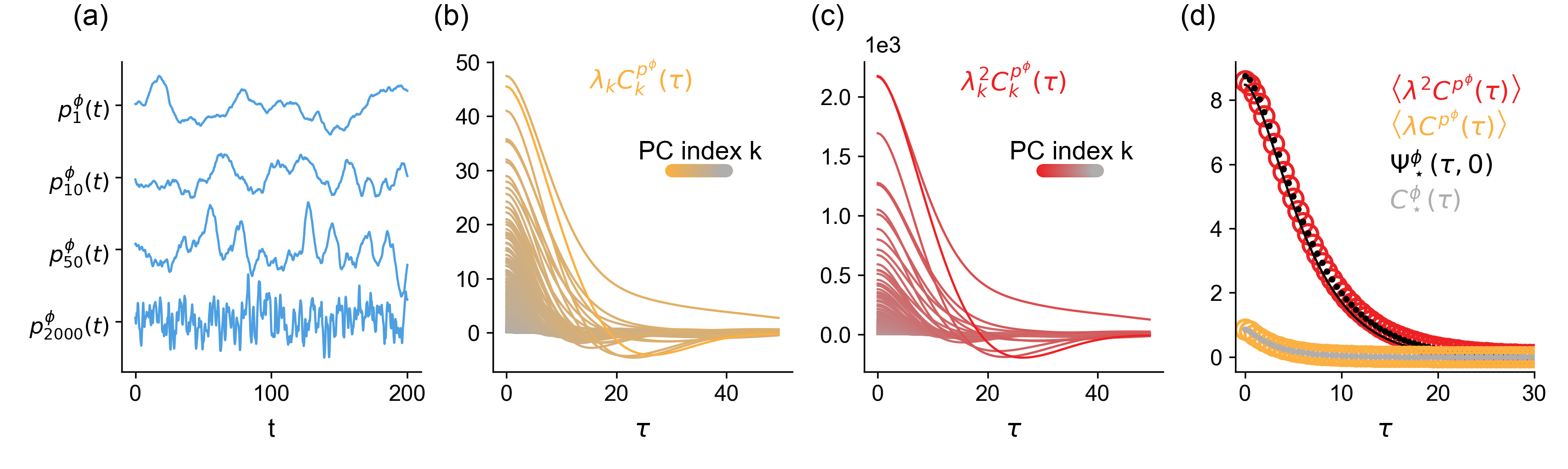}
    \caption{Relationship between principal component timescales and two- and four-point correlation functions. (a) Example activity traces of principal components $p^\phi_k(t)$. (b) Empirical correlation functions for each principal component (PC), weighted by their eigenvalues. (c) Same as (b) but weighted by the squared eigenvalue. (d) Averages over curves in (b) and (c), with comparisons to $C^\phi_\spt(\tau)$ and $\Psi^\phi_\spt(\tau,0)$. Dots, simulation results. Lines, theoretical predictions. 
    All panels describe the same network of $N = 4000$ neurons with i.i.d.\ couplings with variance $g^2/N$ with coupling strength $g = 6$.}
    \label{fig:Psi_timescale_match}
\end{figure*}

The general program, for i.i.d.\ or structured couplings, is as follows:
\begin{enumerate}[itemsep=-0.2em]
    \item {Formulate the field theory.} Begin with the path integral $Z[\bm{J}]$ and average over the connectivity disorder. Introduce auxiliary fields representing two-point functions [such as $C^\phi(t_1,t_2)$] along with their conjugate partners. Use integral representations of $\delta$ functions to enforce field definitions, factorizing the exponential over extensive dimensions. This yields a statistical field theory with action $N$ times an order-one function of the auxiliary fields.

    \item {Solve the saddle-point equations.} Find the saddle point by setting action derivatives to zero. This produces self-consistent equations for the two-point functions, completing the standard DMFT analysis.
    
    \item {Compute the Hessian.} Calculate the time$^2$-by-time$^2$ Hessian matrix, which depends on four time variables $(t_1,t_2,t_3,t_4)$ and characterizes fluctuations around the saddle point.

    \item {Apply temporal separation limits.} Impose the temporal limits [\eref{eq:temporal-limits}], making the Hessian blocks translation-invariant functions of time differences $(\tau_1, \tau_2)$. Transform to Fourier space to obtain a low dimensional, frequency-dependent Hessian whose elements are functions of $(\omega_1, \omega_2)$.

    \item {Invert the frequency-space Hessian.} Perform matrix inversion to directly obtain $\Psi^a_\spt(\bm{\omega})$. If needed, apply inverse Fourier transform to recover $\Psi^a_\spt(\bm{\tau})$.
\end{enumerate}

Here we consider i.i.d.\ $\bm{J}$ with mean zero and variance $g^2/N$. After performing the Gaussian integration over $\bm{J}$, we introduce an auxiliary field $C^\phi(t_1, t_2)$, defined by \eref{eq:nac}, and its conjugate $\hat{C}^\phi(t_1, t_2)$ to factorize the action across the neuron index $i$.
This leads to a partition function for a statistical field theory involving $C^\phi(t_1,t_2)$ and $\hat{C}^\phi(t_1, t_2)$:
\begin{equation}
    Z = \pin{C}^\phi \pinh{C}^\phi \exp\left(-N\mathcal{S}[C^\phi, \hat{C}^\phi]\right),
\end{equation}
where the intensive action is
\begin{equation}
    \mathcal{S}[C^\phi, \hat{C}^\phi] = -\frac{1}{2}C^\phi \hat{C}^\phi - \log W[C^\phi, \hat{C}^\phi],
\end{equation}
and the single-site path integral is
\begin{equation}
    W[C^\phi, \hat{C}^\phi] = \pin{x}\pinh{x}  
     \exp\left(
    i\hat{x} T[x] - \frac{g^2}{2}\hat{x} C^\phi \hat{x} - \frac{1}{2} \phi \hat{C}^\phi \phi
    \right). \label{eq:sspi}
\end{equation}

\subsection{Two-point functions}

In the limit $N\to \infty$, the saddle point dominates this integral. The derivatives of the action, which are zero at the saddle point, are
\begin{align}
    \ddr{\mathcal{S}}{ C^\phi(t_1, t_2)} &= -\hat{C}^\phi(t_1,t_2) + g^2 \tavg{\hat{x}(t_1)\hat{x}(t_2)}_{W}, \\
    \ddr{\mathcal{S}}{ \hat{C}^\phi(t_1, t_2)} &= -{C}^\phi(t_1,t_2) + \tavg{{\phi}(t_1){\phi}(t_2)}_{W},
\end{align}
where $\tavg{\cdots}_W$ denotes an average within the dynamic process described by $W[C^\phi, \hat{C}^\phi]$. In evaluating derivatives of the action, we follow the rule that derivatives with respect to $C^\phi(t_1, t_2)$ also affect $C^\phi(t_2, t_1)$, and likewise for $\hat{C}^\phi(t_1,t_2)$, due to the symmetry in the action. Using the vanishing of correlation functions involving only the conjugate field $\hat{x}(t)$, the saddle-point conditions yield
\begin{align}
    \hat{C}_{\spt}^\phi(t_1,t_2) &= 0, \\
    C_{\spt}^\phi(t_1, t_2) &= \tavg{\phi(t_1)\phi(t_2)}_{\spt},
\end{align}
where $\tavg{\cdots}_{\spt}$ denotes an average within the dynamic process described by $W[C_{\spt}^\phi, 0]$. This recovers the same single-site process [Eqs.~\eqref{eq:basic_ssp} and \eqref{eq:eta-stats}] and self-consistency condition [\eref{eq:basic_scc}] described in \aref{sec:computing}.

\subsection{Four-point functions}
\label{sec:fptfn}

We now compute $\Psi^a_\spt(\bm{\tau})$ within the path-integral formalism. Fluctuations around the saddle point derived above are governed by the Hessian of the action \citep{crisanti2018path, van2021large}. This Hessian has blocks given by
\begin{subequations}
\begin{align}
    H_{C^\phi C^\phi}(t_1,t_2,t_3,t_4) &= \frac{\delta^2 S}{\delta C^\phi(t_1,t_2) \delta C^\phi(t_3,t_4)} = -g^4 \tavg{\tavg{\hat{x}(t_1) \hat{x}(t_2) , \hat{x}(t_3) \hat{x}(t_4)}}_{\spt}, \\
    H_{C^\phi \hat{C}^\phi}(t_1,t_2,t_3,t_4) &= \frac{\delta^2 S}{\delta C^\phi(t_1,t_2) \delta \hat{C}^\phi(t_3,t_4)} = - I(t_1,t_2,t_3,t_4) - g^2 \tavg{\tavg{\hat{x}(t_1)\hat{x}(t_2), \phi(t_3)\phi(t_4) }}_{\spt}, \\
    H_{\hat{C}^\phi \hat{C}^\phi }(t_1,t_2,t_3,t_4) &= \frac{\delta^2 S}{\delta \hat{C}^\phi (t_1,t_2) \delta \hat{C}^\phi (t_3,t_4)} = -\tavg{\tavg{\phi(t_1) \phi(t_2) , \phi(t_3) \phi(t_4)}}_{\spt},
\end{align} \label{eq:hessian_blocks}
\end{subequations}
where $\tavg{\tavg{A,B}}_{\spt} = \tavg{AB}_{\spt} - \tavg{A}_{\spt} \tavg{B}_{\spt}$ and $I(t_1,t_2,t_3,t_4)
   = \delta(t_1 - t_3)\delta(t_2-t_4) + \delta(t_1 - t_4)\delta(t_2-t_3)$. Expanding the action to second order around the saddle point, the path integral becomes
\begin{equation}
    Z = \exp\left(-N \mathcal{S}[C^\phi_{\spt}, 0] \right) \int \mathcal{D}\delta C^\phi \mathcal{D}\delta \hat{C}^\phi 
    \exp\left(
    -\frac{N}{2}
    \begin{pmatrix}\delta C^\phi \\ \delta \hat{C}^\phi\end{pmatrix}^T
    \begin{pmatrix}H_{C^\phi C^\phi} & H_{C^\phi\hat{C}^\phi} \\ H_{\hat{C}^\phi{C}^\phi}^T & H_{\hat{C}^\phi \hat{C}^\phi}\end{pmatrix} 
    \begin{pmatrix}\delta C^\phi \\ \delta \hat{C}^\phi\end{pmatrix}
    \right), \nonumber 
\end{equation}
revealing the structure of the Gaussian fluctuations around the saddle point. The covariance matrix among the fluctuation variables, $\delta C^\phi(t_1, t_2)$ and $\delta \hat{C}^\phi(t_1, t_2)$, is $1/N$ times the inverse Hessian, yielding
\begin{align}
    N\tavg{\delta C^\phi(t_1,t_2) \delta C^\phi(t_3, t_4)}_{\spt} = \left[H_{C^\phi C^\phi} - H_{C^\phi \hat{C}^\phi} H_{\hat{C}^\phi \hat{C}^\phi}^{-1}H_{\hat{C}^\phi C^\phi}\right]^{-1}(t_1,t_2,t_3,t_4). \nonumber
\end{align}
In principle, we have all the necessary components to evaluate this expression: The single-site dynamics are known via the saddle-point condition, and the Hessian blocks are given in terms of connected correlation functions in this single-site process. We could then take the temporal separation limits of \eref{eq:temporal-limits} to obtain $\Psi^\phi_{\spt}(\bm{\tau})$. The challenge is that each Hessian block is a complicated time$^2$-by-time$^2$ matrix that, for general values of the time variables, has no analytic inverse. To circumvent this problem, we show in \aref{sec:sep_limits} that the temporal limits of \eref{eq:temporal-limits} can be taken \textit{before} taking the inverses. Under these temporal limits, the Hessian blocks are
\begin{align}
    & H_{\hat{C}^\phi \hat{C}^\phi }(t_1,t_2,t_3,t_4) = -C^\phi_{\spt}(\tau_1)C^\phi_{\spt}(\tau_2), \\
    & H_{C^\phi \hat{C}^\phi}(t_1,t_2,t_3,t_4) = - \delta(\tau_1)\delta(\tau_2) +g^2 S^\phi_{\spt}(\tau_1)S^\phi_{\spt}(\tau_2),  \\
    & H_{C^\phi C^\phi}(t_1,t_2,t_3,t_4) = 0,
\end{align}
where we used the fact that multiplication by $-i\hat{x}(t)$ is equivalent to a functional derivative with respect to a source at time $t$.
Since the relevant quantities are now translation invariant, i.e., depend only on $\tau_1$ and $\tau_2$, we can transform to Fourier space. In this representation, each Hessian block becomes a frequency-dependent scalar,
\begin{align}
    H_{\hat{C}^\phi \hat{C}^\phi }(\bm{\omega}) &= -C^\phi_{\spt}(\omega_1)C^\phi_{\spt}(\omega_2), \\
    H_{C^\phi \hat{C}^\phi}(\bm{\omega})
    &= -1 + g^2 S^\phi_{\spt}(\omega_1)S^\phi_{\spt}(\omega_2), \\
    H_{C^\phi C^\phi}(\bm{\omega}) &= 0.
\end{align}
In summary, the full Hessian can be replaced by a $2 \times 2$, frequency-dependent Hessian given by
\begin{equation}
    \bm{H} = \begin{pmatrix}
    0 & -1 + g^2 (S^{\phi}_{12})^* \\
    -1 + g^2 S^\phi_{12} & -C^\phi_{12}
    \end{pmatrix},
    \label{eq:freq-dep-H}
\end{equation}
where we simplified notation by suppressing frequency arguments via the shorthand $C^\phi_{12} = C^\phi_{\spt}(\omega_1) C^\phi_{\spt}(\omega_2)$ and $S^\phi_{12} = S^\phi_{\spt}(\omega_1) S^\phi_{\spt}(\omega_2)$. The Fourier-space function $\Psi_{\spt}^\phi(\bm{\omega})$ is the upper-left element of the inverse of this matrix. Performing the $2\times 2$ matrix inversion gives
\begin{equation}
    \Psi_{\spt}^\phi = \frac{C^{\phi}_{12} }{|1 - g^2 S^\phi_{12} |^2},
    \label{eq:psi_iid}
\end{equation}
which agrees with the two-site cavity result [\eref{eq:psi_phi_sol}].

\subsection{Adding sources}
\label{sec:srcs}

Above, computing $\Psi^\phi_{\spt}(\bm{\tau})$ was simplified by $C^\phi(t_1, t_2)$ appearing naturally in the statistical field theory resulting from integrating out $\bm{J}$. To compute $\Psi^x_\spt(\bm{\tau})$, we need to introduce a source-field term in the action.

Consider a general intensive action $\mathcal{S}[\bm{C}; \mathcal{J}]$, where $\bm{C}$ is a collection of fields (e.g., $\bm{C} = \{C^\phi, \hat{C}^\phi\}$), and $\mathcal{J}$ is a source. Our goal is to compute
\begin{equation}
    \left. \frac{\delta^2}{\delta \mathcal{J}^2} \log Z \right|_{\mathcal{J}=0},
    \label{eq:goal}
\end{equation}
which gives the fluctuations around the mean of the quantity multiplying $\mathcal{J}$ in the action. 
To facilitate this calculation, we introduce a new field $U$ and set it equal to $\mathcal{J}$ using the conjugate $\hat{U}$:
\begin{equation}
    Z[\mathcal{J}] = \pin{\bm{C}}\pin{U}\pinh{U} \exp\left(-N\hat{U}U +N\hat{U}\mathcal{J} - NS[\bm{C}; U] \right).
\end{equation}
Instead of taking the second derivative with respect to $\mathcal{J}$, we set $\mathcal{J} = 0$ and compute the fluctuations of $\hat{U}$, measured by \eref{eq:goal}, in the augmented theory whose path integral is
\begin{equation}
    Z = \pin{\bm{C}} \pin{U} \pinh{U} \exp(-N\tilde{\mathcal{S}}[\bm{C}, U, \hat{U}]),
    \label{eq:augmented_Z}
\end{equation}
where
\begin{equation}
\tilde{\mathcal{S}}[\bm{C},U,\hat{U}] = \hat{U}U + \mathcal{S}[\bm{C};U].
\end{equation}
The Hessian of the augmented action, in terms of the Hessian of the original action, is
\begin{equation}
    \left(
    \begin{array}{c|cc}
        \bm{H} & \frac{\partial^2 S}{\partial \mathcal{J} \partial \bm{C}} & \bm{0} 
        \\[0.25em]
        \hline
        \\[-1em]
        \left(\frac{\partial^2 S}{\partial \mathcal{J} \partial \bm{C}}\right)^T & \frac{\partial^2 S}{\partial \mathcal{J}^2} & 1 \\[0.5em]
        \bm{0}^T & 1 & 0
    \end{array}
    \right),
    \label{eq:augmented_Hessian}
\end{equation}
where
\begin{equation}
    H_{ab} = \frac{\delta^2 S}{\delta C_a \delta C_b}, 
   \quad 
    \left[\frac{\delta^2 S}{\delta \mathcal{J} \delta \bm{C}}\right]_a = \frac{\delta^2 S}{\delta \mathcal{J} \delta C_a}.
\end{equation}
The fluctuations of $\hat{U}$ are given by the bottom-right element of the inverse Hessian evaluated at the saddle point. Using a Schur complement to compute this element, we obtain
\begin{equation}
   \left. \frac{\delta^2}{\delta \mathcal{J}^2} \log Z \right|_{\mathcal{J}=0} = \left[ -\frac{\delta^2 S}{\delta \mathcal{J}^2} + 
    \left(\frac{\delta^2 S}{\delta \mathcal{J} \delta \bm{C}}\right)^T\bm{H}^{-1} \frac{\delta^2 S}{\delta \mathcal{J} \delta \bm{C}}\right]_{\mathcal{J}=0},
    \label{eq:fluctuation_formula}
\end{equation}
with all quantities evaluated at the saddle point. By applying this formula to the action of the i.i.d.\ model with the source term $-\frac{1}{2} x \mathcal{J} x$ added to the single-site path integral [\eref{eq:sspi}], we recover the expression for $\Psi^x_\spt(\bm{\tau})$ from the two-site cavity method [\citet{clark2023dimension}; \eref{eq:Psi_x_combined}]. For actions of the form $\mathcal{S}[\bm{C}; \mathcal{J}] = -\mathcal{J} C_a + \text{$\mathcal{J}$-independent terms}$, this formula reduces to
$
   ({\delta^2}/{\delta \mathcal{J}^2}) \log Z |_{\mathcal{J}=0} = [\bm{H}^{-1}]_{aa},
$
as expected.

We now apply \eref{eq:fluctuation_formula} to the action of the i.i.d.\ connectivity model with a source term for correlations among preactivations. The action is
\begin{equation}
    \mathcal{S}[C^\phi, \hat{C}^\phi] = -\frac{1}{2}C^\phi \hat{C}^\phi
    -\log \pin{x} \pinh{x} \exp\bigg{(}i\hat{x}T[x] -\frac{g^2}{2}\hat{x}C^\phi\hat{x} - \frac{1}{2}\phi \hat{C}^\phi \phi - \frac{1}{2}x\mathcal{J}x \bigg{)}.
    \label{eq:action_with_source}
\end{equation}
Computing the necessary quantities and taking the temporal limits of \eref{eq:temporal-limits}, we obtain, in Fourier space,
\begin{equation}
    \frac{\delta^2 S}{\delta \mathcal{J}^2} = -C_{12}^x, \quad 
    \frac{\delta^2 S}{\delta \mathcal{J} \delta \bm{C}}
    = \begin{pmatrix}
        g^2 \left(S_{12}^x\right)^* \\
        -C_{12}^{x\phi}
    \end{pmatrix},
\end{equation}
where $C^{x\phi}_{12} = C_\spt^{x\phi}(\omega_1)C_\spt^{x\phi}(\omega_2)$ and $S^{x}_{12} = S^x_\spt(\omega_1)S^x_\spt(\omega_2)$.
Substituting these into \eref{eq:fluctuation_formula} and using the frequency-dependent Hessian from \eref{eq:freq-dep-H}, we obtain
\begin{equation}
    \Psi_\spt^x = C^x_{12} + |U|^2 C^\phi_{12} + U C^{x\phi}_{12} + \text{H.c.}, \quad \text{where  } U = \frac{g^2 S^x_{12}}{1 - g^2 S^\phi_{12}},
    \label{eq:Psi_x_result}
\end{equation}
in agreement with the two-site cavity result of Clark \textit{et al.} (Ref.~\cite{clark2023dimension}), given in \eref{eq:Psi_x_combined}.

\subsection{Other kinds of path integrals}
\label{sec:other_kidns_of_path_ints}

Different path-integral formalisms have been applied to neural-network models depending on the underlying dynamics. The Doi-Peliti formalism \cite{buice2007field} derives path integrals from master equations for systems with non-negative integer degrees of freedom (such as spiking neural networks) that evolve according to master equations, using an operator formalism as an intermediate step. In contrast, we use the MSRJD formalism for continuous-time stochastic differential equations, which naturally describes our rate-based network dynamics. For a comprehensive discussion of different path-integral approaches in neural-network theory, see Ref.~\cite{clark2025transient}.

\section{Temporal limits and inverses}
\label{sec:sep_limits}

Here, we demonstrate, for the i.i.d.\ model, that the temporal limits in \eref{eq:temporal-limits} can be taken before the inverses. This allows us to work with a frequency-dependent Hessian. In this section, we use Einstein notation for integrals over time variables (i.e., repeated time indices are integrated over). We would like to evaluate
\begin{equation}
    \Psi^\phi(\tau_1,\tau_2)  
     = U^{-1}(t_1,t_2,s_1,s_2) K(s_1,s_2,s_3,s_4) U^{-1}(t_1+\tau_1,t_2+\tau_2,s_3,s_4) \label{eq:main-equation} 
\end{equation}
in the limits of \eref{eq:temporal-limits}, where $U(t_1,t_2,t_3,t_4) = H_{\hat{C}^\phi C^\phi}(t_1,t_2,t_3,t_4)$ and $K(t_1,t_2,t_3,t_4) = H_{\hat{C}^\phi \hat{C}^\phi }(t_1,t_2,t_3,t_4)$ [note that $H_{C^\phi C^\phi}(t_1,t_2,t_3,t_4) = 0$ due to the vanishing of correlation functions involving only the conjugate field]. Restating \eref{eq:hessian_blocks}, we have
\begin{align}
    U(t_1,t_2,t_3,t_4) \label{eq:U-definition} 
    &= -\left[\delta(t_1-t_3)\delta(t_2-t_4) + \delta(t_1-t_4)\delta(t_2-t_3)\right] 
    - g^2 \langle\langle \phi(t_1)\phi(t_2), \hat{x}(t_3)\hat{x}(t_4) \rangle\rangle_\spt,  \\
    K(t_1,t_2,t_3,t_4) &= \langle\langle\phi(t_1)\phi(t_2),\phi(t_3)\phi(t_4)\rangle\rangle_\spt. \label{eq:K-definition}
\end{align}
We assume temporal stationarity throughout this section and consider only $a=\phi$.

We start by examining the properties of $U(t_1,t_2,t_3,t_4)$. $U$ vanishes if one of the four time points is far [measured on the timescale of $C^\phi_\spt(\tau)$] from the other three. For $t_1$ or $t_2$, this is a consequence of the perturbation associated with $\hat{x}(t_1)$ or $\hat{x}(t_2)$ being far away; for $t_3$ or $t_4$, this is a consequence of $\langle\phi(t_i)\rangle_\spt = 0$ separating from the expectation. $U$ also vanishes if there is more than one time point which is far from the remaining ones by the same arguments. Thus, the time points need to be close in pairs. However, if $t_1,t_2$ are close and $t_3,t_4$ are close, but the pairs are distant, $U$ still vanishes since the perturbations associated with $\hat{x}(t_i)$ are both far. This leaves three possibilities for a nonvanishing $U$: $t_1,t_3$ close and $t_2,t_4$ close, but the pairs are distant; $t_1,t_4$ close and $t_2,t_3$ close, but the pairs are distant; or all time points are close. 

Separating the three possibilities into two subdomains where all time points are close ($c$) and where the time points belong to either of the two pairwise separate ($s$) possibilities, we write $U=U_s+U_c$, where $U_s$ and $U_c$ are defined to vanish on the respective other domain. On the separated domain, $U$ simplifies to 
\begin{multline}
U_s(t_1,t_2,t_3,t_4)=
-\left[\delta(t_1-t_3)\delta(t_2-t_4)+\delta(t_1-t_4)\delta(t_2-t_3)\right] \\
+g^2\left[S^\phi_\spt(t_1-t_3)S^\phi_\spt(t_2-t_4)+S^\phi_\spt(t_1-t_4)S^\phi_\spt(t_2-t_3)\right],
\end{multline}
where we used that the two distant contributions are nonoverlapping to add them.

We assume that we can also decompose the inverse into a contribution for separate pairs and a contribution where all time points are close, $U^{-1}=U_s^{-1}+U_c^{-1}$, where again $U_s^{-1}$ and $U_c^{-1}$ vanish on the respective other domain. We demonstrate that $U_s^{-1}$ is the inverse of $U_s$, and that $U_c^{-1}$ is the inverse of $U_c$, by noting that $U^{-1}(t_1,t_2,s_1,s_2)U(s_1,s_2,t_3,t_4)$ reduces to the contraction of $U_c^{-1}$ and $U_c$ if $t_3$ and $t_4$, and hence $s_1$ and $s_2$, are close; or to the contraction of $U_s^{-1}$ and $U_s$ if $t_3$ and $t_4$, and hence $s_1$ and $s_2$, are distant. 

For \eref{eq:main-equation}, we need $U^{-1}_s$ due to the limits of \eref{eq:temporal-limits}. The fact that $U$ reduces to $U_s$ in the limits of \eref{eq:temporal-limits} justifies taking the limits before inverting $U$. Furthermore, these limits imply that, in \eref{eq:main-equation}, $s_1$ and $s_2$, as well as $s_3$ and $s_4$, are far apart. In this case, $K$ simplifies to
\begin{align}
K_s(s_1,s_2,s_3,s_4)=C^\phi_\spt(s_1-s_3)C^\phi_\spt(s_2-s_4)+C^\phi_\spt(s_1-s_4)C^\phi_\spt(s_2-s_3).
\end{align}
This justifies taking the limits of \eref{eq:temporal-limits} for $K$.

\section{Partially symmetric disorder}
\label{sec:partially_symmetric_appdx}

To demonstrate the flexibility of the path-integral approach to computing the four-point function, we now consider a generalization of the classic i.i.d.\ model in which we introduce correlations between reciprocal couplings, characterized by the following statistics:
\begin{equation}
    \tavg{J_{ij} J_{k\ell}} = \frac{g^2}{N} \delta_{ik}\delta_{j\ell} + \frac{g^2 \rho}{N} \delta_{i\ell}\delta_{jk}.
    \label{eq:correlated_couplings}
\end{equation}
Here, $\rho \in [-1, 1]$ is a parameter that controls the degree of symmetry in the connectivity. When $\rho = 1$, the connectivity is fully symmetric, while $\rho = -1$ corresponds to fully antisymmetric connectivity. The case $\rho = 0$ recovers the i.i.d.\ model. We express this partially symmetric connectivity as a linear combination of two matrices, $\bm{X}$ and $\bm{Y}$, that are i.i.d.\ random with mean zero and variance $1/N$,
\begin{equation}
    J_{ij} = \sqrt{{g^2-|\sigma|}}X_{ij} + \sqrt{\frac{|\sigma|}{2}}[Y_{ij} + \text{sgn}(\sigma)Y_{ji}],
    \label{eq:partially_symmetric_J}
\end{equation}
where $\sigma = \rho g^2$. In terms of $\bm{X}$ and $\bm{Y}$, the path integral for this partially symmetric model is
\begin{multline}
    Z[\bm{X}, \bm{Y}] = \pin{\bm{x}} \pinh{\bm{x}} \exp\bigg(i\sum_{i=1}^N  \hat{x}_iT[x_i] - i\sqrt{{g^2-|\sigma|}}\sum_{i,j=1}^N X_{ij} \hat{x}_i \phi_j 
    -i \sqrt{\frac{|\sigma|}{2}}\sum_{i,j=1}^NY_{ij}[\hat{x}_i\phi_j + \text{sgn}(\sigma) \phi_i \hat{x}_j] \bigg).
    \label{eq:partially_symmetric_Z}
\end{multline}
We integrate out $\bm{X}$ and $\bm{Y}$. To factorize the action over indices spanning extensive dimensions, we introduce
\begin{align}
    C^\phi(t_1, t_2) &= \frac{1}{N}\sum_{i=1}^N  \phi_i(t_1)\phi_i(t_2), 
    &&\conj \hat{C}^\phi(t_1, t_2);  \label{eq:C_phi_def} \\
    S^\phi(t_1, t_2) &= \frac{1}{N}\sum_{i=1}^N  \ddr{\phi_i(t_1)}{I_i(t_2)}, &&\conj \hat{S}^\phi(t_1, t_2). \label{eq:S_phi_def}
\end{align}
We obtain, for the disorder-averaged path integral,
\begin{align}
    Z &= \pin{C}^\phi\pinh{C}^\phi\pin{R}\pinh{R} \exp\left(-N\mathcal{S}[C^\phi, \hat{C}^\phi, S^\phi, \hat{S}^\phi] \right), \label{eq:Z_factorized} \\
    \mathcal{S}[C^\phi, \hat{C}^\phi, S^\phi, \hat{S}^\phi] &= -\frac{1}{2}C^\phi\hat{C}^\phi +\frac{\sigma}{2}S^\phi \hat{S}^\phi - \log W[C^\phi, \hat{C}^\phi, S^\phi, \hat{S}^\phi], \label{eq:S_factorized} \\
    W[C^\phi, \hat{C}^\phi, S^\phi, \hat{S}^\phi] &= \pinh{x}\pin{x}\exp\bigg( i\hat{x}T[x] - \frac{1}{2}\phi\hat{C}^\phi\phi - \frac{g^2}{2}\hat{x}C^\phi\hat{x} -\frac{i\sigma}{2}\phi\hat{S}^\phi \hat{x}- \frac{i\sigma}{2}\hat{x}S^\phi \phi \bigg). \label{eq:W_factorized}
\end{align}

\subsection{Two-point functions}

To obtain the saddle-point solution, we compute the derivatives of the action, under the rule that derivatives with respect to $S^\phi(t_1, t_2)$ also affect $\hat{S}^\phi(t_2, t_1)$, and vice versa, due to the symmetry in the action. This gives
\begin{align}
    \ddr{\mathcal{S}}{ C^\phi(t_1, t_2)} &= -\hat{C}^\phi(t_1,t_2) + g^2 \tavg{\hat{x}(t_1)\hat{x}(t_2)}_{W}, \label{eq:dS_dC} \\
    \ddr{\mathcal{S}}{ \hat{C}^\phi(t_1, t_2)} &= -{C}^\phi(t_1,t_2) + \tavg{{\phi}(t_1){\phi}(t_2)}_{W}, \label{eq:dS_dChat} \\
    \ddr{\mathcal{S}}{ S^\phi(t_1, t_2)} &= \sigma \hat{S}^\phi (t_1,t_2) + i\sigma \tavg{\hat{x}(t_1)\phi(t_2)}_{W}, \label{eq:dS_dS} \\
    \ddr{\mathcal{S}}{ \hat{S}^\phi(t_1, t_2)} &= \sigma S^\phi(t_1,t_2) + i\sigma \tavg{\phi(t_1)\hat{x}(t_2)}_{W}. \label{eq:dS_dShat}
\end{align}
Setting these derivatives to zero yields the saddle-point conditions,
\begin{align}
    \hat{C}^\phi_{\spt}(t_1,t_2) &= 0, \label{eq:Chat_spt} \\
    C^\phi_{\spt}(t_1, t_2) &= \tavg{\phi(t_1)\phi(t_2)}_{\spt}, \label{eq:C_spt} \\
    S^\phi_{\spt}(t_1,t_2) &= \hat{S}^\phi_{\spt}(t_2,t_1) = \tavg{\ddr{\phi(t_1)}{I(t_2)}}_\spt, \label{eq:S_spt}
\end{align}
where $\tavg{\cdots}_{\spt}$ denotes an average within the dynamic process described by $W[C^\phi_{\spt}, 0, S^\phi_\spt, \hat{S}^\phi_\spt]$. The single-site process at the saddle point is described by
\begin{align}
    T[x](t) &= \eta^x(t) + \sigma \left[S^\phi_\spt \circ \phi\right](t), \label{eq:single_site_x} \\
    \eta^x &\sim \mathcal{GP}(0, g^2 C^\phi_\spt),  \label{eq:single_site_eta}
\end{align}
where $\circ$ denotes convolution. The symmetric structure provides a convolutional, nonlinear self-coupling $\sigma [S^\phi_\spt \circ \phi](t)$ in the single-site dynamics. The two-point correlation and response functions $C^\phi_\spt(\tau)$ and $S^\phi_\spt(\tau)$ must be determined self-consistently within this single-site picture. 

\subsection{Four-point functions}

Introducing the notation $C^\phi_k = C^\phi_\spt(\omega_k)$, $S^\phi_k = S^\phi_\spt(\omega_k)$ for $k=1,2$, the frequency-dependent Hessian at the saddle point is
\begin{equation}
\bm{H} = \begin{pmatrix}
0 & -1 + g^2 S^\phi_{12} & 0 & 0 \\
-1 + g^2 (S^\phi_{12})^* & - C^\phi_{12} & \sigma (S^\phi_1)^* C^\phi_2 & \sigma (S^\phi_2)^* C^\phi_1 \\
0 & \sigma S^\phi_1 C^\phi_2 & 0 & \sigma - \sigma^2 S^\phi_1 (S^\phi_2)^* \\
0 & \sigma S^\phi_2 C^\phi_1 & \sigma - \sigma^2 (S^\phi_1)^* S^\phi_2 & 0
\end{pmatrix}.
\label{eq:Hessian_partially_symmetric}
\end{equation}
Inverting this matrix and isolating the upper-left element, we obtain, in agreement with Clark \textit{et al.} (Ref.~\cite{clark2023dimension}),
\begin{equation}
    \Psi^\phi_\spt = \frac{1}{|1 - g^2 S^\phi_{12}|^2} \frac{1 - | \sigma (S^\phi_1)^* S^\phi_2|^2 }{|1 - \sigma (S^\phi_1)^* S^\phi_2|^2} C^\phi_{12}.
    \label{eq:Psi_phi_partially_symmetric}
\end{equation}

\section{Path-integral calculation of two- and four-point functions for the random-mode model}
\label{sec:path_int_approch_to_rmm}

Having demonstrated the path-integral approach to computing two- and four-point functions for i.i.d.\ couplings, we now apply this formalism to the random-mode model. In terms of the mode matrices $\bm{L}$ and $\bm{R}$, the path integral is
\begin{equation}
    Z[\bm{L}, \bm{R}] = \pin{\bm{x}} \pinh{\bm{x}} 
     \exp\left(i\sum_{i=1}^N  \hat{x}_i T[x_i] 
    - i\sum_{i,j=1}^N  \left[\bm{L}\bm{D}\bm{R}\right]_{ij} \hat{x}_i \phi_j \right).
\end{equation}
We would like to integrate out $\bm{L}$ and $\bm{R}$, but this is complicated by them appearing together in a quadratic term in the action. To simplify the integration, we introduce a set of $M$ latent variables,
\begin{equation}
    z_a(t) = D_a \sum_{i=1}^N  R_{ia} \phi_i(t),
    \label{eq:z_def}
\end{equation}
and their conjugates $\hat{z}_a(t)$. We use indices $i,j$ for neurons and $a,b$ for latent variables. The introduction of these latent variables makes the action linear in $\bm{L}$ and $\bm{R}$,
\begin{equation}
    Z[\bm{L}, \bm{R}] = \pin{\bm{x}} \pinh{\bm{x}} \pin{\bm{z}} \pinh{\bm{z}} 
     \exp\Big(i\sum_{i=1}^N  \hat{x}_i T[x_i]
    + i\sum_{a=1}^M  z_a \hat{z}_a 
    - i\sum_{i,a} L_{ia} \hat{x}_i z_a - i\sum_{i,a}D_a R_{ia} \phi_i \hat{z}_a\Big).
\end{equation}
We can now easily integrate out $\bm{L}$ and $\bm{R}$. To factorize the action over indices spanning extensive dimensions, we introduce the field $C^\phi(t_1, t_2)$ [\eref{eq:nac}] with conjugate $\hat{Q}(t_1, t_2)$ as well as
\begin{equation}
    Q(t_1, t_2) = \frac{1}{N}\sum_{a=1}^M  z_a(t_1) z_a(t_2)
\end{equation}
with conjugate $\hat{C}^\phi(t_1, t_2)$. The disorder-averaged path integral is
\begin{align}
    Z &= \pin{C}^\phi \pinh{C}^\phi \pin{Q} \pinh{Q} \exp\left(-N \mathcal{S}[C^\phi, \hat{C}^\phi, Q, \hat{Q}]\right), \\
\mathcal{S}[C^\phi, \hat{C}^\phi, Q, \hat{Q}] &= -\frac{1}{2}C^\phi \hat{Q} - \frac{1}{2}Q\hat{C}^\phi
- \log W^x[Q, \hat{Q}] - \alpha \tavg{\log W^z_D[C^\phi, \hat{C}^\phi]}_D, \\
 W^x[Q, \hat{Q}] &= \pin{x}\pinh{x}\exp\left(i\hat{x}T[x]
-\frac{1}{2}\phi \hat{Q} \phi - \frac{1}{2}\hat{x}Q\hat{x}
\right), \label{eq:nrn_sspi} \\
W^z_D[C^\phi, \hat{C}^\phi] &= \pin{z}\pinh{z}\exp\left(i\hat{z}z
-\frac{1}{2} z\hat{C}^\phi z - \frac{1}{2}D^2 \hat{z}C^\phi \hat{z}
\right).
\end{align}
Thus, analysis of the path integral is made tractable by recasting the problem as interacting neurons and latent variables. The neuronal single-site picture is described by $W^x[Q, \hat{Q}]$, and the latent-variable single-site picture is described by $W^z_D[C^\phi, \hat{C}^\phi]$, where the subscript $D$ reflects that each latent variable has its own component strength. This formulation introduces an extensive set of latent variables $z_a(t)$, in contrast to the finite set of latent variables, conventionally denoted by $\kappa_a(t)$, in low-rank neural networks \cite{mastrogiuseppe2018linking, beiran2021shaping}. The statistics of the Gaussian input to each single-site process (neurons or latent variables) are determined by the statistics of the other process, creating a bipartite, mutually referential structure that also arises in the cavity treatment of the problem.

\subsection{Two-point functions}
\label{sec:two_pt_fn_rmm}

Following the steps outlined at the end of \secref{sec:fptfn}, we first compute the DMFT by calculating the saddle point. The derivatives of the action are
\begin{align}
    \ddr{\mathcal{S}}{ C^\phi(t_1, t_2)} &= -\hat{Q}^\phi(t_1, t_2) + \alpha \tavg{D^2\tavg{ \hat{z}(t_1)\hat{z}(t_2)}_{W^z_D}}_D , \\
    \ddr{\mathcal{S}}{ \hat{C}^\phi(t_1, t_2)} &= -Q(t_1, t_2) + \alpha \tavg{\tavg{ z(t_1)z(t_2)}_{W^z_D}}_D, \\
    \ddr{\mathcal{S}}{ Q(t_1, t_2)} &= -\hat{C}^\phi(t_1, t_2) + \tavg{\hat{x}(t_1)\hat{x}(t_2)}_{W^x}, \\
    \ddr{\mathcal{S}}{ \hat{Q}(t_1, t_2)} &= -C^\phi(t_1, t_2) +  \tavg{\phi(t_1)\phi(t_2)}_{W^x},
\end{align}
where $\tavg{\cdots}_{W^x}$ and $\tavg{\cdots}_{W^z_D}$ are averages within the dynamic processes described by $W^x[Q, \hat{Q}]$ and $W^z_D[C^\phi, \hat{C}^\phi]$, respectively. Setting these derivatives to zero yields
\begin{align}
    \hat{C}^\phi_{\spt}(t_1, t_2) &= \hat{Q}_{\spt}(t_1, t_2) = 0, \\
    C^\phi_{\spt}(t_1, t_2) &= \tavg{\phi(t_1)\phi(t_2)}_{\spt}, \label{eq:rmm_nrn_scc} \\
    Q_{\spt}(t_1, t_2) &= \alpha \tavg{\tavg{ z(t_1)z(t_2)}_{\spt D}}_D, \label{eq:Q0-cond}
\end{align}
where $\tavg{\cdots}_{\spt}$ and $\tavg{\cdots}_{\spt D}$ are averages within the dynamic processes described by $W^x[Q_{\spt}, 0]$ and $W^z_D[C^\phi_{\spt}, 0]$, respectively. These single-site processes are
\begin{align}
    T[x](t) &= \eta^x(t) , \label{eq:rmm_nrn_ssp} \\ 
    z(t) &= \eta^z(t) , \label{eq:z_single_site} 
\end{align}
where $\eta^x(t)$ and $\eta^z(t)$ are Gaussian fields,
\begin{align}
    \eta^x &\sim \mathcal{GP}(0, Q_\spt) , \\  
    \eta^z &\sim \mathcal{GP}(0, D^2 C^\phi_\spt) ,\label{eq:z_ss_stats}
\end{align}
showing that the neuronal and latent-variable single-site processes have mutually referential statistics \footnote{Note that the variance of $\eta^z(t)$ is scaled by $D^2$; equivalently, we could factor out $D^2$ from the variance and include $D$ as a multiplicative factor in the latent-variable single-site dynamics [\eref{eq:z_single_site}]}. 

Solving for $Q_\spt(t_1, t_2)$ by combining Eqs.~\eqref{eq:Q0-cond}, \eqref{eq:z_single_site}, and \eqref{eq:z_ss_stats}, we obtain
$
Q_{\spt}(t_1,t_2) = \alpha r_2 C^\phi_{\spt}(t_1, t_2),
$
with $r_n$ defined by \eref{eq:D_moment}. Consolidating these results yields \eref{eq:twopt-rmm}.

\subsection{Four-point functions}
\label{sec:rmm_hes_subsec}

\begin{figure*}
    \centering
    \includegraphics[width=7in]{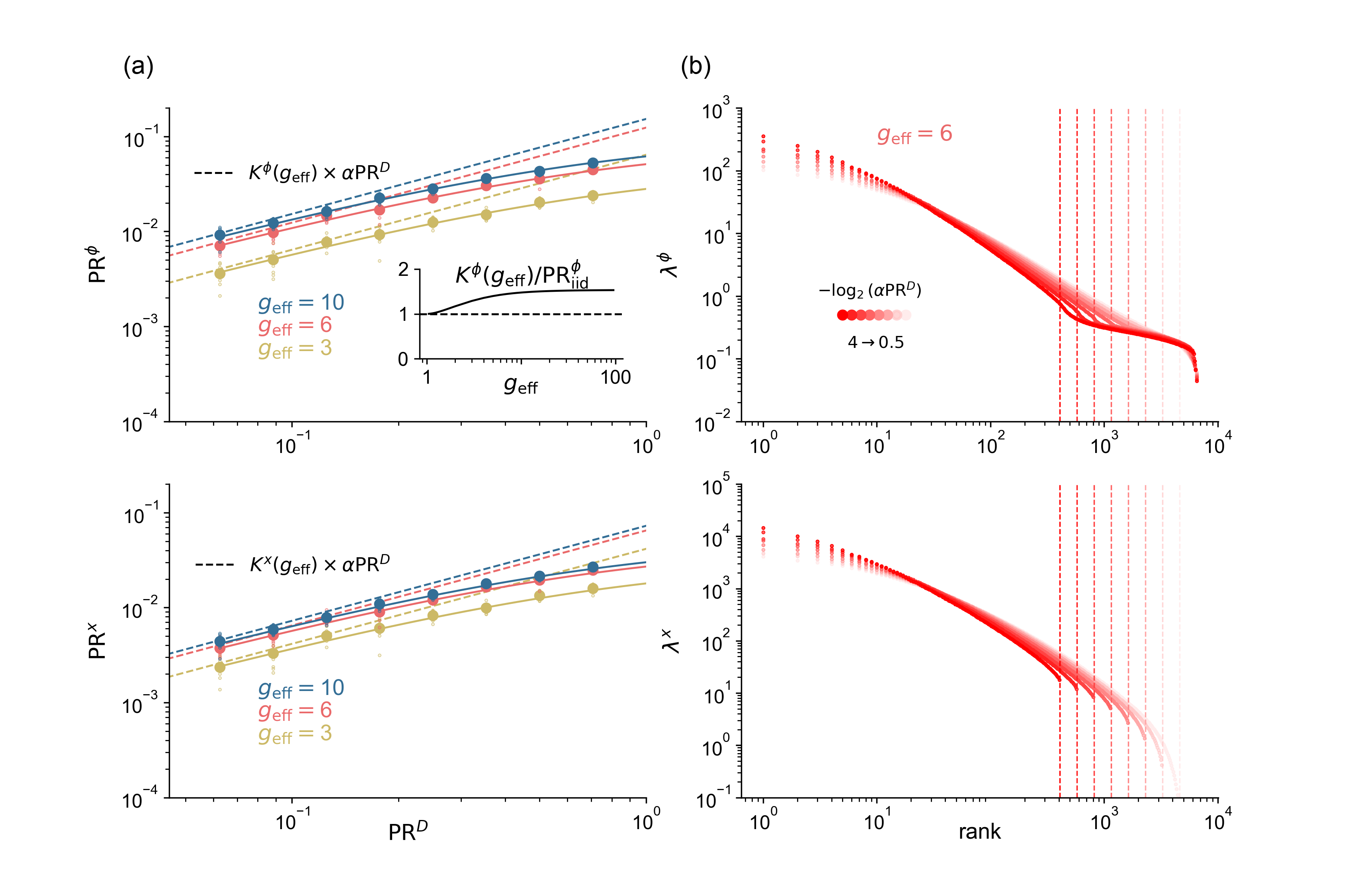}
    \caption{Analysis of dimension of activity and covariance spectrum in the low-effective-rank limit. (a) Top: dimension of activity $\text{PR}^\phi$ versus effective rank $\alpha \text{PR}^D$ ($\alpha = 1$) for various coupling strengths $g_{\text{eff}}$. Component strengths $D_a$ are constant for $a/N \leq \text{PR}^D$ and $0$ otherwise. Small dots, individual simulations; large dots: means over ten simulations; solid lines, theoretical predictions; dotted lines, small-$\text{PR}^D$ expansion, $K^\phi(g_\text{eff}) \mathrm{PR}^D$. Inset: ratio of proportionality factor $K^\phi(g_\text{eff})$ to $\text{PR}^\phi_{\text{i.i.d.}}$ versus $g_\text{eff}$. Bottom: same analysis for preactivations. (b) Top: rank-ordered eigenvalue plots for activation covariance matrix $C^{\phi}_{ij}(0)$. Dotted vertical lines indicate $N\,\text{PR}^D$. Bottom: same for preactivations. All simulations use $N = 6500$ neurons.}
    \label{fig:spectrum_shapes}
\end{figure*}

The frequency-dependent saddle-point Hessian is
\begin{equation}
\bm{H} = \left(
\begin{array}{cc|cc}
0 & \alpha r_2 & 0 & -1 \\
\alpha r_2 & -\alpha r_4 C_{12}^\phi &  -1 & 0 \\
\hline
0 &  -1 & 0 & S_{12}^\phi \\
-1 & 0 & (S_{12}^\phi)^* & -C_{12}^\phi 
\end{array}
\right).
\label{eq:rmm_hessian}
\end{equation}
The upper-left block, corresponding to the latent variables, is related to the Hessian of an i.i.d.\ network with linear dynamics and a heterogeneous distribution of single-neuron scale factors $D_a$; the lower-right block, corresponding to the neurons, is related to the Hessian of an i.i.d.\ network of nonlinear neurons [\eref{eq:freq-dep-H}]. They are coupled through the simple off diagonal blocks, reflecting the interaction between neurons and latent variables. 

This yields the expression for $\Psi^\phi_{\spt}$ [\eref{eq:path-int-result-rmm}]. To obtain $\Psi^x_{\spt}$, we add $-\frac{1}{2}x\mathcal{J}x$ to the neuronal single-site path integral [\eref{eq:nrn_sspi}] and compute
\begin{equation}
    \frac{\delta^2 S}{\delta \mathcal{J}^2} = -C_{12}^x, \:\:\:
    \frac{\delta^2 S}{\delta \mathcal{J} \delta \bm{C}}
     = \begin{pmatrix}
        0\\
        0\\
        (S_{12}^x)^* \\
        -C_{12}^{x\phi}\\
    \end{pmatrix}.
\end{equation}
Inserting these into \eref{eq:fluctuation_formula} we obtain
\begin{equation}
    \Psi_\spt^x = C^x_{12} + \left(1 + \frac{1}{\alpha\text{PR}^D}\right)|U|^2 C^\phi_{12} 
    + U C^{x\phi}_{12} + \text{H.c.},
    \label{eq:psi_x_rmm}
\end{equation}
where $U = {g_{\text{eff}}^2 S^x_{12}}/({1 - g_{\text{eff}}^2 S^\phi_{12}})$ [\eref{eq:Psi_x_result_cavity_U} with $g = g_\text{eff}$]. Here, we have used the following formula to invert the Hessian:
\renewcommand{\arraystretch}{1.5}
\begin{equation}
\begin{pmatrix}
0 & a & 0 & -1 \\
a^* & b & -1 & 0 \\
0 & -1 & 0 & c \\
-1 & 0 & c^* & d
\end{pmatrix}^{-1}
=\begin{pmatrix}
-\frac{b + |c|^2 d }{|1 - a c|^2} & -\frac{c^*}{1 - (a c)^*} & -\frac{a d + b c^*}{|1 - a c|^2} & -\frac{1}{1 - (a c)^*} \\
-\frac{c}{1 - ac} & 0 & -\frac{1}{1 - ac} & 0 \\
-\frac{a^* d + b c}{|1 - a c|^2} & -\frac{1}{1 - (a c)^*} & -\frac{|a|^2 d + b}{|1 - a c|^2} & -\frac{a^*}{1 - (a c)^*} \\
-\frac{1}{1 - ac} & 0 & -\frac{a}{1 - ac} & 0
\end{pmatrix}.
\label{eq:inv_formula}
\end{equation}

\section{Path-integral calculation of two- and four-point functions for the random-mode model with single-unit heterogeneity}
\label{sec:het_appdx}

We incorporate the generalized nonlinearity $\Phi_{\bm{\theta}_i}(x)$ into the path integral, as well as a source term for the normalized variables, and integrate out $\bm{L}$ and $\bm{R}$. To factorize the action over indices spanning extensive dimensions, we introduce the order parameters
\begin{align}
     C^\Phi(t_1, t_2) &= \frac{1}{N}\sum_{i=1}^N  \Phi_{\bm{\theta}_i}(t_1) \Phi_{\bm{\theta}_i}(t_2),   &&\conj \hat{Q}(t_1, t_2);  \\
    Q(t_1, t_2) &= \frac{1}{N}\sum_{a=1}^M  z_a(t_1) z_a(t_2), &&\conj \hat{C}^\phi(t_1, t_2),
\end{align}
where $\Phi_{\bm{\theta}_i}(t) = \Phi_{\bm{\theta}_i}\bm{(}x_i(t)\bm{)}$. The resulting action and single-site path integrals for this system are
\begin{align}
\mathcal{S}[C^\Phi, \hat{C}^\Phi, Q, \hat{Q}] &= -\frac{1}{2}C^\Phi \hat{Q} - \frac{1}{2}Q\hat{C}^\phi
- \tavg{\log W_{\bm{\theta}}[Q, \hat{Q}; \mathcal{J}]}_{\bm{\theta}} - \alpha \tavg{\log W^z_D[C^\Phi, \hat{C}^\Phi]}_D, \\
 W^x_{\bm{\theta}}[Q, \hat{Q}; \mathcal{J}] &= \pin{x}\pinh{x}\exp\left(i\hat{x}T[x]
-\frac{1}{2}\Phi_{\bm{\theta}} \hat{Q} \Phi_{\bm{\theta}} - \frac{1}{2}\hat{x}Q\hat{x}
- \frac{1}{2} \phi \mathcal{J} \phi
\right), \\
W^z_D[C^\Phi, \hat{C}^\Phi] &= \pin{z}\pinh{z}\exp\left(i\hat{z}z
-\frac{1}{2} z\hat{C}^\Phi z - \frac{1}{2}D^2 \hat{z}C^\Phi \hat{z}
\right).
\end{align}

\subsection{Two-point functions}

The saddle-point calculation proceeds similarly to the non-heterogeneous case of the random-mode model. We obtain
\begin{align}
    \hat{C}^\Phi_\spt(t_1, t_2) &= \hat{Q}_\spt(t_1, t_2) = 0, \\
    C^\Phi_\spt(t_1, t_2) &= \tavg{ \tavg{\Phi_{\bm{\theta}}(t_1)\Phi_{\bm{\theta}}(t_2) }_\spt }_{\bm{\theta}}, \\
    Q_\spt(t_1, t_2) &= \alpha \tavg{ \tavg{z(t_1) z(t_2)}_{\spt D} }_D,
\end{align}
where $\tavg{\cdots}_{\spt}$ and $\tavg{\cdots}_{\spt D}$ are averages with respect to $W^x_{\bm{\theta}}[Q_\spt, 0; 0]$ and $W^z_{D}[C^\Phi_\spt, 0]$, respectively (note that $W^x_{\bm{\theta}}[Q_\spt, 0; 0]$ does not depend on $\bm{\theta}$).
Using $\tavg{z(t_1) z(t_2)}_{\spt D} = D^2 C^\Phi_\spt(t_1, t_2)$ and, consequently, $Q_\spt(t_1, t_2) = \alpha r_2 C^\Phi_\spt(t_1, t_2)$, we obtain the self-consistent single-site dynamic process
\begin{subequations}
\begin{align}
    T[x](t) &= \eta^x(t), \\
    \eta^x &\sim \mathcal{GP}(0, \alpha r_2 C^\Phi_\spt), \\
    C^\Phi_\spt(t_1, t_2) &= \tavg{\tavg{\Phi_{\bm{\theta}}(t_1)\Phi_{\bm{\theta}}(t_2)}_\spt}_{\bm{\theta}}.  \label{eq:snh_tpf}
\end{align}
\end{subequations}
The self-consistency condition \eref{eq:snh_tpf} is identical to that of an i.i.d.\ network, but with an average over $\bm{\theta} \sim P(\bm{\theta})$, in addition to the usual Gaussian average over $\eta^x(t)$, to account for single-neuron heterogeneity.

\subsection{Four-point functions}

To write down the solution for the four-point function, we define new two-frequency correlation functions with an outer average over $\bm{\theta} \sim P(\bm{\theta})$,
\begin{subequations}
\begin{align}
\mathfrak{C}^{\Phi}_{12} &= \tavg{ \tavg{\Phi_{\bm{\theta}}\Phi_{\bm{\theta}}}_{\spt}(\omega_1)\tavg{\Phi_{\bm{\theta}}\Phi_{\bm{\theta}}}_{\spt}(\omega_2)  }_{\bm{\theta}}, \\
\mathfrak{s}^{\Phi}_{12} &= \tavg{ \tavg{\frac{\delta \Phi_{\bm{\theta}}}{\delta I}}_{\spt}(\omega_1)\tavg{\frac{\delta \Phi_{\bm{\theta}}}{\delta I}}_{\spt}(\omega_2)  }_{\bm{\theta}}, \\
\mathfrak{C}^{\Phi\phi}_{12} &= \tavg{\tavg{\Phi_{\bm{\theta}}\phi}_\spt(\omega_1)
\tavg{\Phi_{\bm{\theta}}\phi}_\spt(\omega_2)}_{\bm{\theta}},\end{align}\label{eq:frak_vars}\end{subequations}
along with the usual shorthand $C^\Phi_{12} = C^\Phi_\spt(\omega_1) C^\Phi_\spt(\omega_2)$.

The frequency-dependent Hessian at the saddle point is
\begin{equation}
\bm{H} = \left(
\begin{array}{cc|cc}
0 & \alpha r_2 & 0 & -1 \\
\alpha r_2 & -\alpha r_4 {C_{{12}}^\Phi} &  -1 & 0 \\
\hline
0 &  -1 & 0 & \mathfrak{s}^{\Phi}_{12} \\
-1 & 0 & (\mathfrak{s}^{\Phi}_{12})^* & -\mathfrak{C}^{\Phi}_{12}
\end{array}
\right),
\label{eq:rmm_heterogen_hessian}
\end{equation}
where the new Fraktur variables are defined in \eref{eq:frak_vars}. We invert the Hessian using \eref{eq:inv_formula} and identify the upper-left element as $\Phi_{\bm{\theta}_i}(\bm{\omega})$,
\begin{equation}
    \Psi^\Phi_\spt = \frac{\frac{\mathfrak{C}^\Phi_{12}}{C^\Phi_{12}} + \frac{1}{\alpha\text{PR}^D}|\alpha r_2\mathfrak{s}^\Phi_{12}|^2 }{|1 - \alpha r_2\mathfrak{s}^\Phi_{12}|^2} C^\Phi_{12}. \label{eq:unnorm_het_psi}
\end{equation}

Combining the inverted Hessian with the relevant quantities for computing fluctuations of normalized variables,
\begin{equation}
\frac{\delta^2 S}{\delta \mathcal{J}^2} = -C^\phi_{12}, \:\:\:
\frac{\delta^2 S}{\delta \mathcal{J} \delta \bm{C} }
= \begin{pmatrix}
0 \\
0 \\
(S^{\phi}_{12})^* \\
-\mathfrak{C}^{\Phi \phi}_{12}
\end{pmatrix},
\end{equation}
where 
\begin{align}
    \mathfrak{C}^{\Phi\phi}_{12} &= \tavg{\tavg{\Phi_{\bm{\theta}}\phi}_\spt(\omega_1)
          \tavg{\Phi_{\bm{\theta}}\phi}_\spt(\omega_2)}_{\bm{\theta}}, \\
    S^\phi_{12} &= \tavg{\frac{\delta \phi}{\delta I}}_\spt(\omega_1) \tavg{\frac{\delta \phi}{\delta I}}_\spt(\omega_2),
\end{align}
we obtain for the normalized variables $\phi_i(t)$,
\begin{equation}
    \Psi^\phi_\spt = \Bigg{[}|A|^2\left(\frac{\mathfrak{C}^\Phi_{12}}{C^\phi_{12}} + \frac{1}{\alpha\text{PR}^D}\frac{C^\Phi_{12}}{C^\phi_{12}} \right) 
    + \left(A\frac{\mathfrak{C}^{\Phi\phi}_{12}}{C^\phi_{12}} + \text{H.c.}\right) + 1  \Bigg{]}C^\phi_{12}, \label{eq:norm_het_psi}
\end{equation}
where
\begin{equation}
    A = \frac{\alpha r_2 S^\phi_{12}}{1 -\alpha r_2 \mathfrak{s}^\phi_{12}}.
\end{equation}

\subsection{Four-point functions for firing-rate heterogeneity}
\label{sec:diverse_single_nrn_firing_rates}

Upon specializing the above general results to the case where $\bm{\theta}_i$ contains a single gain parameter $G_i$, we obtain the four-point functions for unnormalized [\eref{eq:psi_G_unnorm}] and normalized [\eref{eq:psi_G_normalized}] variables given in the main text.

Note that, like the distribution over component strengths $D_a$, two-point functions of activity depend only on the second moment of $G_i$, and four-point functions depend only on the second and fourth moments of $G_i$.

For ``random-readout'' gains with the same distribution as the recurrent gains (see main text), the four-point function is
\begin{equation}
    \Psi^{\Phi^{\text{readout}}} = \Bigg{[}\left( \frac{1}{\text{PR}^G} - 1 \right)\left(|1-g_{\text{eff}}^2 S^\phi_{12}|^2 + |g_{\text{eff}}^2 S^\phi_{12}|^2\right) 
    + 1 + \frac{1}{\alpha \text{PR}^D}|g_{\text{eff}}^2 S^\phi_{12}|^2 \Bigg{]}  \frac{C^{\Phi^\text{readout}}_{12}}{|1 - g_{\text{eff}}^2 S^\phi_{12}|^2},
    \label{eq:psi_G_readout}
\end{equation}
where $C^{\Phi^{\text{readout}}}_{12} = C^\Phi_{12} = q_2^2 C^\phi_{12}$

We conclude with a note on gain modulation in recurrent neural networks. It has been shown that modulating single-neuron gains while keeping synaptic weights fixed has great expressive power over the dynamics of recurrent neural networks \citep{stroud2018motor}. Here, we consider random gains rather than judiciously chosen or learned gains. In particular, they are sampled independent of the connectivity. However, our framework could be extended to model gains that are related to the connectivity and, more ambitiously, could be extended to model learned gain parameters (or, as explored in the Discussion, learned connectivity).

\section{Path-integral calculation of two- and four-point functions for the random-mode model with diagonally structured overlaps}
\label{sec:LR_appdx}

To generate matrices $\bm{L}$ and $\bm{R}$ with appropriate correlation structure, we express them in terms of independent Gaussian random matrices $\bm{X}^1, \bm{X}^2,$ and $\bm{Y}$, each with variance $1/N$,
\begin{align}
{L}_{ia} &= \sqrt{{1-|\rho_a|}}{X}^1_{ia} + \sqrt{{|\rho_a|}}{Y}_{ia}, \label{eq:L_decomp} \\
{R}_{ia} &= \sqrt{{1-|\rho_a|}}{X}^2_{ia} + \text{sgn}(\rho_a) \sqrt{{|\rho_a|}}{Y}_{ia}. \label{eq:R_decomp}
\end{align}

We begin with the path integral containing latent variables, insert this parametrization of $\bm{L}$ and $\bm{R}$, and then integrate out $\bm{X}_1$, $\bm{X}_2$, and $\bm{Y}$. To factorize the action over indices spanning extensive dimensions, we introduce
\begin{align}
C^\phi(t_1, t_2) &= \frac{1}{N}\sum_{i=1}^N  \phi_i(t_1) \phi_i(t_2), &&\conj \hat{Q}(t_1, t_2);  \\
S^\phi(t_1, t_2) &= \frac{1}{N}\sum_{i=1}^N  \frac{\delta\phi_i(t_1)}{\delta I_i(t_2)}, &&\conj \hat{R}(t_1, t_2);  \\
Q(t_1, t_2) &= \frac{1}{N}\sum_{a=1}^M  z_a(t_1) z_a(t_2), &&\conj \hat{C}^\phi(t_1, t_2);  \label{eq:Q_def} \\
R(t_1, t_2) &= \frac{1}{N}\sum_{a=1}^M  D_a \rho_a \frac{\delta z_a(t_1)}{\delta I_a(t_2)},  &&\conj \hat{S}^\phi(t_1, t_2). \label{eq:R_def}
\end{align}
The resulting path integral, action, and single-site path integrals are given by
\begin{align}
\centering
Z  = \pin{C}^\phi\pinh{C}^\phi\pin{S}^\phi & \pinh{S}^\phi  \pin{Q}\pinh{Q}\pin{R}\pinh{R} \exp\left( -N\mathcal{S}[C^\phi,\hat{C}^\phi,S^\phi,\hat{S}^\phi,Q,\hat{Q},R,\hat{R}] \right), \label{eq:path_integral} \\
\mathcal{S}[C^\phi,\hat{C}^\phi,S^\phi,\hat{S}^\phi,Q,\hat{Q},R,\hat{R}] &= -\frac{1}{2}C^\phi\hat{Q} - \frac{1}{2}Q\hat{C}^\phi + \frac{1}{2}S^\phi\hat{R} + \frac{1}{2}R\hat{S}^\phi \\
&\hspace{10em} - \log W^x[Q, \hat{Q}, R, \hat{R}] - \alpha \tavg{\log W^z_{D,\rho}[C^\phi, \hat{C}^\phi, S^\phi, \hat{S}^\phi]}_{D,\rho},
\label{eq:action} \\
W^x[Q, \hat{Q}, R, \hat{R}] &= \pin{x}\pinh{x} \exp\left( i\hat{x}T[x] - \frac{1}{2}\hat{x}Q\hat{x} - \frac{1}{2}\phi\hat{Q}\phi - \frac{i}{2}\hat{x}R\phi - \frac{i}{2}\phi \hat{R}\hat{x} \right), \label{eq:W_x} \\
W^z_{D,\rho}[C^\phi, \hat{C}^\phi, S^\phi, \hat{S}^\phi] &= \pin{z}\pinh{z} \exp\left( i\hat{z}z - \frac{1}{2}D^2 \hat{z} C^\phi \hat{z} -\frac{1}{2}z\hat{C}^\phi z - \frac{i}{2}D\rho\hat{z} S^\phi z - \frac{i}{2} D \rho z\hat{S}^\phi\hat{z}  \right). \label{eq:W_z}
\end{align}

\subsection{Two-point functions}

We first calculate the saddle-point conditions by taking the derivatives of the action,
\begin{align}
    \ddr{\mathcal{S}}{ C^\phi(t_1, t_2)} &= -\hat{Q}(t_1, t_2) + \alpha \tavg{D^2\tavg{\hat{z}(t_1)\hat{z}(t_2)}_{W^z_{D,\rho}}}_{D,\rho}, \\
    \ddr{\mathcal{S}}{ \hat{C}^\phi(t_1, t_2)} &= -Q(t_1, t_2) + \alpha\tavg{\tavg{z(t_1)z(t_2)}_{W^z_{D,\rho}}}_{D,\rho}, \\
    \ddr{\mathcal{S}}{ S^\phi(t_1, t_2)} &= \hat{R}(t_1, t_2) + i\alpha\tavg{D\rho \tavg{\hat{z}(t_1) z(t_2)}_{W^z_{D,\rho}}}_{D,\rho}, \\
    \ddr{\mathcal{S}}{ \hat{S}^\phi(t_1, t_2)} &= R(t_1, t_2) + i\alpha\tavg{D\rho \tavg{z(t_1) \hat{z}(t_2)}_{W^z_{D,\rho}}}_{D,\rho}, \\
    \ddr{\mathcal{S}}{ Q(t_1, t_2)} &= -\hat{C}^\phi(t_1, t_2) + \tavg{\hat{x}(t_1)\hat{x}(t_2)}_{W^x}, \\
    \ddr{\mathcal{S}}{ \hat{Q}(t_1, t_2)} &= -C^\phi(t_1, t_2) + \tavg{\phi(t_1)\phi(t_2)}_{W^x}, \\
    \ddr{\mathcal{S}}{ R(t_1, t_2)} &= \hat{S}^\phi(t_1, t_2) + i\tavg{\hat{x}(t_1)\phi(t_2)}_{W^x}, \\
    \ddr{\mathcal{S}}{ \hat{R}(t_1, t_2)} &= S^\phi(t_1, t_2) + i\tavg{\phi(t_1)\hat{x}(t_2)}_{W^x}.
\end{align}
The saddle-point conditions yield
\begin{align}
    \hat{C}^\phi(t_1, t_2) &= \hat{Q}_\spt(t_1, t_2) = 0, \\
    C^\phi_\spt(t_1, t_2) &= \tavg{\phi(t_1)\phi(t_2)}_{\spt}, \label{eq:sym_model_scc_cool}\\
    S^\phi_\spt(t_1, t_2) &= \hat{S}^\phi_\spt(t_2, t_1) = \tavg{\frac{\delta \phi(t_1)}{\delta I(t_2)}}_{\spt},  \label{eq:sym_model_scc_S_cool} \\
    Q_\spt(t_1, t_2) &= \alpha\tavg{\tavg{z(t_1)z(t_2)}_{\spt D,\rho}}_{D,\rho}, \\
    R_\spt(t_1, t_2) &= \hat{R}_\spt(t_2,t_1) = \alpha\tavg{D\rho \tavg{\frac{\delta z(t_1)}{\delta I(t_2)}}_{\spt D,\rho}}_{D,\rho}, 
\end{align}
where $\tavg{\cdots}_{\spt D,\rho}$ and $\tavg{\cdots}_{\spt}$ are averages within the dynamic processes described by $W^z_{D,\rho}[C^\phi_\spt, 0, S^\phi_\spt, \hat{S}^\phi_\spt]$ and $W^x[Q_\spt, 0, R_\spt, \hat{R}_\spt]$, respectively.
This yields neuronal and latent-variable single-site processes described by
\begin{align}
    T[x](t) &= \eta^x(t) + [R_\spt \circ \phi](t) \\
    z(t) &= \eta^z(t) + D\rho[S^\phi_\spt \circ z](t) ,
\end{align}
where the Gaussian fields have statistics
\begin{align}
    \eta^x &\sim \mathcal{GP}(0, Q_\spt) , \\
    \eta^z &\sim \mathcal{GP}(0, D^2 C^\phi_\spt).
\end{align}
This gives rise to the single-site problem stated in \secref{sec:LR_correlations}.

\subsection{Four-point functions}

The frequency-dependent Hessian at the saddle point is
\begin{equation}
\bm{H} =
\left(
\begin{array}{cccc|cccc}
0 & u & 0 & 0 & 0 & -1 & 0 & 0 \\ 
u^* & -v C^\phi_1 C^\phi_2 & x^*_{12} C^\phi_2 & x^*_{21} C^\phi_1 & -1 & 0 & 0 & 0 \\ 
0 & x_{12} C^\phi_2 & 0 & -w^* & 0 & 0 & 0 & 1 \\ 
0 & x_{21} C^\phi_1 & -w & 0 & 0 & 0 & 1 & 0 \\ \hline
0 & -1 & 0 & 0 & 0 & S^\phi_1 S^\phi_2 & 0 & 0 \\ 
-1 & 0 & 0 & 0 & (S^\phi_1 S^\phi_2)^* & -C^\phi_1 C^\phi_2 & (S^\phi_1)^* C^\phi_2 & (S^\phi_2)^* C^\phi_1 \\ 
0 & 0 & 0 & 1 & 0 & S^\phi_1 C^\phi_2 & 0 & -S^\phi_1 (S^\phi_2)^* \\ 
0 & 0 & 1 & 0 & 0 & S^\phi_2 C^\phi_1 & -(S^\phi_1)^* S^\phi_2 & 0 
\end{array}
\right)
\end{equation}
where
\begin{align}
    u &= \alpha\tavg{D^2 \sigma_1 \sigma_2}_{D,\rho}, \\
    v &= \alpha\tavg{D^4 |\sigma_1|^2 |\sigma_2|^2}_{D,\rho}, \\
    w &= \alpha\tavg{D^2 \rho^2 \sigma_1^* \sigma_2}_{D,\rho}, \\
    x_{12} &= \alpha\tavg{D^3 \rho \sigma_1 |\sigma_2|^2}_{D,\rho}, \\
    \sigma_k &= \frac{1}{1 - D\rho S^\phi_k} \:\: \text{for} \:\: k = 1,2.
\end{align}
Isolating the upper-left element of the inverse of this matrix yields \eref{eq:lr_psi_result}. 

\subsection{Sigmoidal parametrization of overlaps}
\label{sec:sigmoid_form}

To model structured left-right mode overlaps, we used the following sigmoidal form:
\begin{equation}
\rho_a = \gamma_\rho \left(1 - \frac{2}{1 + \exp[-\beta_\rho (a/N - u_{\text{flip}})]} \right),
\end{equation}
where $\rho_a = 0$ at $a/N = u_{\text{flip}}$. The parameter $\gamma_\rho$ sets the overall magnitude and sign of the overlaps, while $\beta_\rho$ controls the steepness of the transition. We fixed $\beta_\rho = 16$, varied $\gamma_\rho$, and set $u_{\text{flip}}$ such that $\tavg{D\rho}_{D,\rho} = 0$, thereby eliminating strong self-couplings.

\section{Two-site cavity calculation of four-point functions for the random-mode model}
\label{sec:cav_appdx}

\begin{figure}
\centering
\includegraphics[width=2.5in]{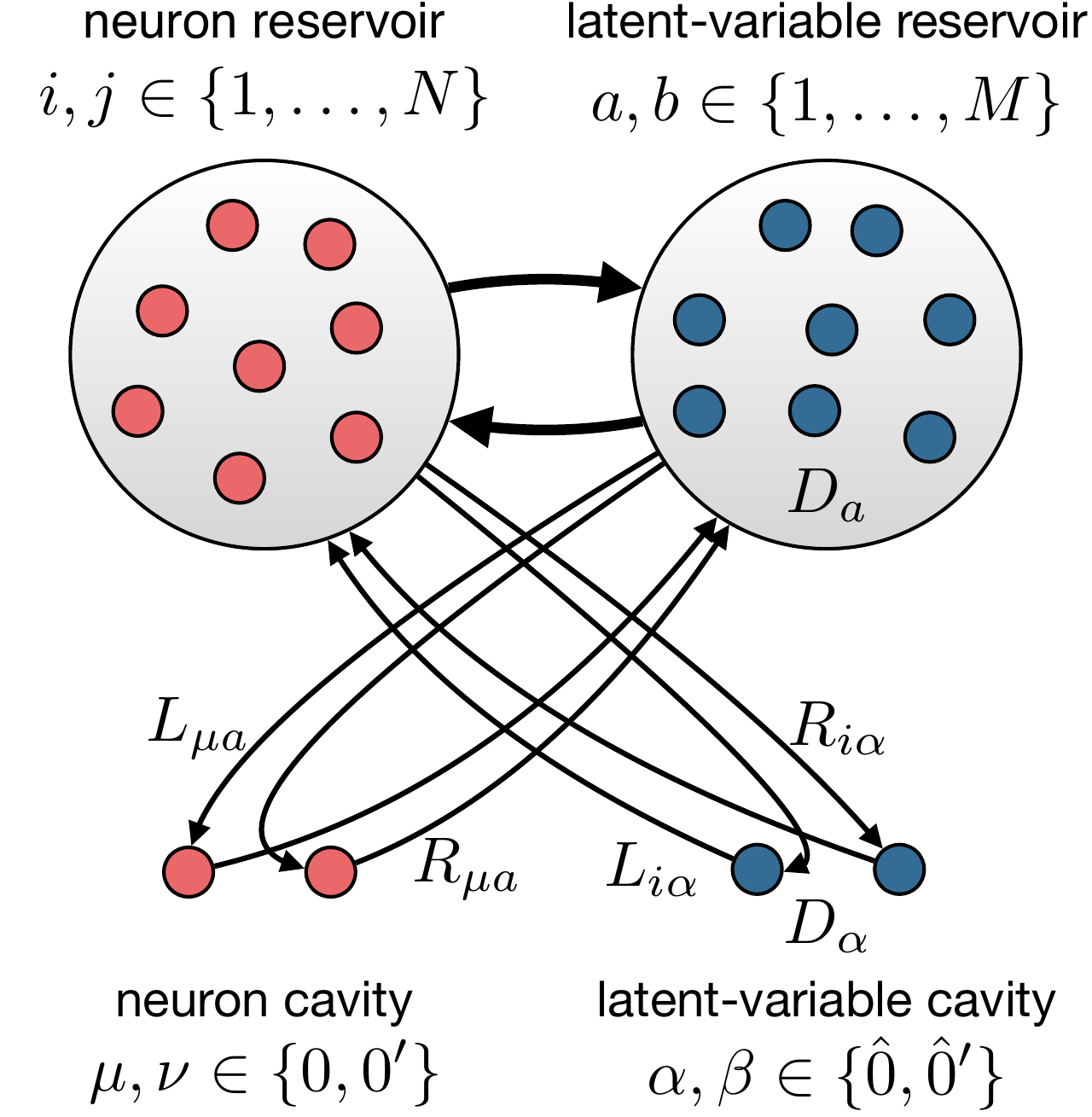}
\caption{Schematic of the two-site cavity approach for the bipartite representation of the random-mode model. Left: neuron cavity, considering the introduction of neurons $x_{0}(t)$ and $x_{0'}(t)$. Right: latent-variable cavity, considering the introduction of latent variables $z_{\hat{0}}(t)$ and $z_{\hat{0}'}(t)$. The cavity pictures are coupled, with averages from one appearing in self-consistent equations for the other.}
\label{fig:cavity_schematic}
\end{figure}

While the path-integral approach provides a systematic framework for analyzing neural-network dynamics, the cavity method offers an intuitive, complementary perspective. This section provides an overview of the cavity analysis for the random-mode model. While these approaches yield equivalent results, they differ in their methodology: The path-integral approach focuses on fluctuations around the saddle point in a field theory, whereas the cavity approach examines perturbations induced by a subset of held-out variables. Here, we compute the four-point function using a two-site cavity approach, noting that the two-point function (which, if desired, could be computed using a single-site cavity method) is the same as that of an i.i.d.\ network with appropriate $g_{\text{eff}}$. 

Leveraging \eref{eq:z_def}, we reformulate the network as a bipartite system of neurons and latent variables and extend the cavity calculation of Clark \textit{et al.}~\cite{clark2023dimension} to both groups. This leads to separate, mutually referential cavity pictures for each group, mirroring the two single-site path integrals (\secref{sec:path_integral_analysis}). As in the path-integral formalism, self-consistent expressions in one picture are defined using averages from the other. Unlike the path-integral approach, which derives $\Psi^\phi(\bm{\tau})$ using its time-by-time definition [\eref{eq:psi-def-time}], the cavity method derives this function using its neuron-by-neuron definition [\eref{eq:psi-def-nrn}].

The cavity method we employ, schematized in Fig.~\ref{fig:cavity_schematic}, has two key features that distinguish it from simpler cavity calculations. First, it uses a two-site structure: We consider the simultaneous removal of two neurons, or two latent variables, from the network, rather than just one. This allows us to study cross-correlations between neurons. This approach was previously used in Clark \textit{et al.}~\cite{clark2023dimension}. Second, it incorporates a bipartite structure: We treat both neurons and latent variables as dynamic objects, creating separate but mutually referential cavity pictures for each group. Such bipartite structure, without the two-site structure, has been used in cavity calculations in the Hopfield model \cite{mezard1989space, mezard2017mean, clark2025transient}. Note that these two features are distinct: The two-site aspect refers to the number of neurons removed in each cavity, while the bipartite aspect refers to the types of variables considered (neurons and latent variables).

To apply cavity techniques, we reformulate the network as a bipartite system of neurons and latent variables,
\begin{align}
    T[x_i](t) &= \sum_{a=1}^M  L_{ia} z_a(t), \label{eq:x_dynamics_bipartite} \\
    z_a(t) &= D_a \sum_{j=1}^N R_{ja} \phi_j(t),
    \label{eq:z_dynamics_bipartite}
\end{align}
as done in the path-integral approach. In the two-site cavity calculation, we keep track of various intermediate quantities to order $1/\sqrt{N}$, leading to an expression for $\Psi^\phi(\bm{\tau})$ accurate to leading order (order one), which we identify as $\Psi^\phi_\spt(\bm{\tau})$.

To distinguish between different types of variables and their indices, we use the following notation. Neurons are indexed by $i,j \in \{1,\ldots,N\}$ and latent variables by $a,b \in \{1,\ldots,M\}$, as usual. For the cavity variables, we introduce special indices: Neuronal cavity variables are indexed by $\mu, \nu \in \{0,0'\}$, while latent cavity variables are indexed by $\alpha, \beta \in \{\hat{0},\hat{0}'\}$.

\subsection{Neuronal cavity}
\label{sec:neuronal_cavity}

We begin by introducing two neurons, $x_{0}(t)$ and $x_{0'}(t)$. The leading-order ($1/\sqrt{N}$) effect on the latent variables is
\begin{equation}
    \delta z_a(t) = \int^t dt' \sum_{b=1}^M S^z_{a b}(t_1, t_2) D_b \sum_{\mu \in \{0, 0'\}} R_{\mu b} \phi_\mu(t'), \label{eq:delta_z}
\end{equation}
where $S^z_{ab}(t_1, t_2)$ is the response function of the latent variables. The dynamic equations for the cavity neurons are
\begin{equation}
    T[x_\mu](t) = \eta_\mu(t) + \sum_{\nu \in \{0, 0'\}} \int^t dt' F_{\mu\nu}(t_1, t_2) \phi_\nu(t'), \label{eq:x_cavity}
\end{equation}
where
\begin{align}
    \eta_\mu(t) &= \sum_{a=1}^M L_{\mu a} z_a(t), \label{eq:eta_mu} \\
    F_{\mu\nu}(t_1, t_2) &= \sqrt{N} \sum_{a,b=1}^M L_{\mu a} D_{b} R_{\nu b} S^z_{ab}(t_1, t_2). \label{eq:F_mu_nu}
\end{align}
Defining the cavity-field time-average cross-covariance as
\begin{equation}
    C^\eta_{00'}(\omega) = \sum_{a,b=1}^M L_{0a} L_{0'b} C^z_{a b}(\omega), \label{eq:C_eta}
\end{equation}
we obtain the following expression for the cross-covariance of the cavity units, up to order $1/\sqrt{N}$:
\begin{equation}
    C^\phi_{00'}(\omega) = |S_\spt^\phi(\omega)|^2 C^{\eta}_{00'}(\omega) + \frac{1}{\sqrt{N}}\left[ \left(S_\spt^\phi(\omega)F_{00'}(\omega)\right)^* + S_\spt^\phi(\omega)F_{0'0}(\omega) \right] C^\phi_{\spt}(\omega). \label{eq:C_phi}
\end{equation}
Our goal is to compute the parameter
\begin{equation}
    \psi^\phi(\bm{\omega}) = N \ldravg{C^\phi_{00'}(\omega_1)C^\phi_{00'}(\omega_2)}, \label{eq:psi_phi_def}
\end{equation}
where $\ldravg{\cdot}$ denotes an average over $\bm{L}$, $\bm{D}$, and $\bm{R}$. To evaluate this, we need to square and disorder average \eref{eq:C_phi}. This requires us to consider several two-frequency correlation functions, which we define as
\begin{align}
    \Gamma_{F_{00'} F_{00'}}(\bm{\omega}) &= \ldravg{F_{00'}(\omega_1) F_{00'}(\omega_2)}, \label{eq:Gamma_FF} \\
    \Gamma_{F^*_{00'} F_{0'0}}(\bm{\omega}) &= \ldravg{F^*_{00'}(\omega_1) F_{0'0}(\omega_2)}, \label{eq:Gamma_FF_star} \\
    \Gamma_{F^*_{00'} C^{\eta}_{00'}}(\bm{\omega}) &= \sqrt{N} \ldravg{F^*_{00'}(\omega_1) C^{\eta}_{00'}(\omega_2)}, \label{eq:Gamma_FC_eta} \\
    \Gamma_{C^{\eta}_{00'} C^{\eta}_{00'}}(\bm{\omega}) &= N \ldravg{C^{\eta}_{00'}(\omega_1) C^{\eta}_{00'}(\omega_2)}. \label{eq:Gamma_C_eta_C_eta}
\end{align}
These have all been scaled to be order one. We can evaluate these $\Gamma_{\cdots}(\bm{\omega})$ functions due to the independence of the couplings and dynamic variables in the expressions defining them, which is a consequence of the cavity construction. Of these, the only nonvanishing ones are
\begin{align}
    \Gamma_{C^{\eta}_{00'} C^{\eta}_{00'}}(\bm{\omega}) 
    &= \alpha \ldravg{C^z_{\hat{0}\hat{0}}(\omega_1) C^z_{\hat{0}\hat{0}}(\omega_2)} + \alpha^2 N \ldravg{C^z_{\hat{0}\hat{0}'}(\omega_1) C^z_{\hat{0}\hat{0}'}(\omega_2)}, \label{eq:Gamma_C_eta_C_eta_nonzero} \\
     \Gamma_{F_{00'} F_{00'}}(\bm{\omega})
     &= \alpha r_2 + \alpha^2 N \ldravg{D_{\hat{0}'}^2 S^z_{\hat{0}\hat{0}'}(\omega_1) S^z_{\hat{0}\hat{0}'}(\omega_2)},
     \label{eq:Gamma_FF_nonzero}
\end{align}
where $\hat{0},\hat{0}'$ denote the indices in the latent-variable cavity picture.

\subsection{Latent-variable cavity}
\label{sec:lv_cavity}

We now introduce two new latent variables, $z_{\hat{0}}(t)$ and $z_{\hat{0}'}(t)$. The leading-order ($1/\sqrt{N}$) effect on the neurons is
\begin{equation}
    \delta \phi_i(t) = \int^t dt' \sum_{j=1}^N S^\phi_{ij}(t_1, t_2) \sum_{\alpha \in \{\hat{0}, \hat{0}'\}} L_{j\alpha} z_\alpha(t'), \label{eq:delta_phi}
\end{equation}
where $S^\phi_{ij}(t_1, t_2)$ is the response function of the neurons. The dynamic equations for the cavity latent variables are
\begin{equation}
    z_\alpha(t) = D_\alpha \left[ \gamma_\alpha(t) + \frac{1}{\sqrt{N}} \sum_{\beta \in \{\hat{0}, \hat{0}'\}} \int^t dt' G_{\alpha\beta}(t_1, t_2) z_\beta(t')\right], \label{eq:z_cavity}
\end{equation}
where
\begin{align}
    \gamma_\alpha(t) &= \sum_{i=1}^N  R_{i\alpha}\phi_i(t), \label{eq:gamma_alpha} \\
    G_{\alpha\beta}(t_1, t_2) &= \sqrt{N}\sum_{i,j=1}^N R_{i\alpha}L_{j\beta}S^\phi_{ij}(t_1, t_2). \label{eq:G_alpha_beta}
\end{align}
Defining the time-average cavity-field cross-covariance as
\begin{equation}
    C^\gamma_{\hat{0}\hat{0}'}(\omega) = \sum_{i,j=1}^N  R_{i\hat{0}} R_{j\hat{0}'} C^\phi_{ij}(\omega), \label{eq:C_gamma}
\end{equation}
we obtain the following expression for the cross-covariance of the cavity latent variables up to order $1/\sqrt{N}$:
\begin{equation} 
    C^z_{\hat{0}\hat{0}'}(\omega) = D_{\hat{0}} D_{\hat{0}'} 
    C^\gamma_{\hat{0}\hat{0}'}(\omega) 
    + \frac{1}{\sqrt{N}}\left[ D_{\hat{0}} G^*_{\hat{0}\hat{0}'}(\omega)C^z_{\hat{0}'\hat{0}'}(\omega) + D_{\hat{0}'} G_{\hat{0}'\hat{0}}(\omega) C^z_{\hat{0}\hat{0}}(\omega)\right]. \label{eq:C_z}
\end{equation}
We aim to compute the parameter
\begin{equation}
    \psi^z(\bm{\omega}) = N \ldravg{C^z_{\hat{0}\hat{0}'}(\omega_1)C^z_{\hat{0}\hat{0}'}(\omega_2)} \label{eq:psi_z_def}
\end{equation}
by squaring and disorder averaging \eref{eq:C_z}. This requires us to consider several two-frequency correlation functions, which we define as
\begin{align}
    \Gamma_{G_{\hat{0}\hat{0}'} G_{\hat{0}\hat{0}'}}(\bm{\omega}) &= \ldrmavg{G_{\hat{0}\hat{0}'}(\omega_1)G_{\hat{0}\hat{0}'}(\omega_2)}, \label{eq:Gamma_GG} \\
    \Gamma_{G^*_{\hat{0}\hat{0}'} G_{\hat{0}'\hat{0}}}(\bm{\omega}) &= \ldrmavg{G^*_{\hat{0}\hat{0}'}(\omega_1) G_{\hat{0}'\hat{0}}(\omega_2)}, \label{eq:Gamma_GG_star} \\
    \Gamma_{G^*_{\hat{0}\hat{0}'} C^{\gamma}_{\hat{0}\hat{0}'}}(\bm{\omega}) &= \sqrt{N} \ldrmavg{G^*_{\hat{0}\hat{0}'}(\omega_1) C^{\gamma}_{\hat{0}\hat{0}'}(\omega_2)}, \label{eq:Gamma_GC_gamma} \\
    \Gamma_{C^{\gamma}_{\hat{0}\hat{0}'} C^{\gamma}_{\hat{0}\hat{0}'}}(\bm{\omega}) &= N \ldrmavg{C^{\gamma}_{\hat{0}\hat{0}'}(\omega_1) C^{\gamma}_{\hat{0}\hat{0}'}(\omega_2)}. \label{eq:Gamma_C_gamma_C_gamma}
\end{align}
Again, we can evaluate these $\Gamma_{\cdots}(\bm{\omega})$ functions due to the independence of the couplings and dynamic variables in the expressions defining them. The nonvanishing ones are
\begin{align}
    \Gamma_{C^{\gamma}_{\hat{0}\hat{0}'} C^{\gamma}_{\hat{0}\hat{0}'}}(\bm{\omega}) &= C^\phi_\spt(\omega_1)C^\phi_\spt(\omega_2)  + N\ldravg{C^\phi_{00'}(\omega_1)C^\phi_{00'}(\omega_2)}, \label{eq:Gamma_C_gamma_C_gamma_nonzero} \\
    \Gamma_{G_{\hat{0}\hat{0}'} G_{\hat{0}\hat{0}'}}(\bm{\omega})
    &=  S^\phi_\spt(\omega_1)S^\phi_\spt(\omega_2)  + N\ldravg{S^\phi_{00'}(\omega_1)S^\phi_{00'}(\omega_2)}, \label{eq:Gamma_GG_nonzero}
\end{align}
where $0,0'$ denote the indices in the neuronal cavity picture.

\subsection{Combining the two two-site cavity pictures}
\label{sec:combining_two_two_site_pics}

To combine the results from both two-site cavity pictures, we use the following relations:
\begin{align}
    C^z_{\hat{0}\hat{0}}(\omega) &= D_{\hat{0}}^2 C^\phi_\spt(\omega), \label{eq:C_z00} \\
    S^z_{\hat{0}\hat{0}'}(\omega) &= \frac{D_{\hat{0}}}{\sqrt{N}} G_{\hat{0}\hat{0}'}(\omega), \label{eq:S_z00'} \\
    S^\phi_{00'}(\omega) &= \frac{1}{\sqrt{N}}[S^\phi_\spt(\omega)]^2 F_{00'}(\omega). \label{eq:S_phi00'}
\end{align}
Using these relations and the definitions of the $\Gamma_{\cdots}(\bm{\omega})$ functions [Eqs.~\eqref{eq:Gamma_FF}--\eqref{eq:Gamma_C_eta_C_eta} and \eqref{eq:Gamma_GG}--\eqref{eq:Gamma_C_gamma_C_gamma}], we obtain
\begin{align}
    \Gamma_{C^{\eta}_{00'} C^{\eta}_{00'}}(\bm{\omega}) 
    &= \alpha r_4 C_\spt^\phi(\omega_1)C_\spt^\phi(\omega_2)
    + \alpha^2 \psi^z(\bm{\omega}), \label{eq:Gamma_C_eta_C_eta_q} \\
     \Gamma_{F_{00'} F_{00'}}(\bm{\omega})
     &= \alpha r_2 \left[1 + \alpha r_2 \Gamma_{G_{\hat{0}\hat{0}'}G_{\hat{0}\hat{0}'} }(\bm{\omega}) \right], \label{eq:Gamma_FF_q} \\
    \Gamma_{C^{\gamma}_{\hat{0}\hat{0}'} C^{\gamma}_{\hat{0}\hat{0}'}}(\bm{\omega}) &=  C_\spt^\phi(\omega_1)C_\spt^\phi(\omega_2)  + \psi^\phi(\bm{\omega}), \label{eq:Gamma_C_gamma_C_gamma_q} \\
    \Gamma_{G_{\hat{0}\hat{0}'} G_{\hat{0}\hat{0}'}}(\bm{\omega})
    &= S_\spt^\phi(\omega_1)S_\spt^\phi(\omega_2)  \left[ 1 + S_\spt^\phi(\omega_1)S_\spt^\phi(\omega_2) \Gamma_{F_{00'}F_{00'}}(\bm{\omega}) \right]. \label{eq:Gamma_GG_q}
\end{align}
We solve Eqs.~\eqref{eq:Gamma_FF_q} and \eqref{eq:Gamma_GG_q} simultaneously, yielding
\begin{align}
    \Gamma_{F_{00'} F_{00'}}(\bm{\omega}) &= \frac{\alpha r_2}{1 - \alpha r_2 S^\phi_\spt(\omega_1)S^\phi_\spt(\omega_2)}, \label{eq:Gamma_FF_sol} \\
    \Gamma_{G_{\hat{0}\hat{0}'} G_{\hat{0}\hat{0}'}}(\bm{\omega}) &= \frac{S^\phi_\spt(\omega_1)S^\phi_\spt(\omega_2)}{1 - \alpha r_2 S^\phi_\spt(\omega_1)S^\phi_\spt(\omega_2)}. \label{eq:Gamma_GG_sol}
\end{align}
With these solutions, we express $\psi^z_\spt(\bm{\omega})$ and $\psi^\phi_\spt(\bm{\omega})$ in terms of the $\Gamma_{\cdots}(\bm{\omega})$ functions. Switching to frequency-suppressed notation,
\begin{align}
    \psi^z_\spt &= r_2^2\Gamma_{C^\gamma_{\hat{0}\hat{0}'}C^\gamma_{\hat{0}\hat{0}'}} + r_2 r_4 \left(\Gamma_{G_{\hat{0}\hat{0}'}G_{\hat{0}\hat{0}'}} + \text{H.c.}\right)C^\phi_{12}, \label{eq:psi_z} \\
    \psi^\phi_\spt &= |S^\phi_{12}|^2 \Gamma_{C^\eta_{00'}C^\eta_{00'}} + \left(\Gamma_{F_{00'}F_{00'}}S^\phi_{12} + \text{H.c.}\right)C^\phi_{12}. \label{eq:psi_phi}
\end{align}
Now, we substitute \eref{eq:psi_z} into \eref{eq:psi_phi}. Using $\Psi^\phi_\spt = \psi^\phi_\spt + C^\phi_{12}$, we obtain
\begin{equation} 
    \Psi^\phi_\spt = |g_{\text{eff}}^2 S^\phi_{12}|^2 \left[\Psi^\phi_\spt + \frac{1}{\alpha\text{PR}^D} \frac{1-|g_{\text{eff}}^2 S^\phi_{12}|^2}{|1-g_{\text{eff}}^2 S^\phi_{12}|^2}C^\phi_{12} \right]  + \frac{1-|g_{\text{eff}}^2 S^\phi_{12}|^2}{|1-g_{\text{eff}}^2 S^\phi_{12}|^2}C^\phi_{12},
    \label{eq:Psi_phi_intermediate}
\end{equation}
where we used the previously defined quantities $g_{\text{eff}}^2 = \alpha r_2$ and $\text{PR}^D = r_2^2 / r_4$. \Eref{eq:Psi_phi_intermediate} can be solved for $\Psi^\phi_\spt$, yielding
\begin{equation}
    \Psi^\phi_\spt = \frac{1 + \frac{1}{\alpha \text{PR}^D} |g_{\text{eff}}^2 S^\phi_{12}|^2}{|1 - g_{\text{eff}}^2 S^\phi_{12}|^2} C^\phi_{12}. \label{eq:Psi_phi_final}
\end{equation}
This result is identical to \eref{eq:path-int-result-rmm} derived using the path-integral approach, demonstrating the consistency between the two methods.

\section{Numerical details}
\label{sec:numerical_details}

\subsection{Numerics for theory}
\label{sec:numerics_for_th}

To validate the theory, we used networks with dynamics defined by $T[x](t) = (1 + \partial_t)x(t)$ and with nonlinearity $\phi(x) = \text{erf}\left({\sqrt{\pi}x}/{2}\right)$. For all connectivity models, except those with structured $\bm{L}$-$\bm{R}$ overlaps (\secref{sec:LR_correlations}), the single-site two-point functions $C^\phi_\spt(\tau)$ and $S^\phi_\spt(\tau)$ depend only on $g_\text{eff}$ and can be numerically calculated as in i.i.d.\ networks (see, e.g., Refs.~\cite{sompolinsky1988chaos, mastrogiuseppe2017dynamics}). In summary:
\begin{enumerate}
    \item Since $\eta(t)$ is Gaussian, so is $x(t)$ for this linear form of $T[\cdot]$, allowing us to write $C^\phi_\spt(\tau)$ in terms of Gaussian integrals over $x(t)$ and $x(t+\tau)$, which have marginal variance $C^\phi_\spt(0)$ and covariance $C^\phi_\spt(\tau)$. Because of our choice of $\phi(x)$, this expression simplifies to
    \begin{align}
        &C^\phi_\spt(\tau) = \frac{2}{\pi}\tan^{-1}\left[\frac{C^x_\spt(\tau)}{ \sqrt{\left(C^x_\spt(0) + \frac{2}{\pi}\right)^2 - C^x_\spt(\tau)^2}  }\right].\label{eq:c_phi_c_x}
    \end{align}
    \item Squaring the single-site picture gives a second-order ordinary differential equation (ODE) $\partial^2_\tau C^x_\spt(\tau) = C^x_\spt(\tau) - g_\text{eff}^2 C^\phi_\spt(\tau)$. Since the rhs depends only on $C^x_\spt(\tau)$ and $C^x_\spt(0)$, we can consider it to be the negative derivative $-\partial V /\partial C^x_\spt(\tau)$ of a potential $V(C^x_\spt(\tau), C^x_\spt(0))$ with an explicit dependence on the initial condition $C^x_\spt(0)$.
    \item The rhs is integrated with respect to $C^x_\spt(\tau)$ to provide an expression for $V(C^x_\spt(\tau), C^x_\spt(0))$, which due to our choice of $\phi(x)$ simplifies to
    \begin{align}
        &V(C^x_\spt(\tau), C^x_\spt(0)) = -\frac{1}{2}C^x_\spt(\tau)^2 + g_
        \text{eff}^2 \left[ \frac{2}{\pi} \sqrt{\left(C^x_\spt(0) + \frac{2}{\pi}\right)^2 - C^x_\spt(\tau)^2} + C^x_\spt(\tau) C^\phi_\spt(\tau)\right].
    \end{align}
    \item Restricting to solutions with $C^x_\spt(\tau) \to 0$ as $\tau \to \infty$ and $\partial_\tau C^x_\spt(\tau)|_{\tau=0} = 0$, enforcing conservation of energy gives $V(C^x_\spt(0), C^x_\spt(0)) = V(0, C^x_\spt(0))$, which can be solved for $C^x_\spt(0)$ by numerical root finding.
    \item Finally, Euler integration of the $[C^x_\spt(\tau),\partial_\tau C^x_\spt(\tau)]$ dynamics with these initial conditions gives its full time course, along with $C^\phi_\spt(\tau)$ via \eref{eq:c_phi_c_x}.
\end{enumerate}

We integrated the Newtonian ODE from $\tau = 0$ to $\tau_\text{max} = 200$ with step size $d\tau = 0.025$. We then use an FFT to obtain the frequency-space representation. The frequency-space representation of the response function $S^\phi_\spt(\omega)$ was computed directly as
\begin{equation}
    S^\phi_\spt(\omega) = \frac{\tavg{\phi'}_\spt}{1 + i\omega},
\end{equation}
with $\tavg{\phi'}_\spt = 1/\sqrt{1 + (\pi/2)C^x_\spt(0)}$. Various expressions for $\Psi^a_\spt(\omega_1, \omega_2)$ were then computed in frequency space before transforming back to the time domain by a two-dimensional inverse FFT.

For structured $\bm{L}$-$\bm{R}$ overlaps, the $\mathcal{O}(1)$ self-coupling in the single-site process makes $x(t)$ non-Gaussian, preventing the use of Gaussian integrals in the DMFT solution. We therefore obtain the solution via standard numerical methods, enforcing self-consistency among the $\eta^x(t)$ correlation function $Q_\spt(\tau)$, the self-coupling kernel $R_\spt(\tau)$, the response function $S^\phi_\spt(\tau)$, and the $\phi(t)$ correlation function $C^\phi_\spt(\tau)$. In summary:
\begin{enumerate}
    \item Seed values for kernels are set as $C^\phi_0(\tau) = \exp[-(\tau/15)^2] + 10^{-4}\delta(\tau)$ and $S^\phi_0(\tau) = 0$.
    \item $Q_0(\tau)$ and  $R_0(\tau)$ are computed in Fourier space using the current values of $C^\phi_0(\tau)$ and $S^\phi_0(\tau)$ via \eref{eq:sym_R_expr}, with the average taken over $(D,\rho)$.
    \item A set of $N_\text{sample} = 10^4$ instances of $\eta^x$ are sampled according to the Gaussian process kernel $Q_0(\tau)$ by standard spectral methods.
    \item Each of $N_\text{sample}$ single-site processes $x_s(t)$ is Euler integrated according to \eref{eq:sym_ssp} with the input $\eta_s^x(t)$ and self-feedback by the convolution between $R_0(\tau)$ and $\phi\bm{(}x_s(t)\bm{)}$.
    \item These time series $\phi\bm{(}x_s(t)\bm{)}$ and $\eta^x_s(t)$ are used to compute empirical values $(C^\phi_1(\tau), S^\phi_1(\tau))$ for the autocovariance and response. The latter is computed by the Furutsu-Novikov theorem: $S^\phi(\omega) = C^{\phi \eta}(
    \omega)/C^\eta(\omega)$. The numerator is estimated using these time series, while $C^\eta(\omega)$ is simply $Q_0(\omega)$.
    \item After $N_\text{iter} = 300$ repetitions of steps (1)-(5) with smoothed updates
    \begin{align}
    &C^\phi_m(\tau) \leftarrow 0.2C^\phi_{m-1}(\tau) + 0.8C^\phi_m(\tau)\\
    &S^\phi_m(\tau) \leftarrow 0.2S^\phi_{m-1}(\tau) + 0.8S^\phi_m(\tau)
    \end{align}
    as new seed values for the upcoming $m$th iteration, the final numerical estimates for $C^\phi_\spt(\tau)$ and $S^\phi_\spt(\tau)$ are computed as the averages over the last $50$ iterations.
\end{enumerate}
We used a temporal discretization $\tau_\text{max} = 120$, $d\tau = 0.04$. With these quantities, the various components of \eref{eq:lr_psi_result} and the expression for $\Psi^\phi_\spt$ are straightforward to compute. We took $\tavg{\cdot}_{D,\rho}$ averages in this case by averaging over every other discrete value of $D_a$ and $\rho_a$.

\subsection{Simulation details}
\label{sec:sim_details}

Each network was integrated via Runge-Kutta with $dt = 0.05$. The empirical lagged covariance matrix was computed by averaging over $N_\text{loop}$ individual covariance estimates, each of which is based on $T = 2000$ discrete, evenly spaced time samples $t^{i_\text{loop}}_k$, with $t^{i_\text{loop}}_k - t^{i_\text{loop}}_{k-1} = 1$,
\begin{equation}
    \widehat{C^\phi_{ij}(\tau)} = \tavg{\frac{1}{T-1}\sum_{k=1}^T \phi_i(t^{i_\text{loop}}_k)\phi_j(t^{i_\text{loop}}_k - \tau)}_{i_\text{loop}=1}^{N_\text{loop}}.
\end{equation}
The value of $N_\text{loop}$ varied but was generally $\geq 50$ and always chosen to ensure saturated dimension values, i.e.\ $TN_\text{loop} \gg N$. Specific quantities of interest, such as $\text{PR}^\phi$ or $\Psi^\phi(\tau, 0)$, were computed using this estimate.

\end{widetext}

\bibliography{refs}

\end{document}